\documentclass[twocolumn,prd,superscriptaddress,preprintnumbers,nofootinbib]{revtex4}[11pt]
\pdfoutput=1
\usepackage{amsmath,amssymb,graphicx}
\graphicspath{{figs/}}
\usepackage{epsf,verbatim}
\usepackage{hyperref}
\usepackage{comment}
\usepackage{color}
\usepackage{slashed}
\usepackage{subfigure}
\usepackage[usenames,dvipsnames]{xcolor}
\usepackage{comment}
\usepackage{mdwlist, paralist}
\usepackage{rotating}
\usepackage{multirow}

\usepackage[utf8]{inputenc}

\usepackage{url}

\newcommand{\Br}{{\mathrm{Br}}}
\newcommand{\bwt}{\begin{widetext}}
\newcommand{\ewt}{\end{widetext}}

\def\to{\rightarrow}
\def\ord{\mathcal{O}}
\def\TeV{~{\mbox{TeV}}}

\def\fb{~{\mbox{fb}}}
\def\GeV{~{\mbox{GeV}}}

\def\mL{\mathcal{L}}

\def\TOT{\text{TOT}}

\def\mt{\text{matter}}

\begin{document}

\title{Phenomenology of A Little Higgs Pseudo-Axion}

\preprint{NCTS-PH/1812}

\affiliation{Physics Division, National Center for Theoretical Sciences, Hsinchu, Taiwan 300}
\affiliation{Department of Physics, National Tsing Hua University, Hsinchu 300, Taiwan}
\affiliation{Division of Quantum Phases and Devices, School of Physics, Konkuk University, Seoul 143-701, Republic of Korea}
\affiliation{Institute of Theoretical Physics \& State Key Laboratory of Nuclear Physics and Technology, Peking University, Beijing 100871, China}
\affiliation{Center for Future High Energy Physics \& Theoretical Physics Division,
Institute of High Energy Physics, Chinese Academy of Sciences, Beijing 100049, China}
\affiliation{Kavli IPMU (WPI), UTIAS, The University of Tokyo, Kashiwa, Chiba 277-8583, Japan}

\author{Kingman Cheung}
\thanks{cheung@phys.nthu.edu.tw}
\affiliation{Physics Division, National Center for Theoretical Sciences, Hsinchu, Taiwan 300}
\affiliation{Department of Physics, National Tsing Hua University, Hsinchu 300, Taiwan}
\affiliation{Division of Quantum Phases and Devices, School of Physics, Konkuk University, Seoul 143-701, Republic of Korea}

\author{Shi-Ping He}
\thanks{sphe@pku.edu.cn}
\affiliation{Institute of Theoretical Physics \& State Key Laboratory of Nuclear Physics and Technology, Peking University, Beijing 100871, China}

\author{Ying-nan Mao}
\thanks{ynmao@cts.nthu.edu.tw}
\affiliation{Physics Division, National Center for Theoretical Sciences, Hsinchu, Taiwan 300}
\affiliation{Center for Future High Energy Physics \& Theoretical Physics Division,
Institute of High Energy Physics, Chinese Academy of Sciences, Beijing 100049, China}

\author{Po-Yan Tseng}
\thanks{poyen.tseng@ipmu.jp}
\affiliation{Kavli IPMU (WPI), UTIAS, The University of Tokyo, Kashiwa, Chiba 277-8583, Japan}

\author{Chen Zhang}
\thanks{czhang@cts.nthu.edu.tw}
\affiliation{Physics Division, National Center for Theoretical Sciences, Hsinchu, Taiwan 300}

\begin{abstract}
In models where the Higgs is realized as a pseudo-Nambu-Goldstone boson (pNGB) of some global
symmetry breaking, there are often remaining pNGBs of some $U(1)$ groups (called ``pseudo-axions''), which could lead to smoking gun
signatures of such scenarios and provide important clues on the electroweak symmetry breaking mechanism.
As a concrete example, we investigate the phenomenology of the pseudo-axion in the anomaly-free Simplest
Little Higgs (SLH) model. After clarifying a subtle issue related to the effect of symmetric vector-scalar-scalar (VSS) vertices (e.g. $Z_\mu(H\partial^\mu\eta+\eta\partial^\mu H)$), we show that for natural region in the parameter space, the SLH pseudo-axion is top-philic, decaying almost exclusively to a pair of top quarks. The direct and indirect (i.e. via heavy particle decay) production of such a pseudo-axion at the $14\TeV$ (HL-)LHC turn out to suffer from either large backgrounds or small rates, making its detection quite challenging. A $pp$ collider with higher energy and luminosity, such as the $27\TeV$ HE-LHC, or even the $100\TeV$ FCC-hh or SppC, is therefore motivated to capture the trace of such a pNGB.
\end{abstract}

\maketitle

\setcounter{equation}{0} \setcounter{footnote}{0}

\section{Introduction}
\label{sec:intro}

Despite the great success of the Standard Model (SM), marked by the discovery
of the $125\GeV$ Higgs-like boson~\cite{Aad:2012tfa,Chatrchyan:2012xdj} and
the on-going measurements of its properties, how the SM is embedded into a
larger theory still remains a mystery. Since the Higgs boson mass parameter
is in general not protected under radiative correction, a naive embedding would signal
a high sensitivity of infrared (IR) parameters (the electroweak scale and the Higgs boson mass)
to ultraviolet (UV) parameters (i.e. physical parameters defined at a high scale). Although this fine-tuned
situation is logically possible, or might be explained to some extent by
anthropic reasoning~\cite{Schellekens:2013bpa,Donoghue:2016tjk}, it is nevertheless
natural to conjecture the existence of some systematic mechanism which protects the
Higgs boson mass parameter from severe radiative instability. A well-known example
of such systematic mechanism is supersymmetry, which has the merit of being weakly-coupled
and thus offers better calculability compared to scenarios based on strong dynamics.
However supersymmetry requires the introduction of a large number of new degrees of
freedom, and a large number of new parameters associated with them, making the model
quite cumbersome. None of the new degrees of freedom have been observed. It is therefore
well-motivated to consider alternative but simpler mechanisms with weakly-coupled
dynamics in their range of validity.

One candidate of such alternative is the Little Higgs mechanism~\cite{ArkaniHamed:2001nc,
ArkaniHamed:2002pa,ArkaniHamed:2002qx,ArkaniHamed:2002qy}\footnote{We refer the reader
to ref.~\cite{Schmaltz:2005ky,Perelstein:2005ka} for reviews of Little Higgs
models and ref.~\cite{Dercks:2018hgz,Reuter:2012sd,Reuter:2013iya,Han:2013ic}
for some recent phenomenological analyses of Little Higgs models.}, in which the Higgs boson
is a Goldstone boson of some spontaneous global symmetry breaking. The global symmetry is
also explicitly broken in a collective manner\footnote{More specifically, the global symmetry is
completely (explicitly) broken by a collection of spurions but not by any single spurion~\cite{ArkaniHamed:2002qy}.}
such that the Higgs boson acquires a mass and at the same time the model is radiatively more stable.
A very simple implementation of this collective symmetry breaking (CSB) idea is the Simplest
Little Higgs (SLH) model~\cite{Kaplan:2003uc,Schmaltz:2004de}, in which the electroweak gauge group
is enlarged to $SU(3)_L\times U(1)_X$, and two scalar triplets are introduced to realize
the global symmetry breaking pattern
\begin{align}
& [SU(3)_1\times U(1)_1]\times[SU(3)_2\times U(1)_2] \nonumber \\
& \to[SU(2)_1\times U(1)_1]\times[SU(2)_2\times U(1)_2]
\label{eq:gsb}
\end{align}
The global symmetry is also explicitly broken by gauge and Yukawa interactions, but in a
collective manner to improve the radiative stability of the scalar sector. The particle
content is quite economical. Especially in the low energy scalar sector, there exists only
two physical degrees of freedom, one of which (denoted $H$) could be identified with the $125\GeV$ Higgs-like
particle, while the other is a CP-odd scalar $\eta$ which is referred to as a pseudo-axion
in the literature~\cite{Kilian:2004pp,Kilian:2006eh}.

In the SLH, the pseudo-axion $\eta$ is closely related to the electroweak symmetry
breaking (EWSB) and therefore studying its phenomenology is well-motivated. According to
the hidden mass relation derived in ref.~\cite{Cheung:2018iur}, the $\eta$ mass $m_\eta$
is anti-correlated with the top partner mass $m_T$, which is
in turn related to the degree of fine-tuning in the model. The hidden mass relation is derived
within an approach consistent with the continuum effective field theory (CEFT) and does not rely
on the assumption on the contribution from the physics at the cutoff. Although phenomenology of
the $\eta$ particle has been studied by quite a few papers (e.g.~\cite{Kilian:2004pp,Kilian:2006eh,
Cheung:2006nk,Cheung:2008zu,Han:2013ic}), their treatment was not based on the hidden mass relation,
and also most of the papers were written before the $125\GeV$ boson was discovered. It is thus timely to revisit
the status of $\eta$ phenomenology in light of the discovery of the $125\GeV$ boson, taking into
account the properly derived hidden mass relation and focusing on the parameter space favored by
naturalness considerations.

There is another important reason that warrants a reanalysis of the $\eta$ phenomenology. The SLH
is usually written as a gauged nonlinear sigma model, in which the EWSB can be parametrized
through vacuum misalignment. However, the vacuum misalignment also leads to the fact that,
in the usual parametrization of the two scalar triplets, there exist scalar kinetic terms
that are not canonically-normalized, and vector-scalar two-point transitions that are
``unexpected''~\cite{He:2017jjx}. A further field rotation, including an appropriate gauge-fixing
procedure, is thus required to properly diagonalize the vector-scalar sector of the SLH model.
This subtlety had been overlooked in all related papers before ref.~\cite{He:2017jjx}, and if
one carries out a proper diagonalization of the bosonic sector of the SLH, some of the $\eta$-related
couplings will turn out to be different from what has been obtained in previous literature.
This is the case for both the $ZH\eta$ coupling and the coupling of $\eta$ to a pair of SM fermions.
The occurrence of the mass eigenstate antisymmetric $ZH\eta$ vertex
(i.e. $Z_\mu(H\partial^\mu\eta-\eta\partial^\mu H)$)  is postponed to $\ord(\xi^3)$
(with $\xi\equiv\frac{v}{f}$, $v\approx 246\GeV$ and $f$ is the global symmetry breaking scale
of Eq.~\eqref{eq:gsb}), and the couplings of $\eta$ to a pair of SM charged leptons, and to
$b\overline{b},c\overline{c},u\overline{u}$ are found to vanish to all order in $\xi$. This leads
to significant changes in the $\eta$ phenomenology, which will be studied in detail in this work.

When one tries to derive the $\eta$-related Lagrangian in the SLH, symmetric vector-scalar-scalar (VSS) vertices,
e.g. $Z_\mu(H\partial^\mu\eta+\eta\partial^\mu H)$ naturally appear, which is a feature that is often
present in models based on a nonlinearly-realized scalar sector. The effects of such symmetric VSS
vertices contain some subtleties which, to our knowledge, have not been discussed before in literature.
Therefore, we devote one section to the analysis of symmetric VSS vertices, which could also be helpful
to clarify similar situations in other nonlinearly-realized models.

In this work we do not aim to give a complete characterization of the $\eta$ phenomenology, which could
be very complicated in certain corner of parameter space. Instead, we focus our attention on the parameter
space favored by naturalness considerations. More specifically, we will consider $\eta$ mass in the
region $2m_t\lesssim m_\eta\lesssim 1\TeV$, which is favored by naturalness. We then calculate the $\eta$ decay and production
at future high energy hadron colliders in various channels. It turns out at the $14\TeV$ (HL)-LHC the detection of $\eta$ is quite
challenging due to various suppression mechanisms. A $pp$ collider with higher energy and luminosity, such as the $27\TeV$ HE-LHC, or even the $100\TeV$ FCC-hh or SppC, is therefore motivated to capture the trace of such a pNGB.

The paper is organized as follows. In Section~\ref{sec:slh} we review the basic ingredients of
the SLH, including the crucial hidden mass relation obtained from a CEFT analysis, and present
the mass eigenstate Lagrangian relevant for phenomenological studies. In Section~\ref{sec:vss}
we clarify the effect of symmetric VSS vertices. Then in Section~\ref{sec:ewpt} we derive important
constraints from electroweak precision observables relevant for the pseudo-axion phenomenology.
Section~\ref{sec:prod} is dedicated to the study of $\eta$ decay and production at hadron colliders.
In Section~\ref{sec:dnc} we present the discussion and conclusions.

\section{The Simplest Little Higgs}
\label{sec:slh}

\subsection{Overview of the Simplest Little Higgs}

In the SLH, the electroweak gauge group is enlarged to $SU(3)_L\times U(1)_X$. Two scalar
triplets $\Phi_1,\Phi_2$ are introduced to realize the spontaneous global symmetry breaking
pattern in Eq.~\eqref{eq:gsb}. They are parameterized as
\begin{align}
\Phi_1=\exp\left(\frac{i\Theta'}{f}\right)
\exp\left(\frac{it_\beta\Theta}{f}\right)
\begin{pmatrix}
0 \\ 0 \\ fc_\beta
\end{pmatrix} \label{eq:phi1} \\
\Phi_2=\exp\left(\frac{i\Theta'}{f}\right)
\exp\left(-\frac{i\Theta}{ft_\beta}\right)
\begin{pmatrix}
0 \\ 0 \\ fs_\beta
\end{pmatrix} \label{eq:phi2}
\end{align}
Here we have introduced the shorthand notation
$s_\beta\equiv\sin\beta,c_\beta\equiv\cos\beta,t_\beta\equiv\tan\beta$.
$f$ is the Goldstone decay constant. $\Theta$ and $\Theta'$ are $3\times 3$ matrix fields,
parameterized as
\begin{align}
\Theta=\frac{\eta}{\sqrt{2}}+
\begin{pmatrix}
\textbf{0}_{2\times 2} & h \\
h^\dagger & 0
\end{pmatrix},\quad
\Theta'=\frac{\zeta}{\sqrt{2}}+
\begin{pmatrix}
\textbf{0}_{2\times 2} & k \\
k^\dagger & 0
\end{pmatrix}
\label{eq:theta}
\end{align}
$\eta$ is the pseudo-axion, and $h$ and $k$ are parameterized as ($v$ denotes
the vacuum expectation value (vev) of the Higgs doublet)
\begin{align}
h & =\begin{pmatrix} h^0 \\ h^- \end{pmatrix},\quad
h^0=\frac{1}{\sqrt{2}}(v+H-i\chi) \label{eq:hdoub} \\
k & =\begin{pmatrix} k^0 \\ k^- \end{pmatrix},\quad
k^0=\frac{1}{\sqrt{2}}(\sigma-i\omega) \label{eq:kdoub}
\end{align}
For future convenience, we introduce the notation
\begin{equation}
\hat{h}\equiv(h^\dagger h)^{1/2}
\end{equation}
We note that the spontaneous global symmetry breaking Eq.~\eqref{eq:gsb}
should deliver 10 Goldstone bosons, which are parameterized here
in $\Theta$ and $\Theta'$. The electroweak gauge group $SU(3)_L\times U(1)_X$
will eventually break to $U(1)_{EM}$, and therefore 8 Goldstone
bosons will be eaten to make the associated gauge bosons massive.
Only two Goldstone bosons remain physical, parameterized here as
$h$ and $\eta$. The parametrization of these Goldstone fields
actually has some freedom, which we refer the reader to ref.~\cite{Cheung:2018iur}
for explanation.

In the SLH, under the full gauge group $SU(3)_C\times SU(3)_L\times U(1)_X$,
$\Phi_1$ and $\Phi_2$ have quantum number $(\textbf{1},\textbf{3})_{-\frac{1}{3}}$.
The gauge kinetic term of $\Phi_1$ and $\Phi_2$ can thus be written as\footnote{
We note that Eq.~\eqref{eq:lgk} automatically satisfies the requirement of CSB.}
\begin{equation}
\mL_{gk}=(D_\mu\Phi_1)^\dagger(D^\mu\Phi_1)+
(D_\mu\Phi_2)^\dagger(D^\mu\Phi_2)
\label{eq:lgk}
\end{equation}
in which the covariant derivative can be expressed as
\footnote{In this paper our convention agrees with Ref.~\cite{delAguila:2011wk}
but differs from Ref.~\cite{Han:2005ru}. The conversion between the two
conventions is discussed in Appendix~\ref{sec:cc}.}
\begin{equation}
D_\mu=\partial_\mu-igA_\mu^a T^a+ig_xQ_xB_\mu^x,\quad
g_x=\frac{gt_W}{\sqrt{1-t_W^2/3}}
\end{equation}
In the above equation, $A_\mu^a$ and $B_\mu^x$ denote $SU(3)_L$
and $U(1)_X$ gauge fields, respectively. $g$ and $g_x$ denote the coupling
constants of $SU(3)_L$ and $U(1)_X$ gauge groups, respectively.
It is convenient to trade $g_x$ for $t_W\equiv\tan\theta_W$. $T^a=\frac{\lambda^a}{2}$ where
$\lambda^a,a=1,...,8$ denote the Gell-Mann matrices. For $\Phi_1,\Phi_2$,
$Q_x=-\frac{1}{3}$. Following ref.~\cite{delAguila:2011wk}, we
parameterize the $SU(3)_L$ gauge bosons as
\begin{align}
A_\mu^a T^a &=\frac{A_\mu^3}{2}
\begin{pmatrix}
1 & 0 & 0 \\
0 & -1 & 0 \\
0 & 0 & 0
\end{pmatrix} \nonumber \\
& +\frac{A_\mu^8}{2\sqrt{3}}
\begin{pmatrix}
1 & 0 & 0 \\
0 & 1 & 0 \\
0 & 0 & -2
\end{pmatrix}
+\frac{1}{\sqrt{2}}
\begin{pmatrix}
0 & W_\mu^+ & Y_\mu^0 \\
W_\mu^- & 0 & X_\mu^- \\
Y_\mu^{0\dagger} & X_\mu^+ & 0
\end{pmatrix}
\end{align}
with the \textit{first-order} neutral gauge boson mixing relation ($c_W\equiv\cos\theta_W,s_W\equiv\sin\theta_W$)
\begin{align}
\begin{pmatrix}
A^3 \\ A^8 \\ B^x
\end{pmatrix}
=
\begin{pmatrix}
0 & c_W & -s_W \\
\sqrt{1-\frac{t_W^2}{3}} & \frac{s_W t_W}{\sqrt{3}} & \frac{s_W}{\sqrt{3}} \\
-\frac{t_W}{\sqrt{3}} & s_W\sqrt{1-\frac{t_W^2}{3}} & c_W\sqrt{1-\frac{t_W^2}{3}}
\end{pmatrix}
\begin{pmatrix}
Z' \\ Z \\ A
\end{pmatrix}
\label{eq:gbmixing}
\end{align}

Since the electroweak gauge group is enlarged to $SU(3)_L\times U(1)_X$,
it is also necessary to enlarge the fermion sector in order that fermions
transform properly under the enlarged group. We adopt the elegant anomaly-free
embedding proposed in ref.~\cite{Kong:2003vm,Schmaltz:2004de,Kong:2004cv}.
In the lepton Yukawa sector, the SM left-handed lepton doublets
are enlarged to $SU(3)_L$ triplets $L_m=(\nu_L,\ell_L,iN_L)_m^T$
with $Q_x=-\frac{1}{3}$ ($m=1,2,3$ is the family index). There
are also right-handed singlet lepton fields $\ell_{Rm}$ with $Q_x=-1$
and $N_{Rm}$ with $Q_x=0$. The lepton Yukawa Lagrangian can
be written as~\cite{delAguila:2011wk}
\begin{equation}
\mL_{LY}=i\lambda_N^m\bar{N}_{Rm}\Phi_2^\dagger L_m
+\frac{i\lambda_\ell^{mn}}{\Lambda}\bar{\ell}_{Rm}\epsilon_{ijk}\Phi_1^i\Phi_2^j L_n^k
+\text{h.c.}
\label{eq:lly}
\end{equation}
In the quark sector, we have the following field content
\begin{align}
Q_1 & =(d_L,-u_L,iD_L)^T, \quad d_R,\quad u_R,\quad D_R \\
Q_2 & =(s_L,-c_L,iS_L)^T, \quad s_R,\quad c_R,\quad S_R \\
Q_3 & =(t_L,b_L,iT_L)^T, \quad t_R,\quad b_R,\quad T_R
\end{align}
Here $Q_1,Q_2$ transform under $\bar{\textbf{3}}$ representation
of $SU(3)_L$ with $Q_x=0$. $Q_3$ transforms under $\textbf{3}$
representation of $SU(3)_L$ with $Q_x=\frac{1}{3}$. The right-handed
quark fields are all $SU(3)_L$ singlets with various $U(1)_X$ charges.
More specifically, $u_R,c_R,t_R,T_R$ carry $Q_x=\frac{2}{3}$ while
$d_R,s_R,b_R,D_R,S_R$ carry $Q_x=-\frac{1}{3}$. The quark
Yukawa Lagrangian can be written as~\cite{delAguila:2011wk}
\begin{align}
\mL_{QY} & = i\lambda_1^t\bar{u}_{R3}^1\Phi_1^\dagger Q_3
+i\lambda_2^t\bar{u}_{R3}^2\Phi_2^\dagger Q_3  \nonumber \\
& +i\frac{\lambda_b^m}{\Lambda}\bar{d}_{Rm}\epsilon_{ijk}\Phi_1^i\Phi_2^jQ_3^k
+i\lambda_1^{dn}\bar{d}_{Rn}^1Q_n^T\Phi_1  \nonumber \\
& +i\lambda_2^{dn}\bar{d}_{Rn}^2Q_n^T\Phi_2
+i\frac{\lambda_u^{mn}}{\Lambda}\bar{u}_{Rm}\epsilon_{ijk}\Phi_1^{*i}\Phi_2^{*j}Q_n^k
+\text{h.c.}
\label{eq:lqy}
\end{align}
In the above equation, $n=1,2$ is the family index for
the first two generations of quark triplets. $d_{Rm}$ runs over
$(d_R,s_R,b_R,D_R,S_R)$ and $u_{Rm}$ runs over
$(u_R,c_R,t_R,T_R)$. $u_{R3}^1,u_{R3}^2$ are linear combinations
of $t_R$ and $T_R$. $d_{Rn}^1,d_{Rn}^2$ are linear combinations
of $d_R$ and $D_R$ for $n=1$ and of $s_R$ and $S_R$ for $n=2$.
It is worth noting that in the dimension-4 part of the Eq.~\eqref{eq:lly}
and Eq.~\eqref{eq:lqy} CSB is formally preserved. In contrast,
in Eq.~\eqref{eq:lly} and Eq.~\eqref{eq:lqy}, the dimension-5 part
formally violates CSB. Nevertheless the amount of violation is proportional
to light fermion Yukawas and is thus negligible.

We now turn to the scalar potential. Using a CEFT approach and combining tree level\footnote{At tree level
we don't include a $(\Phi_1^\dagger\Phi_2)^2+\text{h.c.}$ term because it formally violates
CSB. We note that introducing such a term may lead to spontaneous CP violation~\cite{Mao:2017hpp}.
Furthermore, if both the $(\Phi_1^\dagger\Phi_2)^2+\text{h.c.}$ term and Majorana mass terms
for $N_R$'s are introduced, the SLH light neutrino masses can be radiatively generated~\cite{delAguila:2005yi}.} and one-loop
contributions, the scalar effective potential in the SLH is calculated to be~\cite{Cheung:2018iur}
\begin{equation}
V=-\mu^2(\Phi_1^\dagger\Phi_2+\Phi_2^\dagger\Phi_1)
+\lambda |\Phi_1^\dagger\Phi_2|^2
+\Delta(\hat{h})\hat{h}^4
\label{eq:vrc}
\end{equation}
$\mu^2$ and $\lambda$ could be regarded as parameters to be determined from
experiments, while $\Delta(\hat{h})$ is automatically finite, and could be
expressed from Lagrangian parameters in the model
\begin{align}
\Delta(\hat{h}) & =\frac{3}{16\pi^2}\Bigg\{\lambda_t^4
\left[\ln\frac{M_T^2}{m_t^2(\hat{h})}-\frac{1}{2}\right] \nonumber \\
& -\frac{1}{8}g^4\left[\ln\frac{M_X^2}{m_W^2(\hat{h})}-\frac{1}{2}\right] \nonumber \\
& -\frac{1}{16}g^4(1+t_W^2)^2\left[\ln\frac{M_{Z'}^2}{m_Z^2(\hat{h})}-\frac{1}{2}\right]
\Bigg\}
\label{eq:dh}
\end{align}
$\lambda_t$ is defined as
\begin{equation}
\lambda_t\equiv\frac{\lambda_1^t\lambda_2^t}
{\sqrt{\lambda_1^{t2} c_\beta^2+\lambda_2^{t2}s_\beta^2}}
\label{eq:lt1}
\end{equation}
where $\lambda_1^t,\lambda_2^t$ are the two
Yukawa couplings in the top sector, introduced in
Eq.~\eqref{eq:lqy}. $M_T^2,M_X^2,M_{Z'}^2$ are defined as
\begin{align}
M_T^2 & \equiv(\lambda_1^{t2} c_\beta^2+\lambda_2^{t2}s_\beta^2)f^2 \\
M_X^2 & \equiv\frac{1}{2}g^2 f^2 \\
M_{Z'}^2 & \equiv\frac{2}{3-t_W^2}g^2 f^2 \label{eq:zpmass}
\end{align}
They are related to physical mass squared of the relevant particles
as follows
\begin{align}
M_T^2 & =m_T^2+m_t^2 \label{eq:MT} \\
M_X^2 & =m_X^2+m_W^2 \\
M_{Z'}^2 & =m_{Z'}^2+m_Z^2
\end{align}
in which $m_T,m_t$ denote the physical mass of the heavy top $T$
and the top quark $t$, $m_X,m_W$ denote the physical mass of
the $X$ boson and $W$ boson, $m_{Z'},m_Z$ denote the physical mass
of the $Z'$ boson and $Z$ boson, respectively. $m_t^2(\hat{h}),
m_W^2(\hat{h}),m_Z^2(\hat{h})$ are field-dependent mass squared,
which we use the following leading order (LO) expression
\begin{align}
m_t^2(\hat{h}) & =\lambda_t^2\hat{h}^2 \\
m_W^2(\hat{h}) & =\frac{1}{2}g^2\hat{h}^2 \\
m_Z^2(\hat{h}) & =\frac{1}{2}g^2(1+t_W^2)\hat{h}^2
\end{align}
With the above expressions for the scalar effective potential we are
able to compute the electroweak vev, Higgs mass, pseudo-axion mass, etc.
as a function of $\mu^2,\lambda$ and other Lagrangian parameters in the model.

Finally we note that there of course exist gauge-invariant kinetic Lagrangian
for the $SU(3)_L\times U(1)_X$ gauge fields and the fermion fields in the model,
according to their representations.

\subsection{Hidden Mass Relation, Unitarity and Naturalness}

Before starting the phenomenological analysis in the SLH, it is important
to notice that there exist certain constraints that we have to take into account~\cite{Cheung:2018iur}.

First, there exists a hidden mass relation which follows from an analysis of
the scalar effective potential Eq.~\eqref{eq:vrc}. This is because if we consider
$g,t_W,\lambda_t$ as fixed, then the scalar effective potential Eq.~\eqref{eq:vrc}
is fully determined by 5 parameters, say $\mu^2,\lambda,f,t_\beta,m_T$. Requiring
electroweak vev to be $246\GeV$ and the CP-even Higgs mass to be $125\GeV$ should
eliminate two parameters, leaving only three parameters as independent. For instance,
we may choose $f,t_\beta,m_T$ as the three independent parameters, then any other observable
could be expressed in terms of these three parameters. Especially, the pseudo-axion mass $m_\eta$ is
determined from the following hidden mass relation derived in ref.~\cite{Cheung:2018iur}
\begin{align}
m_\eta^2=[m_h^2-v^2\Delta_A(3-2\theta t_{2\theta}^{-1})
+v^2 A(5-2\theta t_{2\theta}^{-1})]s_\theta^{-2}
\label{eq:mr2}
\end{align}
Here $t_{2\theta}^{-1}\equiv\frac{1}{\tan(2\theta)},
s_\theta^{-2}\equiv\frac{1}{\sin^2\theta}$, and $\theta,A,\Delta_A$ are defined by
\begin{align}
\theta\equiv\frac{v}{\sqrt{2}fs_\beta c_\beta}
\end{align}
\begin{align}
A\equiv\frac{3}{16\pi^2}
\left[\lambda_t^4-\frac{g^4}{8}-\frac{g^4}{16}(1+t_W^2)^2\right]
\label{eq:A}
\end{align}
\begin{align}
\Delta_A & \equiv\frac{3}{16\pi^2}\Bigg[\lambda_t^4
\ln\frac{M_T^2}{m_t^2}-\frac{g^4}{8}\ln\frac{M_X^2}{m_W^2} \nonumber \\
& -\frac{g^4}{16}(1+t_W^2)^2\ln\frac{M_{Z'}^2}{m_Z^2}\Bigg]
\end{align}
The basic feature of this mass relation is that the pseudo-axion mass is
anti-correlated with the top partner mass.

Second, the SLH is meant to be only an effective field theory valid up
to some energy scale, which could be revealed by an analysis of
partial wave unitarity. This is done in ref.~\cite{Cheung:2018iur}
and the unitarity cutoff is determined to be
\begin{align}
\Lambda_U=\sqrt{8\pi}\times\min\{fc_\beta,fs_\beta\}
\end{align}
Apart from the lepton Yukawa part, the SLH Lagrangian is manifestly
symmetric with respect to the exchange $\Phi_1\leftrightarrow\Phi_2$
(with the corresponding exchange of all related coefficients), therefore
without loss of generality we may restrict to $t_\beta\geq 1$. The resulting
formulae have the $t_\beta\leftrightarrow\frac{1}{t_\beta}$ invariance.
Nevertheless, the lepton Yukawa Lagrangian Eq.~\eqref{eq:lly} does not
share this exchange symmetry, and the $t_\beta\leftrightarrow\frac{1}{t_\beta}$
invariance could be lost. However, if we do not deal directly with lepton-related
vertices, the $t_\beta\leftrightarrow\frac{1}{t_\beta}$ invariance violation
could only come from input parameter corrections, which are all suppressed
by $\frac{v^2}{f^2}$~\cite{Cheung:2018iur}, which is a very small quantity if we consider current bound
on $f$. Therefore in the following, unless otherwise specified, we will assume $t_\beta\geq 1$.
(Moreover, in Section~\ref{sec:ewpt} we will show that the $t_\beta<1$ case is disfavored by electroweak
precision measurements for natural region of parameter space.) Then we can express the unitarity cutoff as
\begin{align}
\Lambda_U=\sqrt{8\pi}fc_\beta
\end{align}
and we require all particle masses be less than $\Lambda_U$. We note that
since $\Lambda_U$ is determined by the smaller of the triplet vevs, while
$m_{Z'}$ is determined by the quadrature of the triplet vevs, therefore
requiring $m_{Z'}\leq\Lambda_U$ leads to an upper bound on $t_\beta$
(besides our assumption $t_\beta\geq1$)
\begin{align}
1\leq t_\beta\leq\sqrt{\frac{4\pi(3-t_W^2)}{g^2}-1}\approx 8.9
\label{eq:tbrange}
\end{align}

Third, the parameter $M_T$ has a lower bound derived simply from the structure of
the Yukawa Lagrangian~\cite{Han:2005ru}
\begin{equation}
M_T\geq \sqrt{2}\frac{m_t}{v}fs_{2\beta}\approx fs_{2\beta}
\label{eq:MTmin}
\end{equation}
where $s_{2\beta}\equiv\sin(2\beta)$. $M_T$ is also bounded from above by either
$\Lambda_U$ or the requirement that $m_\eta^2$ obtained from Eq.~\eqref{eq:mr2}
should be positive.

Finally, from the LHC search of $Z'$ boson in the dilepton channel~\cite{Aaboud:2017buh,Sirunyan:2018exx}, we estimate
the lower bound on $f$ as~\cite{Mao:2017hpp}
\begin{align}
f\gtrsim 7.5\TeV
\end{align}
We note that when combined with Eq.~\eqref{eq:MTmin} and Eq.~\eqref{eq:tbrange} this also leads to
a lower bound on the top partner mass of around $1.7\TeV$, which is much more stringent
than constraints from top partner searches at the LHC.

It is remarkable that the naturalness issue can also be analyzed in a CEFT approach,
which is done in ref.~\cite{Cheung:2018iur}. We define the total degree of fine-tuning
at a certain parameter point as
\begin{align}
\Delta_{\TOT}=\max\{\Delta_{\TOT}^{\mu^2},\Delta_{\TOT}^\lambda\}
\end{align}
where $\Delta_{\TOT}^{\mu^2},\Delta_{\TOT}^\lambda$ are defined by
\begin{align}
\Delta_{\TOT}^\lambda\equiv\left|\frac{\lambda_U}{m_h^2}\frac{\partial m_h^2}{\partial \lambda_U}\right|,\quad
\Delta_{\TOT}^{\mu^2}\equiv\left|\frac{\mu_U^2}{m_h^2}\frac{\partial m_h^2}{\partial \mu_U^2}\right|
\end{align}
Here $\lambda_U,\mu_U^2$ denote the $\lambda,\mu^2$ parameters defined at the unitarity cutoff. The above
definitions obviously reflect how the IR parameters (e.g. $m_h^2$) are sensitive to UV parameters (e.g. $\lambda_U,
\mu_U^2$), and thus may serve as a measure of the degree of fine-tuning in the allowed parameter space.
We may follow ref.~\cite{Cheung:2018iur} to compute the degree of fine-tuning, and find several general
features. One feature which is easy to understand is, generally speaking, with smaller $f$ and $m_T$ we
could get smaller degree of fine-tuning.
\begin{figure}[ht]
\begin{center}
\includegraphics[width=2.2in]{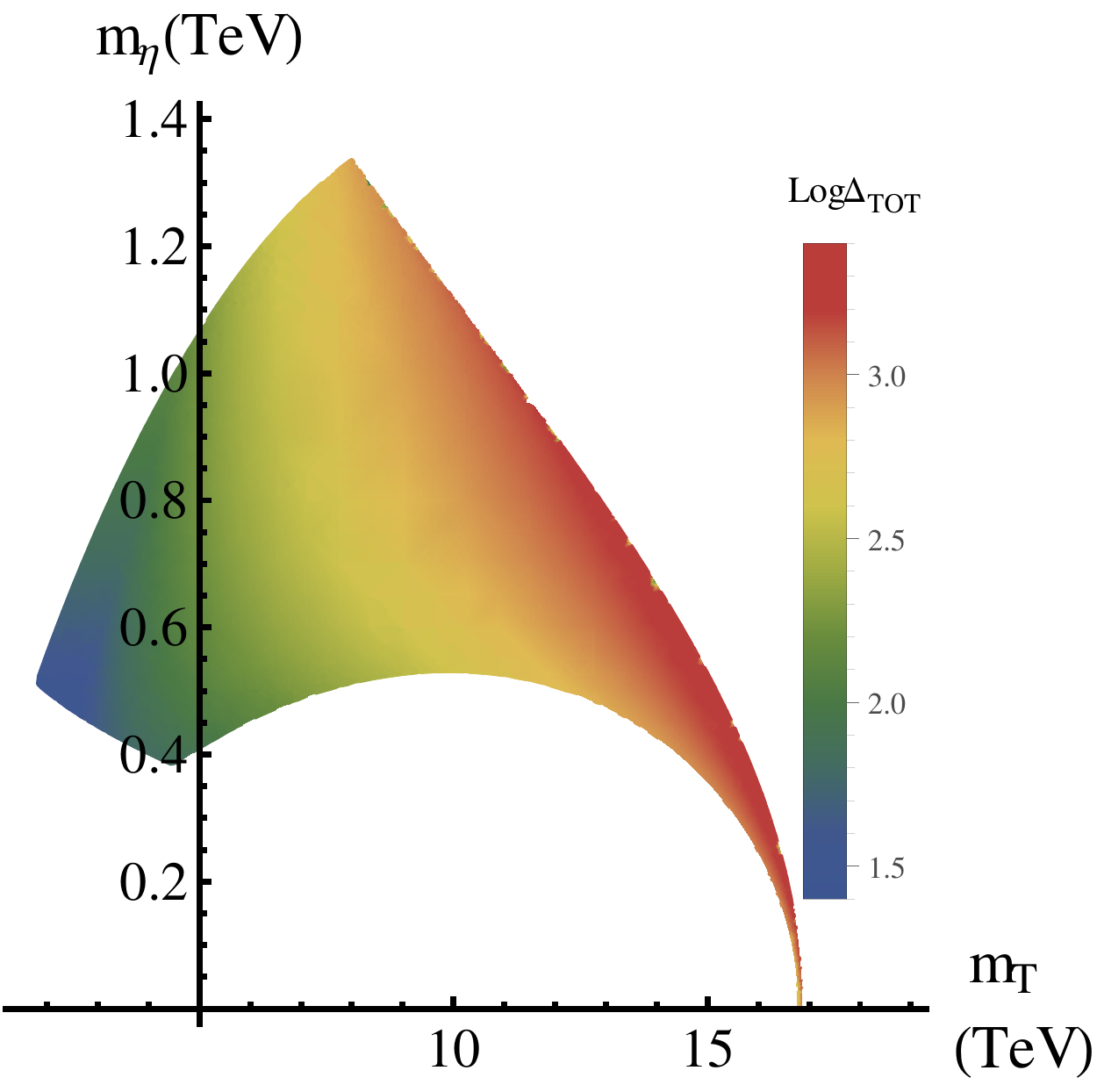}
\end{center}
\caption{\label{fig:dft8} Density plot of $\text{Log}\Delta_\TOT$ in the
$m_\eta-m_T$ plane for $f=8\TeV$. $\text{Log}$ means $\log_{10}$.}
\end{figure}

In Figure~\ref{fig:dft8} we present the density plot of
$\text{Log}\Delta_\TOT$ in the $m_\eta-m_T$ plane for $f=8\TeV$.
Only the colored region is allowed by various constraints. From the
figure it is clear that the parameter region favored by naturalness
considerations is featured by a small $m_T$, with $m_\eta$ around $500\GeV$.
A light $\eta$, with a mass less than $2m_t$, is unfortunately disfavored.

\subsection{Fermion Mass Diagonalization and Flavor Assumption}

Fermion mass diagonalization has been studied in ref.~\cite{Han:2005ru,delAguila:2011wk}.
In the lepton sector, the fermion mass matrices can be diagonalized by the following
field rotations:
\begin{align}
\begin{pmatrix}
N_{Ln} \\ \nu_{Ln}
\end{pmatrix}
\rightarrow
\begin{pmatrix}
c_\delta & s_\delta \\
s_\delta & -c_\delta
\end{pmatrix}
\begin{pmatrix}
N_{Ln} \\ \nu_{Ln}
\end{pmatrix}, \nonumber \\
\quad n=1,2,3,\quad \delta\equiv\frac{v}{\sqrt{2}ft_\beta}
\end{align}
\begin{align}
\begin{pmatrix}
e_L \\ \mu_L \\ \tau_L
\end{pmatrix}
\rightarrow U_l
\begin{pmatrix}
e_L \\ \mu_L \\ \tau_L
\end{pmatrix},\quad
\begin{pmatrix}
e_R \\ \mu_R \\ \tau_R
\end{pmatrix}
\rightarrow W_l
\begin{pmatrix}
e_R \\ \mu_R \\ \tau_R
\end{pmatrix}
\end{align}
where $U_l,W_l$ are both $3\times 3$ unitary matrices. In this
work, for simplicity we will assume $U_l,W_l$ are both identity
matrices. This leads to simplification of some Feynman rules
associated with the heavy neutrino $N$.

In the quark sector, first of all we perform field rotations
in the right-handed sector as follows
\begin{align}
u_{R3}^1=\frac{-\lambda_2^t s_\beta t_R+\lambda_1^t c_\beta T_R}
{\sqrt{\lambda_1^{t2}c_\beta^2+\lambda_2^{t2}s_\beta^2}},\quad
u_{R3}^2=\frac{\lambda_1^t c_\beta t_R+\lambda_2^t s_\beta T_R}
{\sqrt{\lambda_1^{t2}c_\beta^2+\lambda_2^{t2}s_\beta^2}}
\end{align}
\begin{align}
d_{R1}^1=\frac{-\lambda_2^d s_\beta d_R+\lambda_1^d c_\beta D_R}
{\sqrt{\lambda_1^{d2}c_\beta^2+\lambda_2^{d2}s_\beta^2}},\quad
d_{R1}^2=\frac{\lambda_1^d c_\beta d_R+\lambda_2^d s_\beta D_R}
{\sqrt{\lambda_1^{d2}c_\beta^2+\lambda_2^{d2}s_\beta^2}}
\end{align}
\begin{align}
d_{R2}^1=\frac{-\lambda_2^s s_\beta s_R+\lambda_1^s c_\beta S_R}
{\sqrt{\lambda_1^{s2}c_\beta^2+\lambda_2^{s2}s_\beta^2}},\quad
d_{R2}^2=\frac{\lambda_1^s c_\beta s_R+\lambda_2^s s_\beta S_R}
{\sqrt{\lambda_1^{s2}c_\beta^2+\lambda_2^{s2}s_\beta^2}}
\end{align}
For simplicity, the phenomenological studies done in this work will be carried out
under the following flavor assumptions on the quark Yukawa Lagrangian
Eq.~\eqref{eq:lqy}
\begin{equation}
\lambda_u^{Tu}=\lambda_u^{Tc}=\lambda_u^{12}=\lambda_u^{21}
=\lambda_u^{31}=\lambda_u^{32}=0
\label{eq:fa1}
\end{equation}
\begin{equation}
\lambda_b^D=\lambda_b^S=\lambda_b^1=\lambda_b^2=0
\label{eq:fa2}
\end{equation}
These flavor assumptions turn off all the generation-crossing quark flavor
transitions and lead to a trivial CKM matrix, i.e. $V_{CKM}=\textbf{1}_{3\times 3}$,
which is not realistic. Nevertheless, in this paper we are concerned with
the direct production of new physics particles at high energy colliders rather than
quark flavor observables. Also, for the parameter region which we are interested in,
the phenomenology is not sensitive to the flavor assumptions adopted here, if
the $\lambda$'s in Eq.~\eqref{eq:fa1} and Eq.~\eqref{eq:fa2}, which characterize
the generation-crossing quark flavor changing effects, are small.

With the above flavor assumptions, it is then straightforward to show, up to $\ord(\frac{v}{f})$,
after right-handed sector field rotations we only need to perform the following field
rotations in the left-handed sector to diagonalize the quark mass matrices
\begin{align}
\begin{pmatrix}
t_L \\ T_L
\end{pmatrix}
& \rightarrow
\begin{pmatrix}
1 & -\delta_t \\
\delta_t & 1
\end{pmatrix}
\begin{pmatrix}
t_L \\ T_L
\end{pmatrix} \\
\begin{pmatrix}
d_L \\ D_L
\end{pmatrix}
& \rightarrow
\begin{pmatrix}
1 & -\delta_{Dd} \\
\delta_{Dd} & 1
\end{pmatrix}
\begin{pmatrix}
d_L \\ D_L
\end{pmatrix} \\
\begin{pmatrix}
s_L \\ S_L
\end{pmatrix}
& \rightarrow
\begin{pmatrix}
1 & -\delta_{Ss} \\
\delta_{Ss} & 1
\end{pmatrix}
\begin{pmatrix}
s_L \\ S_L
\end{pmatrix}
\label{eq:dss}
\end{align}
In the above equations, the field rotation parameters $\delta_t,\delta_{Dd},\delta_{Ss}$
can be expressed using $f,\beta$ and the corresponding heavy fermion mass as follows\footnote{Our
expression for $\delta_t,\delta_{Dd},\delta_{Ss}$ differs from the corresponding expression in
Eq.(2.63) of ref.~\cite{delAguila:2011wk}. The expressions of $\delta_t,\delta_{Dd},\delta_{Ss}$
given by ref.~\cite{delAguila:2011wk} are not consistent with their counterparts in ref.~\cite{Han:2005ru}.
Our calculation agrees with ref.~\cite{Han:2005ru}.}
\begin{align}
\delta_t & =\frac{v}{2\sqrt{2}fs_\beta c_\beta}\left(
s_\beta^2-c_\beta^2\pm\sqrt{1-8\frac{m_t^2}{v^2}\frac{f^2}{M_T^2}s_\beta^2 c_\beta^2}\right) \\
\delta_{Dd} & =-\frac{v}{2\sqrt{2}fs_\beta c_\beta}\left(
s_\beta^2-c_\beta^2\pm\sqrt{1-8\frac{m_d^2}{v^2}\frac{f^2}{M_D^2}s_\beta^2 c_\beta^2}\right) \\
\delta_{Ss} & =-\frac{v}{2\sqrt{2}fs_\beta c_\beta}\left(
s_\beta^2-c_\beta^2\pm\sqrt{1-8\frac{m_s^2}{v^2}\frac{f^2}{M_S^2}s_\beta^2 c_\beta^2}\right)
\label{eq:signchoice}
\end{align}
Note in the above equations, before the square root, both the plus sign and minus sign give
possible solutions, which leads to a total of eight sign combinations. When we refer to the sign
combination in these equations, we will list according to the order $\delta_t,\delta_{Dd},\delta_{Ss}$,
as e.g. $(+,+,+),(+,+,-)$, etc. $m_d,m_s,M_D,M_S$ correspond to the mass of $d,s,D,S$, respectively.
In the following we will simply neglect the small $m_d,m_s$, then the expressions of
$\delta_{Dd},\delta_{Ss}$ become identical, apart from a possible sign difference before the
square root. Then we obtain the simple expression
\begin{align}
\delta_{Dd}^{+}=\delta_{Ss}^{+}=-\frac{vt_\beta}{\sqrt{2}f},\quad
\delta_{Dd}^{-}=\delta_{Ss}^{-}=\frac{v}{\sqrt{2}ft_\beta}
\label{eq:sc12}
\end{align}
where the superscripts indicate the sign choice for the corresponding rotation parameter.
The rotation parameters $\delta_t,\delta_{Dd},\delta_{Ss}$ are important since they appear
directly in the coefficients of various interaction vertices which affect the $\eta$ phenomenology,
as we will see.

\subsection{Lagrangian in the Mass Basis}

We are now prepared to present the Lagrangian in the mass basis which is relevant
for the investigation of $\eta$ phenomenology. However, let us first note that there
is a subtle issue regarding the diagonalization in the bosonic sector. After EWSB, it can be shown
that the CP-odd sector scalar kinetic matrix in terms of the $\eta,\zeta,\chi,\omega$ fields
are not canonically-normalized. Also, there exist ``unexpected'' two-point vector-scalar
transition terms like $Z^\mu\partial_\mu\eta$ after expanding the covariant derivative terms
of the scalar fields. Therefore, a further field rotation (including a proper gauge-fixing)
is needed to diagonalize the bosonic sector. This subtle issue had been overlooked for a long time
in the literature, and was only remedied in a recent paper~\cite{He:2017jjx}. In ref.~\cite{He:2017jjx},
an expression for the fraction of mass eigenstate $\eta$ field contained in the $\eta,\zeta,\chi,\omega$ fields
originally introduced in the parametrization Eq.~\eqref{eq:theta}, Eq.~\eqref{eq:hdoub},Eq.~\eqref{eq:kdoub}
was obtained, valid to all orders in $\xi\equiv\frac{v}{f}$, as follows (we collect
the four fraction values into a four-component column vector $\Upsilon$)
\begin{align}
\Upsilon=\begin{pmatrix}
c_{\gamma+\delta}^{-1} \\
\\
-c_{\gamma+\delta}^{-1}(s_\delta^2 t_\beta-s_\gamma^2 t_\beta^{-1}) \\
\\
\frac{v}{\sqrt{2}f}c_{\gamma+\delta}^{-1}(c_{2\delta}t_\beta-c_{2\gamma}t_\beta^{-1}) \\
\\
\frac{1}{2}c_{\gamma+\delta}^{-1}(s_{2\delta}t_\beta+s_{2\gamma}t_\beta^{-1})
\end{pmatrix}
\label{eq:upsilon}
\end{align}
where
\begin{align}
\gamma\equiv\frac{vt_\beta}{\sqrt{2}f},\quad \delta\equiv\frac{v}{\sqrt{2}ft_\beta}
\end{align}
The $\Upsilon$ vector is involved in the derivation of all $\eta$-related mass eigenstate
vertices. Especially, from the expression of $\Upsilon$ we see there is an $\ord(\xi)$
component of mass eigenstate $\eta$ contained in $\chi$. This has the following consequences.
If we parameterize the mass eigenstate $ZH\eta$ vertex as follows
\begin{align}
\mathcal{L}_{ZH\eta} & =c^{as}_{ZH\eta}Z^\mu(\eta\partial_\mu H-H\partial_\mu\eta) \nonumber \\
& +c^{s}_{ZH\eta}Z^\mu(\eta\partial_\mu H+H\partial_\mu\eta)
\label{eq:lzhe}
\end{align}
where $c^{as}_{ZH\eta}$ denotes the coefficient of the anti-symmetric $ZH\eta$ vertex, and
$c^{s}_{ZH\eta}$ denotes the coefficient of the symmetric $ZH\eta$ vertex, then it is
shown in ref.~\cite{He:2017jjx} that
\begin{align}
c^{as}_{ZH\eta}=-\frac{g}{4\sqrt{2}c_W^3 t_{2\beta}}\xi^3+\ord(\xi^5)
\label{eq:casv}
\end{align}
\bwt
\begin{align}
c^{s}_{ZH\eta}=\frac{g}{\sqrt{2}c_W t_{2\beta}}\xi
+\frac{g}{24\sqrt{2}c_W s_{2\beta}}\left[\frac{8}{s_{2\beta}t_{2\beta}}+3c_{2\beta}\left(8+\frac{6}{c_W^2}-\frac{1}{c_W^4}\right)\right]\xi^3
+\ord(\xi^5)
\label{eq:csv}
\end{align}
\ewt
We see that the anti-symmetric $ZH\eta$ vertex only shows up from $\ord(\xi^3)$, in contrast to
the results presented in ref.~\cite{Kilian:2004pp,Kilian:2006eh} which claimed the
existence of anti-symmetric $ZH\eta$ vertex at $\ord(\xi)$ due to the lack of an appropriate
diagonalization in the bosonic sector.

This subtle issue of diagonalization in the bosonic sector also has impact on the $\eta$ coupling
to fermions. For instance, if we consider the expansion of $\epsilon_{ijk}\Phi_1^i\Phi_2^j$,
with the help of the expression for the $\Upsilon$ vector in Eq.~\eqref{eq:upsilon}, we could
find the following result for the neutral component
\begin{align}
\epsilon_{ijk}\Phi_1^i\Phi_2^j\supset
-if\begin{pmatrix}
0 \\
fs_\beta c_\beta s_{\gamma+\delta}+\frac{1}{\sqrt{2}}c_{\gamma+\delta}H \\
0
\end{pmatrix}
\end{align}
An important message from this is that $\epsilon_{ijk}\Phi_1^i\Phi_2^j$ does not contain
any fraction of mass eigenstate $\eta$ field, to all orders in $\xi$. Therefore, from Eq.~\eqref{eq:lly}
we immediately conclude that $\eta$ does not couple to a pair of charged leptons to all orders
in $\xi$. This point has been overlooked by previous studies~\cite{Cheung:2008zu,Kim:2011bv} which rely on $\eta\rightarrow\tau\tau$.

In the following let us collect the other mass eigenstate vertices that are relevant for $\eta$ phenomenology, to
the first nontrivial order in $\xi$.
In the Yukawa sector, we have the following couplings of $H$ and $\eta$ to a pair of fermions:
\bwt
\begin{enumerate}
\item $H$ and $\eta$ couplings to lepton sector:
\begin{align}
\mL_{LY} & \supset-\sum_{n=1}^3\frac{m_{ln}}{\sqrt{2}fs_\beta c_\beta t_{\gamma+\delta}}
H\bar{l}_{Rn}l_{Ln}+\sum_{n=1}^3\frac{M_{Nn}}{\sqrt{2}ft_\beta}H\bar{N}_{Rn}\nu_{Ln} \nonumber \\
& -i\sum_{n=1}^3\frac{M_{Nn}}{\sqrt{2}ft_\beta}c_{\gamma+\delta}\eta\bar{N}_{Rn}N_{Ln}
-i\sum_{n=1}^3\frac{M_{Nn}}{\sqrt{2}ft_\beta}s_{\gamma+\delta}\eta\bar{N}_{Rn}\nu_{Ln}+\text{h.c.}
\label{eq:lYu}
\end{align}
\item $H$ and $\eta$ couplings to up-type quark sector:
\begin{align}
\mL_{QY} & \supset-\frac{m_u}{v}H\bar{u}_R u_L-\frac{m_c}{v}H\bar{c}_R c_L \nonumber \\
& -\frac{m_t}{v}H\bar{t}_R t_L+\frac{m_t}{v}\left(\frac{\sqrt{2}v}{ft_{2\beta}}+\delta_t \right)H\bar{t}_R T_L
+\frac{M_T}{v}\delta_t H\bar{T}_R t_L+\frac{m_t^2}{vM_T}H\bar{T}_R T_L \nonumber \\
& -i\frac{m_t}{v}\delta_t\eta\bar{t}_R t_L-i\frac{m_t}{v}\eta\bar{t}_R T_L
+i\frac{M_T}{v}\left(\frac{v^2}{2f^2}+\delta_t^2\right)\eta\bar{T}_R t_L
+i\frac{M_T}{v}\delta_t\eta\bar{T}_R T_L \nonumber \\
&+\text{h.c.}
\end{align}
\item $H$ and $\eta$ couplings to down-type quark sector:
\begin{align}
\mL_{QY} & \supset-\frac{m_b}{v}H\bar{b}_R b_L \nonumber \\
& -\frac{m_d}{v}H\bar{d}_R d_L+\frac{m_d}{v}\left(-\frac{\sqrt{2}v}{ft_{2\beta}}+\delta_{Dd}\right)H\bar{d}_R D_L
+\frac{M_D}{v}\delta_{Dd}H\bar{D}_R d_L+\frac{m_d^2}{vM_D}H\bar{D}_R D_L \nonumber \\
& -\frac{m_s}{v}H\bar{s}_R s_L+\frac{m_s}{v}\left(-\frac{\sqrt{2}v}{ft_{2\beta}}+\delta_{Ss}\right)H\bar{s}_R S_L
+\frac{M_S}{v}\delta_{Ss}H\bar{S}_R s_L+\frac{m_s^2}{vM_S}H\bar{S}_R S_L \nonumber \\
& -i\frac{m_d}{v}\delta_{Dd}\eta\bar{d}_R d_L-i\frac{m_d}{v}\eta\bar{d}_R D_L
+i\frac{M_D}{v}\left(\frac{v^2}{2f^2}+\delta_{Dd}^2\right)\eta\bar{D}_R d_L
+i\frac{M_D}{v}\delta_{Dd}\eta\bar{D}_R D_L \nonumber \\
& -i\frac{m_s}{v}\delta_{Ss}\eta\bar{s}_R s_L-i\frac{m_s}{v}\eta\bar{s}_R S_L
+i\frac{M_S}{v}\left(\frac{v^2}{2f^2}+\delta_{Ss}^2\right)\eta\bar{S}_R s_L
+i\frac{M_S}{v}\delta_{Ss}\eta\bar{S}_R S_L \nonumber \\
& +\text{h.c.}
\label{eq:dYu}
\end{align}
\end{enumerate}
\ewt
In the above equations, $m_{ln},n=1,2,3$ denote the masses of $e,\mu,\tau$ leptons,
$M_{Nn},n=1,2,3$ denote the masses of the three heavy neutral leptons $N_n$. $m_u,m_c$
denote the masses of the $u,c$ quarks, respectively. $\eta$ can also be a decay product
of the heavy fermions $N,T,D,S$, therefore we also list the relevant Lagrangian for
the heavy fermion gauge interaction which enters the heavy fermion decays
\bwt
\begin{align}
\mL_{\mt} & \supset\frac{gv}{2ft_\beta}W_\mu^+\bar{N}_{Lm}\gamma^\mu l_{Lm}
-\frac{gv}{2\sqrt{2}c_W ft_\beta}Z_\mu\bar{N}_{Lm}\gamma^\mu\nu_{Lm} \nonumber \\
& -\frac{g\delta_t}{\sqrt{2}}W_\mu^+\bar{T}_L\gamma^\mu b_L
-\frac{g\delta_t}{2c_W}Z_\mu\bar{T}_L\gamma^\mu t_L \nonumber \\
& -\frac{g\delta_{Dd}}{\sqrt{2}}W_\mu^+\bar{u}_L\gamma^\mu D_L
+\frac{g\delta_{Dd}}{2c_W}Z_\mu\bar{d}_L\gamma^\mu D_L
-\frac{g\delta_{Ss}}{\sqrt{2}}W_\mu^+\bar{c}_L\gamma^\mu S_L
+\frac{g\delta_{Ss}}{2c_W}Z_\mu\bar{s}_L\gamma^\mu S_L \nonumber \\
& +\text{h.c.}
\end{align}
A further interesting possibility is that $\eta$ might come from the
decay of a $Z'$ boson. The $Z'$-related parts of interaction Lagrangian are listed below:
\begin{enumerate}
\item $Z'$ couplings to leptons:
\begin{align}
\mL_{\mt} & \supset g\frac{1-t_W^2}{2\sqrt{3-t_W^2}}\bar{l}_{Ln}\gamma^\mu l_{Ln}Z'_\mu
-g\frac{t_W^2}{\sqrt{3-t_W^2}}\bar{l}_{Rn}\gamma^\mu l_{Rn}Z'_\mu \nonumber \\
& +g\frac{1-t_W^2}{2\sqrt{3-t_W^2}}\bar{\nu}_{Ln}\gamma^\mu\nu_{Ln}Z'_\mu
-g\frac{1}{\sqrt{3-t_W^2}}\bar{N}_{Ln}\gamma^\mu N_{Ln}Z'_{\mu}
\label{eq:zpl}
\end{align}
\item $Z'$ couplings to 3rd generation quarks:
\begin{align}
\mL_{\mt} & \supset -g\frac{3-2t_W^2}{3\sqrt{3-t_W^2}}\bar{T}_L\gamma^\mu T_LZ'_\mu
+g\frac{2t_W^2}{3\sqrt{3-t_W^2}}\bar{T}_R\gamma^\mu T_RZ'_\mu \nonumber \\
& +g\frac{3+t_W^2}{6\sqrt{3-t_W^2}}\bar{t}_L\gamma^\mu t_LZ'_\mu
+g\frac{2t_W^2}{3\sqrt{3-t_W^2}}\bar{t}_R\gamma^\mu t_RZ'_\mu \nonumber \\
& +g\frac{3+t_W^2}{6\sqrt{3-t_W^2}}\bar{b}_L\gamma^\mu b_LZ'_\mu
-g\frac{t_W^2}{3\sqrt{3-t_W^2}}\bar{b}_R\gamma^\mu b_RZ'_\mu
\label{eq:zp3q}
\end{align}
\item $Z'$ couplings to 1st and 2nd generation quarks:
\begin{align}
\mL_{\mt} & \supset g\frac{\sqrt{3-t_W^2}}{3}\bar{D}_L\gamma^\mu D_LZ'_\mu
-g\frac{t_W^2}{3\sqrt{3-t_W^2}}\bar{D}_R\gamma^\mu D_RZ'_\mu \nonumber \\
& -g\frac{\sqrt{3-t_W^2}}{6}\bar{d}_L\gamma^\mu d_LZ'_\mu
-g\frac{t_W^2}{3\sqrt{3-t_W^2}}\bar{d}_R\gamma^\mu d_RZ'_\mu \nonumber \\
& -g\frac{\sqrt{3-t_W^2}}{6}\bar{u}_L\gamma^\mu u_LZ'_\mu
+g\frac{2t_W^2}{3\sqrt{3-t_W^2}}\bar{u}_R\gamma^\mu u_RZ'_\mu \nonumber \\
& +\text{terms with}\quad u\rightarrow c, d\rightarrow s, D\rightarrow S
\label{eq:zp12q}
\end{align}
\item $Z'$ couplings to bosons (relevant for $Z'$ decay):
\begin{align}
\mL_{\text{gauge}} & \supset -ig\sqrt{3-t_W^2}(1-t_W^2)\frac{v^2}{8f^2}\Big\{
(\partial_\mu Z'_\nu)(W^{-\mu}W^{+\nu}-W^{+\mu}W^{-\nu}) \nonumber \\
& +Z'^\mu[(\partial_\mu W_\nu^+)W^{-\nu}-(\partial_\mu W_\nu^-)W^{+\nu}]
+Z'^\nu[(\partial_\mu W_\nu^-)W^{+\mu}-(\partial_\mu W_\nu^+)W^{-\mu}]\Big\}
\end{align}
\begin{align}
\mL_{gk} & \supset -\frac{\sqrt{2}gv}{\sqrt{3-t_W^2}ft_{2\beta}}Z'^\mu(\eta\partial_\mu H-H\partial_\mu\eta)
-\frac{\sqrt{3-t_W^2}gv}{\sqrt{2}ft_{2\beta}}Z'^\mu(\eta\partial_\mu H+H\partial_\mu\eta) \nonumber \\
& -\frac{g^2 v}{2c_W^2\sqrt{3-t_W^2}}\eta Z'^\mu\frac{(Y_\mu^{0\dagger}+Y_\mu^0)}{\sqrt{2}}
+\frac{g^2 vc_{2W}}{2c_W^3\sqrt{3-t_W^2}}HZ'^\mu Z_\mu
\end{align}
\end{enumerate}
\ewt

\section{Symmetric VSS Vertices}
\label{sec:vss}
In the derivation of SLH Lagrangian in the mass basis we obtain the $ZH\eta$ vertex in the form
of Eq.~\eqref{eq:lzhe}, which contains two parts: the antisymmetric part ($Z^\mu(\eta\partial_\mu H-H\partial_\mu\eta)$)
and the symmetric part ($Z^\mu(\eta\partial_\mu H+H\partial_\mu\eta)$)\footnote{The Hermiticity requirement on the Lagrangian
does not forbid the symmetric part. $Z_\mu,H,\eta$ are all real fields. $\partial_\mu$
does not lead to an additional minus sign under Hermitian conjugate because
in quantum field theory $x^\mu$'s are labels, not operators. This is not to
be confused with the situation in ordinary quantum mechanics.}. An antisymmetric VSS vertex often
appears in models based on a linearly-realized scalar sector, such as the usual two-Higgs-doublet model(2HDM). It is natural
to ask whether the symmetric VSS vertices can have any physical effect. We note that in a Lorentz-invariant
$ZH\eta$ vertex, the $\partial_\mu$ may act on any of the three fields ($Z^\mu,H,\eta$). However
because a total derivative term $\partial_\mu(Z^\mu H\eta)$ has no physical effects, we therefore expect
at most two independent contributions from the interaction of one vector fields with two scalar fields.
If symmetric VSS vertices are allowed and present in a general theory and could lead to distinct physical
effects, it would mean that a vector field could interact with two scalar fields in a manner different
from the usually expected antisymmetric pattern, which may further reveal interesting features of the enlarged
scalar sector.

Let us first note that the symmetric VSS Lagrangian $Z^\mu(\eta\partial_\mu H+H\partial_\mu\eta)$ can be
written as
\begin{eqnarray}
Z_\mu\partial^\mu(H\eta)
\label{eqn:ZdelhA}
\end{eqnarray}
via Leibniz rule and is therefore (via integration by parts)
equivalent to
\begin{eqnarray}
-(\partial^\mu Z_\mu)(H\eta)
\label{eqn:delZhA}
\end{eqnarray}
in the Lagrangian formulation of the theory. A reflective reader
might at this moment wonder whether terms like ~\eqref{eqn:delZhA}
indeed contribute to S-matrix elements if canonical quantization
is adopted. Note that what matters in canonical quantization is
the interaction Hamiltonian in the interaction picture (denoted $H_I^{\text{int}}$), and if
$Z^\mu$ is a massive spin-1 field, then the corresponding
interaction picture field operator $Z_{I}^\mu$ (the subscript
``$I$'' denotes interaction picture) will
automatically satisfy~\cite{Weinberg:1995mt}
\begin{eqnarray}
\partial_\mu Z_{I}^\mu=0
\label{eqn:delZ0}
\end{eqnarray}
It is tempting to arrive at the conclusion that terms like
~\eqref{eqn:delZhA} cannot contribute to S-matrix elements
due to Eq.~\eqref{eqn:delZ0}. Actually this is not quite correct.
The correct procedure from the classical Lagrangian to
the interaction Hamiltonian in the interaction picture $H_I^{\text{int}}$
is first identify appropriate canonical coordinates and their
conjugate momenta, then perform a Legendre transformation
to obtain the Hamiltonian and express it in terms of
canonical coordinates and their conjugate momenta, then
promote the canonical variables to field operators satisfying
appropriate canonical communtation relations, and finally
split the Hamiltonian into a free part and an interaction part
and replace the Heisenberg-picture quantities with their
interaction-picture counterparts~\cite{Weinberg:1995mt}. If
this procedure is strictly followed, we would find that only
the spatial components of $Z^\mu$ can be treated as independent
canonical coordinates while $Z^0$ is dependent because no matter
we start with Eq.~\eqref{eqn:ZdelhA} or Eq.~\eqref{eqn:delZhA}
the derivative of the Lagrangian with respect to $\dot{Z}_0$
cannot be made to satisfy canonical commutation relations. To
avoid the appearance of $\partial^0 Z_0$ in the Hamiltonian
we could start with Eq.~\eqref{eqn:ZdelhA} and then the
problem turns out to be what has been treated in Section 7.5
of Ref.~\cite{Weinberg:1995mt}. Using the results there
we could see that Eq.~\eqref{eqn:ZdelhA} leads to a term
\begin{eqnarray}
-Z^\mu_I\partial_\mu(h_I A_I)
\label{eqn:Hint}
\end{eqnarray}
in the interaction Hamiltonian in the interaction-picture
(barring a Lorentz non-covariant term which is not shown
here). This will certainly lead to a vertex Feynman rule
\begin{eqnarray}
-k^\mu
\end{eqnarray}
where $k^\mu$ is the $Z$ momentum flowing into the vertex.
This vertex Feynman rule could also be derived from
Eq.~\eqref{eqn:ZdelhA} via the path-integral method.
Notice that it is not legitimate to perform integration by
parts in the interaction-picture Hamiltonian $H_I^{\text{int}}$ to obtain
\begin{eqnarray}
(\partial_\mu Z^\mu_I)(h_I A_I)
\end{eqnarray}
from Eq.~\eqref{eqn:Hint}\footnote{More specifically, integration by parts
for spatial components of $H_I^{\text{int}}$ should be fine if
the fields are assumed to satisfy certain boundary condition, which is usually
the case. However, integration by parts for the temporal component of $H_I^{\text{int}}$
is problematic since in the expression for the scattering operator
$S=T\exp(-i\int_{-\infty}^{+\infty}H_I^{\text{int}}dt)$ the temporal integration
is actually twisted by the time-ordering. No such problem exists if we adopt
the path-integral method.}.

The appearance of $\partial_\mu Z^\mu$ in
Eq.~\eqref{eqn:delZhA} is reminiscent of covariant
gauge-fixing in gauge field theories. Eq.~\eqref{eqn:delZhA}
is not gauge-invariant, nevertheless at this moment
let us suppose that it can be deduced from a
gauge-invariant operator. Because we are
dealing with quantum field theories it is important
not to be confused with the case of classical
field theories. In a classical gauge field theory
a gauge-fixing condition (such as the Landau gauge
condition $\partial_\mu Z^\mu=0$) is employed so that
the solutions of the equation of motion are required
to also satisfy the gauge-fixing condition. In quantum
field theory all classical field configurations,
regardless of whether they satisfy the classical
equation of motion, are to be integrated over in the
path-integral. The usually-adopted covariant gauge,
the general $R_\xi$ gauge, actually corresponds to a Gaussian
smearing of a class of covariant gauge conditions and
does not strictly force the classical field to satisfy
a simple gauge-fixing equation. However, the limit
$\xi\rightarrow 0$ makes the gauge-fixing functional
act like a delta-function imposing the Landau gauge
condition $\partial_\mu Z^\mu=0$~\cite{Weinberg:1995mt}.
Therefore it is heuristic to guess that in the Landau gauge,
symmetric VSS vertices do not
contribute to S-matrix of the theory. However, we should
not forget that in the Landau gauge it is necessary to
take into account the Goldstone contribution to the S-matrix,
and also the associated ghost contribution when we go
beyond tree level in perturbation theory. This observation
suggests that at tree level, processes involving
symmetric VSS vertices can be seen as purely Goldstone-mediated.

\begin{figure}[ht]
\begin{center}
\includegraphics[width=2.6in]{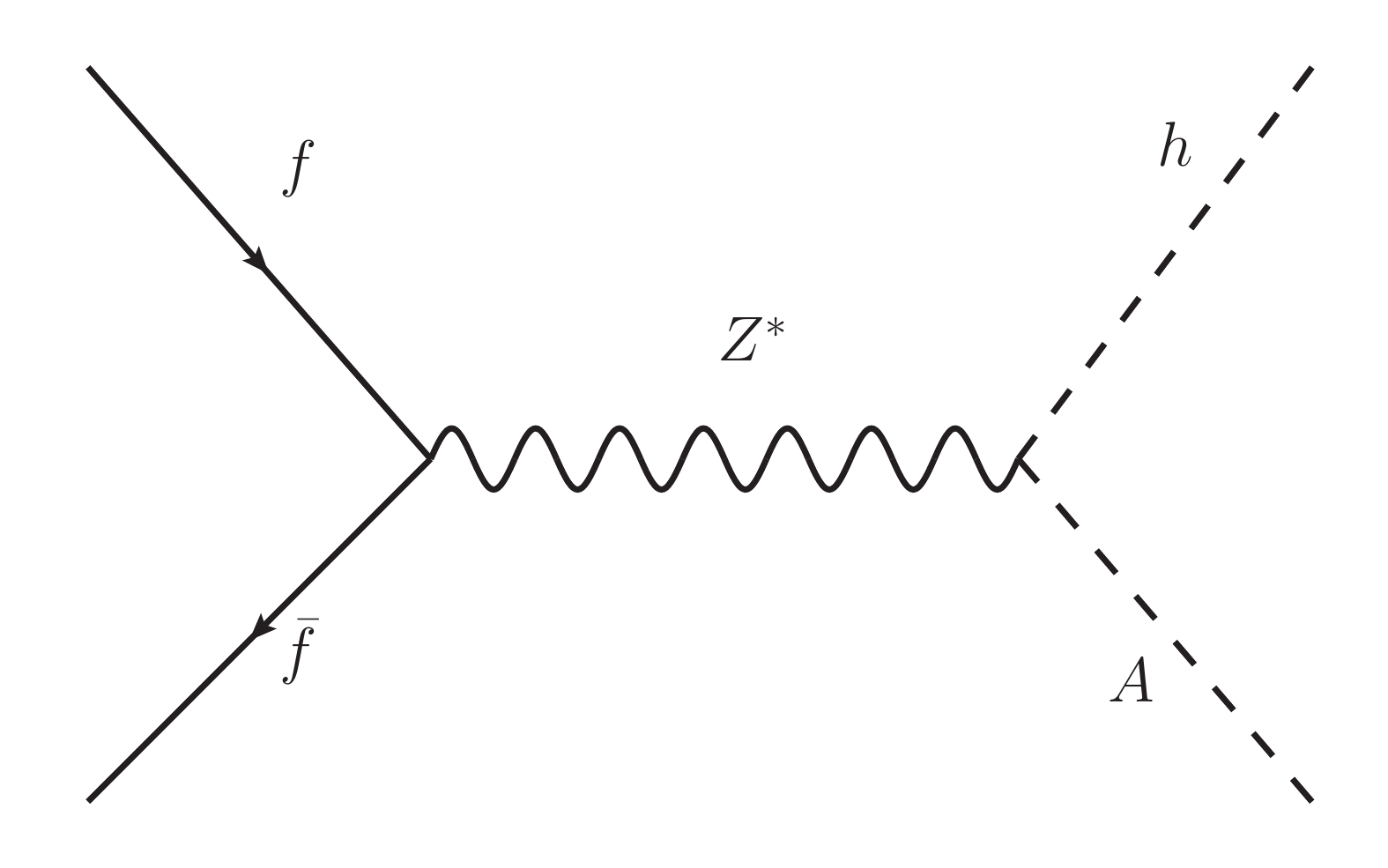}
\end{center}
\caption{Associated production of $h$ and $A$.
\label{fig:hAproduction}}
\end{figure}

Physical effects of antisymmetric VSS vertices
have been well-studied in the literature. For example, in 2HDM,
a benchmark process which embodies the effect of antisymmetric VSS
vertices is
\begin{eqnarray}
f+\bar{f}\rightarrow A+h
\end{eqnarray}
where $A$ and $h$ denote a generic CP-odd and CP-even 2HDM Higgs boson,
respectively. The corresponding Feynman diagram is shown in Fig.~\ref{fig:hAproduction}
in unitarity gauge. Now suppose we replace the antisymmetric VSS $ZhA$ vertex
in Fig.~\ref{fig:hAproduction} by a completely symmetric VSS $ZhA$ vertex.
It is obvious that if the $Z$ boson is on-shell, then the amplitude should
vanish since for an on-shell massive vector-boson we have the relation $p\cdot\epsilon=0$
for its momentum and polarization vector. It is tempting to proceed with the
case that $Z$ boson is off-shell. The amplitude in this case can be examined from two perspectives.
First, we can perform the calculation in unitarity gauge. In this gauge, the result of
dotting the $Z$ momentum $p$ at the $ZhA$ vertex into its s-channel propgator is again proportional
to the $Z$ momentum $p$ at the $Zf\bar{f}$ vertex. It is then obvious that
only the axial-vector part of the $Zf\bar{f}$ vertex contributes to the amplitude,
with a contribution proportional to the fermion mass $m_f$. Alternatively, we may
perform the calculation in Landau gauge ($\xi=0$), in which the diagram shown in
Fig.~\ref{fig:hAproduction} does not contribute to the amplitude, nevertheless we
need to take into account the s-channel Goldstone-mediated amplitude, which again
gives an contribution proportional to the fermion mass $m_f$.

\begin{figure}[ht]
\begin{center}
\includegraphics[width=3.5in]{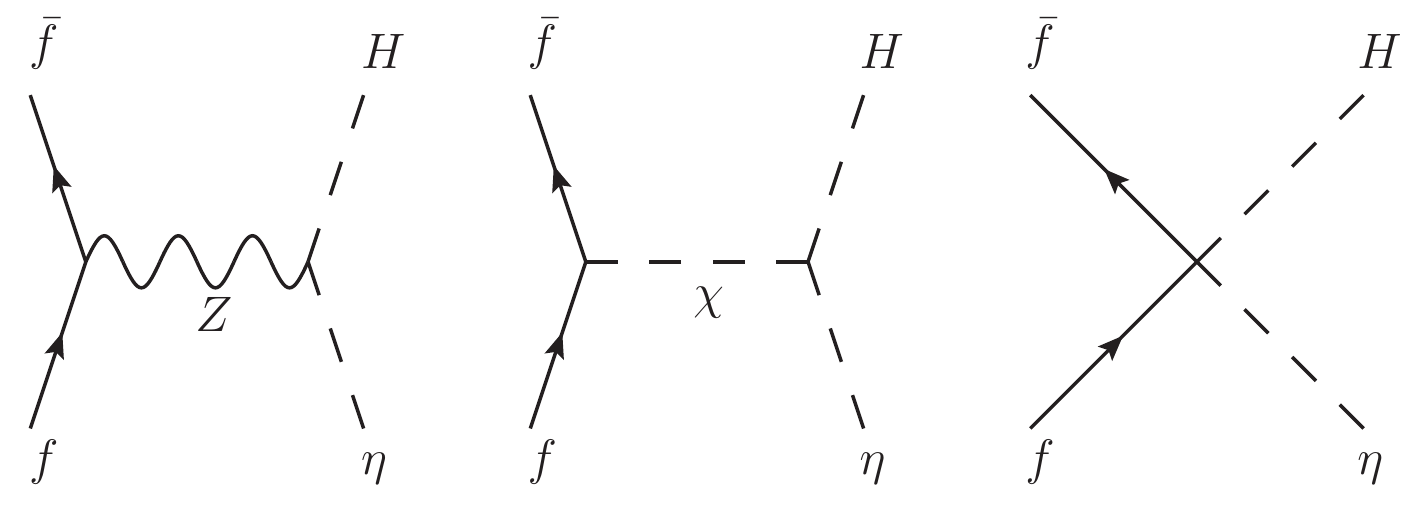}
\end{center}
\caption{Feynman diagrams in the SLH for $f\bar{f}\rightarrow H\eta$ in $R_\xi$ gauge.
\label{fig:ffHeta}}
\end{figure}

Although usually $f$ is a light fermion with negligible mass effects, we might be interested in the case
that $f$ is heavy with important mass effects, e.g. the top quark. If in this case the symmetric
VSS vertex could lead to physical effects, we would seem to produce a paradox in the SLH. In the SLH
there exists a symmetric $ZH\eta$ vertex, however if we consider a linearly-realized SLH as a UV completion,
then it cannot lead to symmetric VSS vertices and hence there will be no related physical effects. Since the usual nonlinearly-realized
SLH can be related to a linearly-realized SLH via an appropriate field redefinition, the above discussion
seems to cause violation of the field redefinition invariance of the S-matrix element\footnote{The radial mode does not help since
it does not have the required CP property.}. We can turn the argument around
to use the field redefinition invariance to infer the existence of additional contribution in the SLH which also
contributes to the $f\bar{f}\rightarrow H\eta$ process such that the field redefinition invariance is
maintained. In fact, if we examine the Yukawa part of the SLH Lagrangian, we would find the following
four-point contact vertex ($m_f$ denotes the mass of $f$)
\begin{align}
\mL\supset i\frac{2\sqrt{2}g_A}{ft_{2\beta}}\frac{m_f}{v}H\eta\bar{f}\gamma^5 f
\end{align}
Here $g_A$ is the axial coupling of the fermion $f$ which also appears in its interaction to $Z$ boson and
the associated Goldstone $\chi$ as
\begin{align}
\mL\supset \frac{g}{2c_W}Z^\mu\bar{f}\gamma_\mu(g_V+g_A\gamma^5)f
+i\frac{2g_A m_f}{v}\bar{f}\gamma^5 f\chi
\end{align}
Now if we compute the amplitude for $f\bar{f}\rightarrow H\eta$ in $R_\xi$ gauge,
we need to include three contributions: s-channel $Z$ exchange, s-channel $\chi$ exchange,
and $ffH\eta$ contact interaction, as shown in Fig.~\ref{fig:ffHeta}.
The amplitudes corresponding to these three diagrams are computed to be (from left to right):
\begin{align}
i\mathcal{M}_I & =\frac{\sqrt{2}}{vft_{2\beta}}\frac{-\xi m_Z^2}{q^2-\xi m_Z^2}2g_A m_f\bar{v}(p_{\bar{f}})\gamma^5 u(p_f) \\
i\mathcal{M}_{II} & =\frac{\sqrt{2}}{vft_{2\beta}}\frac{q^2}{q^2-\xi m_Z^2}2g_A m_f\bar{v}(p_{\bar{f}})\gamma^5 u(p_f) \\
i\mathcal{M}_{III} & =-\frac{\sqrt{2}}{vft_{2\beta}}2g_A m_f\bar{v}(p_{\bar{f}})\gamma^5 u(p_f)
\end{align}
Here $p_f$ and $p_{\bar{f}}$ are the four-momenta of $f$ and $\bar{f}$, respectively and
$q\equiv p_f+p_{\bar{f}}$. When we add the three contributions, we find
\begin{align}
i\mathcal{M}_I+i\mathcal{M}_{II}+i\mathcal{M}_{III}=0
\end{align}
which is exactly what we would expect from field redefinition invariance. Moreover, we see that the $Z$ and $\chi$
contributions add to be gauge-independent, while the contact interaction contribution itself is gauge-independent.

Here we would like to mention a further subtle point related to the symmetric VSS vertex. It might still be
somewhat counter-intuitive the contribution from the symmetric $ZH\eta$ vertex is cancelled by the contribution
from $ffH\eta$ contact vertex, since the former contribution should know the position of $Z$ pole and therefore
vanish for an on-shell $Z$ boson while the latter certainly does not ``feel'' the $Z$ pole. To illustrate this
issue, we can include the effect of $Z$ boson width $\Gamma_Z$ so that the $Z$ boson propagator in the unitarity gauge
is written as
\begin{align}
\frac{-g^{\mu\nu}+\frac{q^\mu q^\nu}{m_Z^2}}{q^2-m_Z^2+im_Z\Gamma_Z}
\label{eq:prop}
\end{align}
When this propagator is dotted into $q_\nu$ coming from the symmetric VSS Feynman rule, at $q^2=m_Z^2$
it will vanish, which seems quite plausible given our previous argument that symmetric VSS vertex
does not contribute to the process in which the related vector boson is on-shell. However, this immediately
leads to the paradoxical situation that near on-shell region the field redefinition invariance
is again violated since the contribution from $ffH\eta$ contact vertex certainly does not know
about the $Z$ pole.

The resolution of this paradox consists in the treatment of particle width in its propagator.
The naive treatment in Eq.~\eqref{eq:prop} is actually not quite correct and will in general
lead to results that violate the Ward-Takahashi identities. A proper treatment can be made
by e.g. employing the complex mass scheme which properly retains gauge invariance. The final result
is, of course, no exotic structure appears near $Z$ pole and the field redefinition invariance
is maintained.

\section{Constraints from Electroweak Precision Observables}
\label{sec:ewpt}

As discussed in Section~\ref{sec:slh} in the study of the pseudo-axion phenomenology there are
eight sign combinations for the rotation parameters $\delta_t,\delta_{Dd},\delta_{Ss}$. Moreover,
when lepton sector is relevant, either $t_\beta\geq1$ or $t_\beta<1$ could be possible, leading to
further complication. Nevertheless, as will be shown in this section, the number of possibilities
greatly reduces if we require
\begin{enumerate}
\item The parameter space under consideration is favored by naturalness consideration and thus embodies
(to some extent) the original motivation of the SLH model.
\item The parameter space under consideration is allowed by electroweak precision measurements.
\end{enumerate}
As discussed in Section~\ref{sec:slh} the first requirement points to the region characterized by
a small top partner mass. In the SLH, currently the lower bound on top partner mass is derived from
Eq.~\eqref{eq:MTmin} where $f$ is stringently constrained by dilepton resonance searches. Constraints from
direct searches for top partner production is not as competitive at the moment. For given $f$,
a small top partner mass could be obtained by requiring a large $t_\beta$ (or $t_\beta^{-1}$ for $t_\beta<1$), which
is in turn bounded by unitarity consideration. To summarize, the first requirement points to the region
characterized by a small $f$ and large $t_\beta$ (or $t_\beta^{-1}$ for $t_\beta<1$).

As to the second requirement, in the present work we consider the following electroweak observables
\begin{enumerate}
\item The $W$ boson mass $m_W$.
\item $R$ observables measured at the $Z$-pole: $R_b,R_c,R_e,R_{\mu},R_{\tau}$, which are defined
by
\begin{align}
R_b & \equiv\Gamma(b\bar{b})/\Gamma(\text{had}),R_c\equiv\Gamma(c\bar{c})/\Gamma(\text{had}),\nonumber \\
R_l & \equiv\Gamma(\text{had})/\Gamma(l^+l^-),l=e,\mu,\tau
\end{align}
in which $\Gamma(\text{had})$ denotes the total hadronic width of the $Z$ boson, and $\Gamma(b\bar{b}),
\Gamma(c\bar{c}), \Gamma(l^+l^-)$ denote the $Z$ boson partial widths into $b\bar{b},c\bar{c},l^+l^-$ channels.
\end{enumerate}
To set up the calculation we choose the fine structure constant $\alpha_{\text{em}}\equiv\frac{e^2}{4\pi}$ (defined
at $Z$-pole), Fermi constant $G_F$ and $Z$ boson mass $m_Z$ as the input parameters. Expressed with the SM
quantities we have the tree level relations
\begin{align}
e=g_{SM}s_{W,SM},\quad \frac{G_F}{\sqrt{2}}=\frac{g_{SM}^2}{8m_{W,SM}^2}
\end{align}
\begin{align}
m_Z^2=\frac{g_{SM}^2 {v_{SM}^2}}{4c_{W,SM}^2},\quad m_{W,SM}^2=\frac{1}{4}g_{SM}^2 v_{SM}^2
\end{align}
These relations get modified in the SLH to be
\begin{align}
e=gs_W,\quad \frac{G_F}{\sqrt{2}}=\frac{g^2}{8m_{W,SLH}^2}\left(1
-\frac{v^2}{4f^2 t_\beta^2}\right)^2
\end{align}
\begin{align}
& m_Z^2=\frac{g^2 v^2}{4c_W^2}+\frac{g^2}{32c_W^2}\left[c_W^{-2}(3-t_W^2)
-\frac{4}{3}s_\beta^{-2}c_\beta^{-2}\right]\frac{v^4}{f^2} \\
& m_{W,SLH}^2=\frac{1}{4}g^2 v^2+\frac{1}{24}g^2(3-s_\beta^{-2}c_\beta^{-2})\frac{v^4}{f^2}
\end{align}
Here we note that in the above equations, as in Section~\ref{sec:slh},
$g,v,s_W$ represent quantities in the SLH and are thus different from the SM
quantities $g_{SM},v_{SM},s_{W,SM}$. From the above two set of relations
we may derive
\begin{align}
\frac{m_{W,SLH}^2}{m_{W,SM}^2}=1+\frac{1}{8}\left(1-t_{W,SM}^2+\frac{1-c_{W,SM}^2}{2c_{W,SM}^2-1}
\frac{4}{t_\beta^2}\right)\frac{v_{SM}^2}{f^2}
\end{align}
\begin{align}
\frac{s_W^2}{s_{W,SM}^2}=1-\frac{1}{8}\left(1-t_{W,SM}^2+\frac{c_{W,SM}^2}{2c_{W,SM}^2-1}
\frac{4}{t_\beta^2}\right)\frac{v_{SM}^2}{f^2}
\label{eq:swc}
\end{align}
To calculate the $R$ observables in the SLH we also need the modified $Z$ couplings
to light fermions. Although the corrections relative to the SM come in in the $\frac{v^2}{f^2}$
order, they are still relevant since the $R$ observables have been measured to a few per mille
precision. In such a case the diagonal entries in the rotational matrices in Eq.~\eqref{eq:dss}
should be understood as $1-\frac{1}{2}\delta_{Dd}^2$ and $1-\frac{1}{2}\delta_{Ss}^2$, respectively.
Then the modified $Z$ couplings to light fermions in the SLH can be written as
\begin{align}
g'_{L,Z,f} & =g_{L,Z,f}+\delta_Z g_{L,Z',f}, \nonumber \\
g'_{R,Z,f} & =g_{R,Z,f}+\delta_Z g_{R,Z',f}, \nonumber \\
& \text{for  } f=u,c,b,e,\mu,\tau
\label{eq:zc1}
\end{align}
In the above equations, $\delta_Z$ is the $\ord\left(\frac{v^2}{f^2}\right)$ $Z-Z'$ mixing angle,
appearing in the mixing relation
\begin{align}
Z'=Z'_m+\delta_Z Z_m,\quad Z=Z_m-\delta_Z Z'_m
\end{align}
Here $Z_m,Z'_m$ denote the final mass eigenstates after the $\ord\left(\frac{v^2}{f^2}\right)$
rotation while $Z,Z'$ denote the states before the $\ord\left(\frac{v^2}{f^2}\right)$
rotation, as define via Eq.~\eqref{eq:gbmixing}. In the process of gauge boson mass diagonalization,
$\delta_Z$ is computed to be
\begin{align}
\delta_Z=-\frac{(1-t_W^2)\sqrt{3-t_W^2}}{8c_W}\frac{v^2}{f^2}
\label{eq:deltaZ}
\end{align}
In Eq.~\eqref{eq:zc1}, $g_{L,Z,f}=\frac{g}{c_W}(T_3^f-Q_f s_W^2),
g_{R,Z,f}=-\frac{g}{c_W}Q_f s_W^2$ are leading-order coefficients of the Lagrangian
terms $\bar{f}_L\gamma^\mu f_L Z_\mu,\bar{f}_R\gamma^\mu f_R Z_\mu$ and $T_3^f,Q_f$
denote the third component of the isospin and the electric charge of $f$, respectively.
$g_{L,Z',f},g_{R,Z',f}$ are leading-order coefficients of the Lagrangian
terms $\bar{f}_L\gamma^\mu f_L Z'_\mu,\bar{f}_R\gamma^\mu f_R Z'_\mu$, which are
given in Eq.~\eqref{eq:zpl},Eq.~\eqref{eq:zp3q} and Eq.~\eqref{eq:zp12q}. $g'_{L,Z,f},g'_{R,Z,f}$
in Eq.~\eqref{eq:zc1} denote the coefficients of the Lagrangian
terms $\bar{f}_L\gamma^\mu f_L Z_\mu,\bar{f}_R\gamma^\mu f_R Z_\mu$ and $T_3^f,Q_f$
, to the $\ord\left(\frac{v^2}{f^2}\right)$ precision.

For $f=d$ the modified $Z$ couplings in the SLH turn out to be
\begin{align}
g'_{L,Z,d} & =g_{L,Z,d}+\delta_Z g_{L,Z',d}+\delta_{Dd}^2 (g_{L,Z,D}-g_{L,Z,d}), \nonumber \\
g'_{R,Z,d} & =g_{R,Z,d}+\delta_Z g_{R,Z',d}
\label{eq:zc2}
\end{align}
Obviously the additional correction is due to the left-handed $D-d$ mixing. The corresponding
formulae for $f=s$ can be obtained by the replacement $d\rightarrow s,D\rightarrow S$.
$g_{L,Z,D},g_{L,Z,S}$ are leading-order coefficients of the Lagrangian
terms $\bar{D}_L\gamma^\mu D_L Z_\mu,\bar{S}_R\gamma^\mu S_R Z_\mu$
\begin{align}
g_{L,Z,D}=g_{L,Z,S}=\frac{1}{3}gs_W t_W
\end{align}
Now we have all the SLH couplings that are necessary to calculate the $R$ observables. It should
be noted that in the above coupling formulae, $s_W,c_W,t_W$ are quantities in the SLH and are
therefore different from their SM counterparts $s_{W,SM},c_{W,SM},t_{W,SM}$, see Eq.~\eqref{eq:swc}. Therefore, the modification
of $Z$ couplings to light fermions relative to the SM is caused by three factors: $Z-Z'$ mixing,left-handed
$D-d,S-s$ mixing, and correction of the weak-mixing angle.

A $95\%$ CL level constraint can be obtained in the $f-t_\beta$ plane by performing a $\chi^2$-fit of
the five $R$ observables. The $\chi^2$ is defined by
\begin{align}
\chi^2=\sum_{f=b,c,e,\mu,\tau} \frac{(R_{f,SLH}-R_f)^2}{\delta_{R_f}^2+\delta_{R_{f,SM}}^2}
\end{align}
\begin{figure*}[ht]
\includegraphics[width=2.2in]{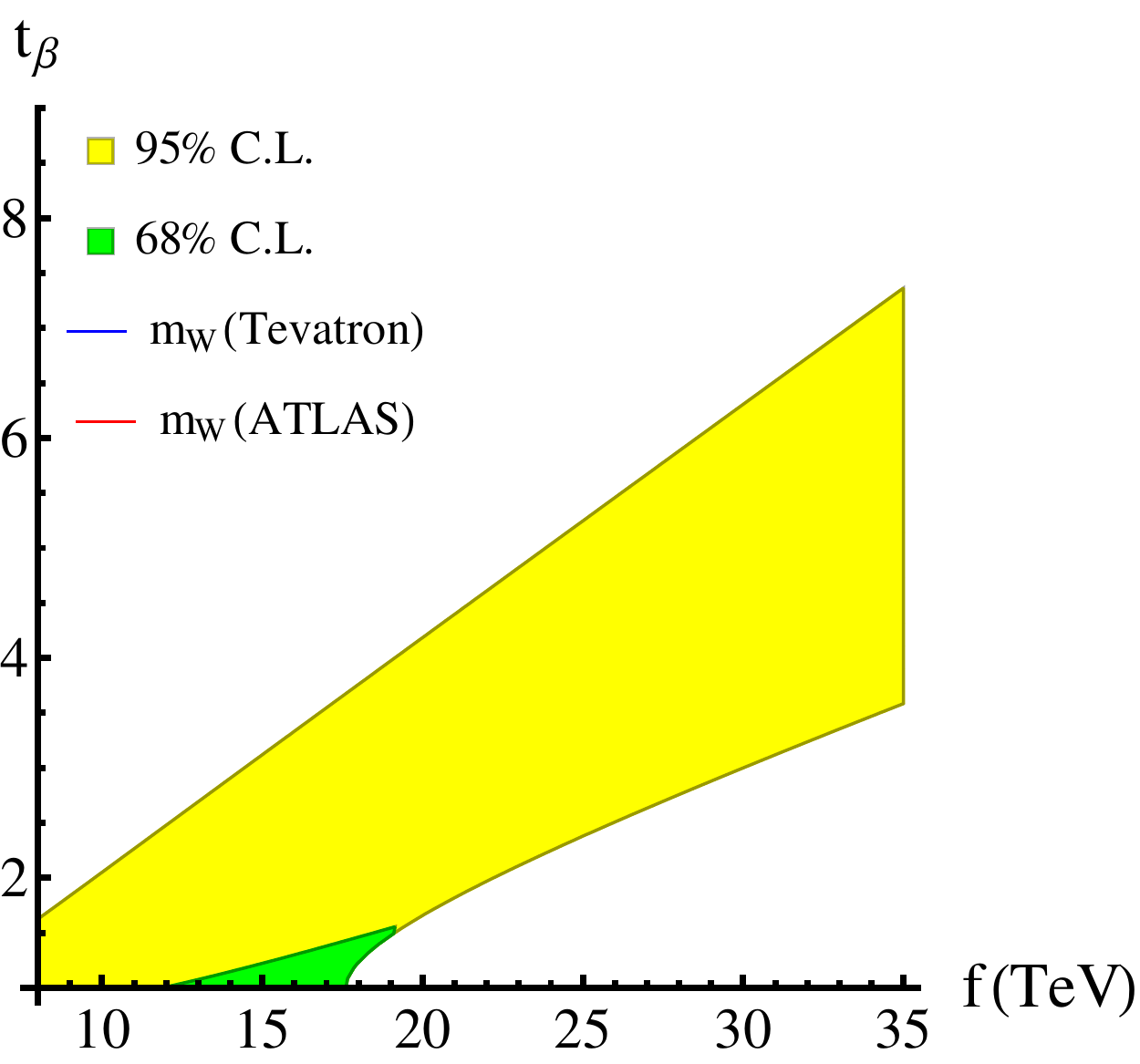}
\includegraphics[width=2.2in]{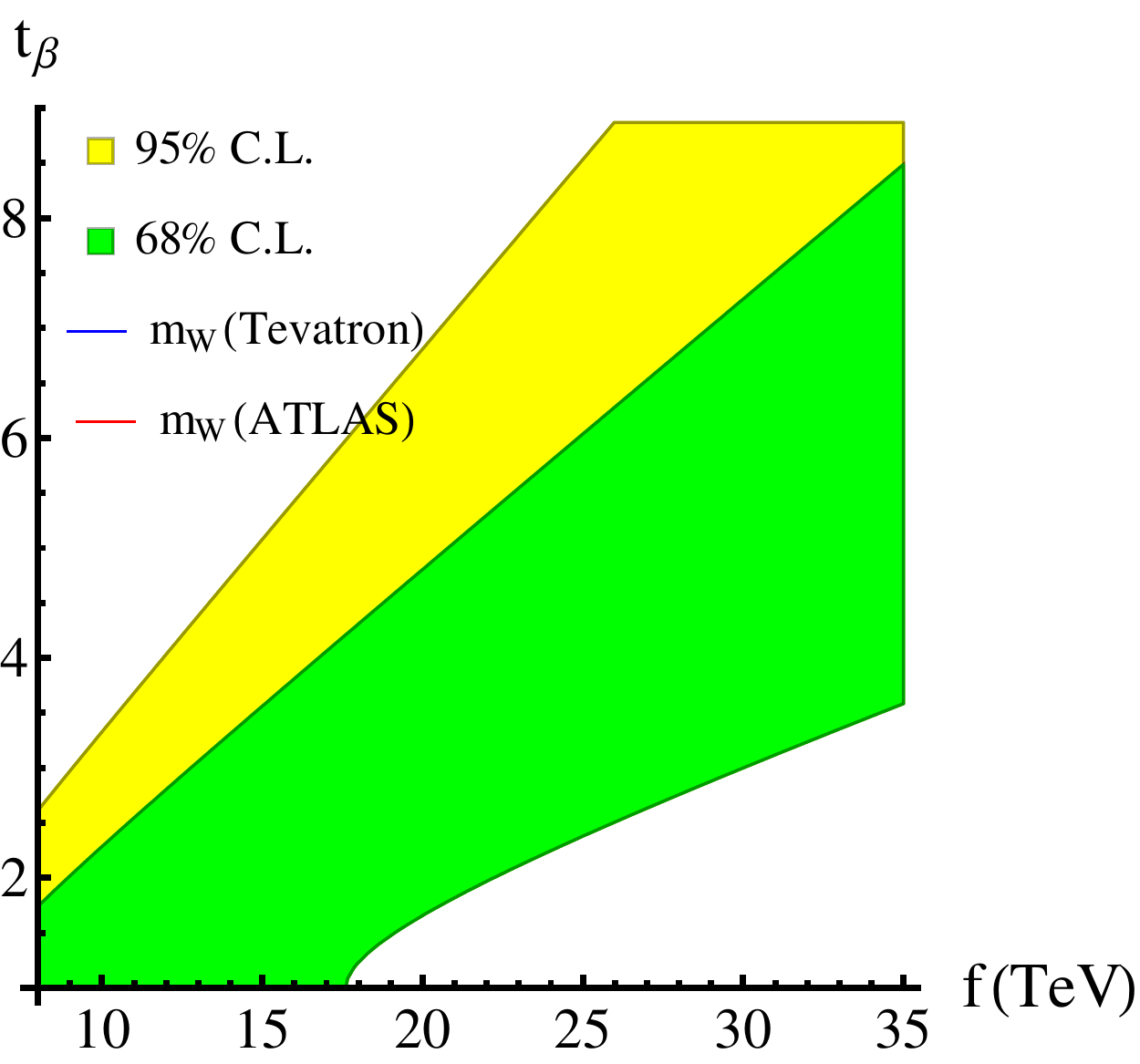}
\includegraphics[width=2.2in]{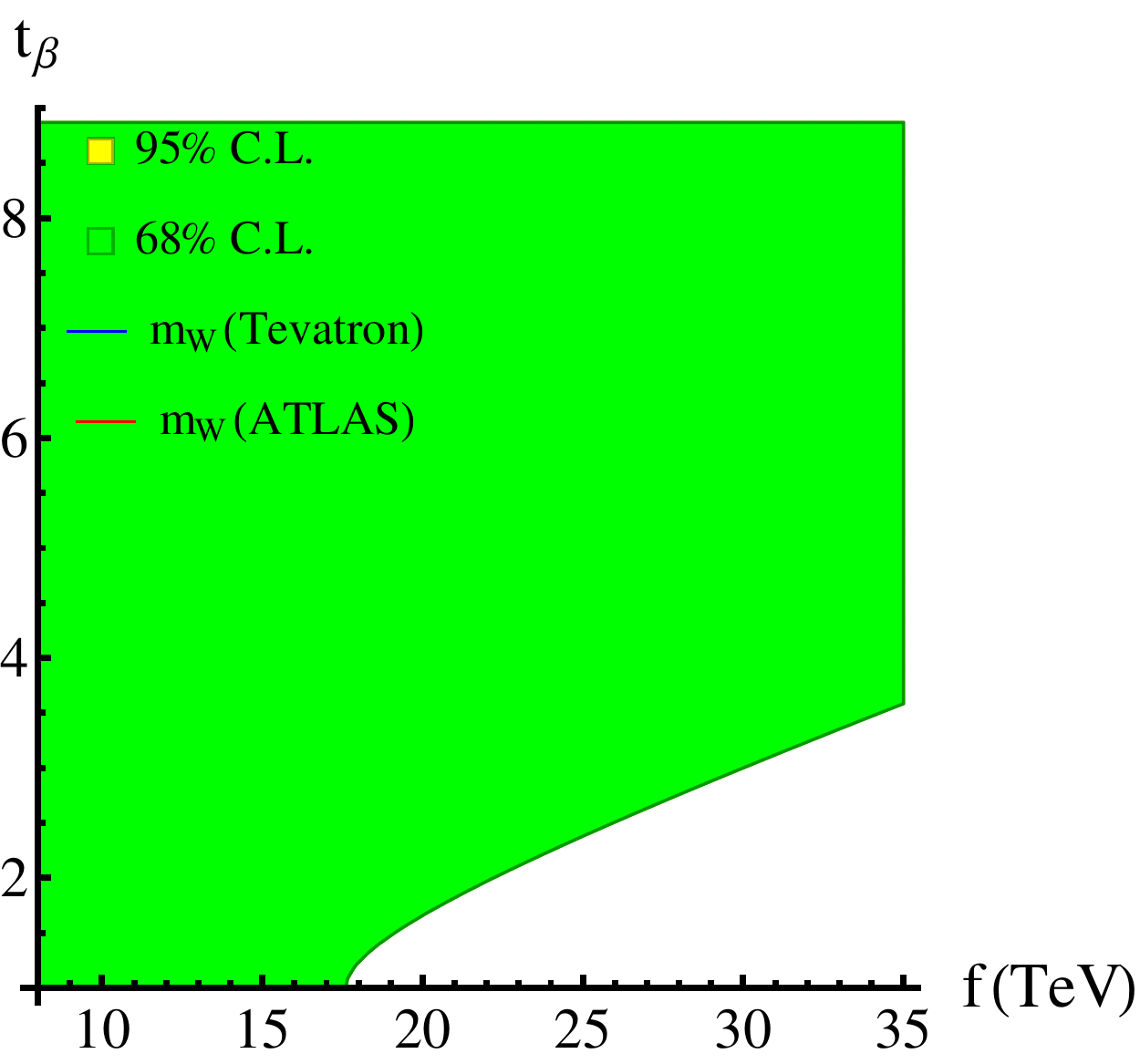} \\
\vspace{0.5cm}
\includegraphics[width=2.2in]{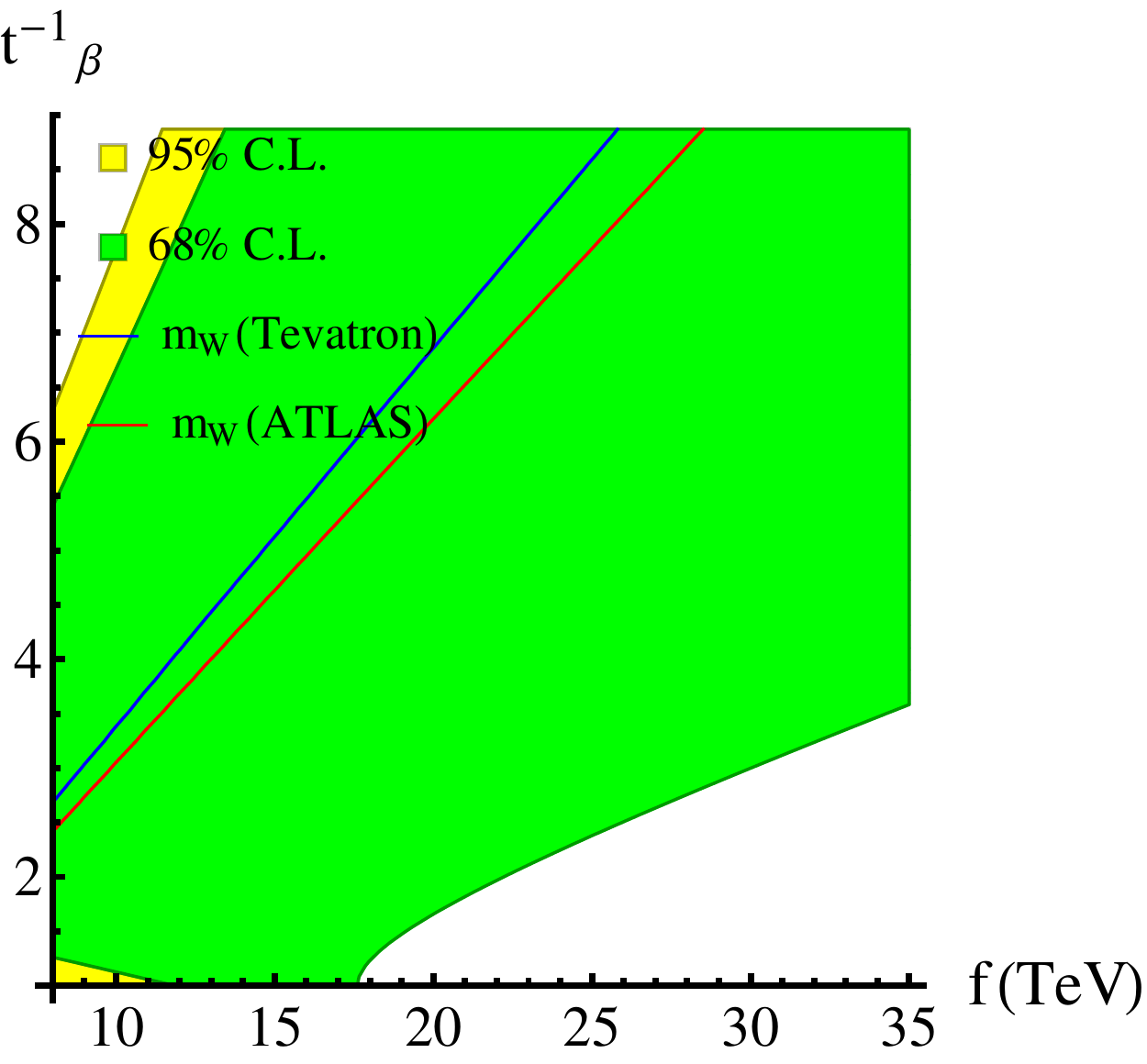}
\includegraphics[width=2.2in]{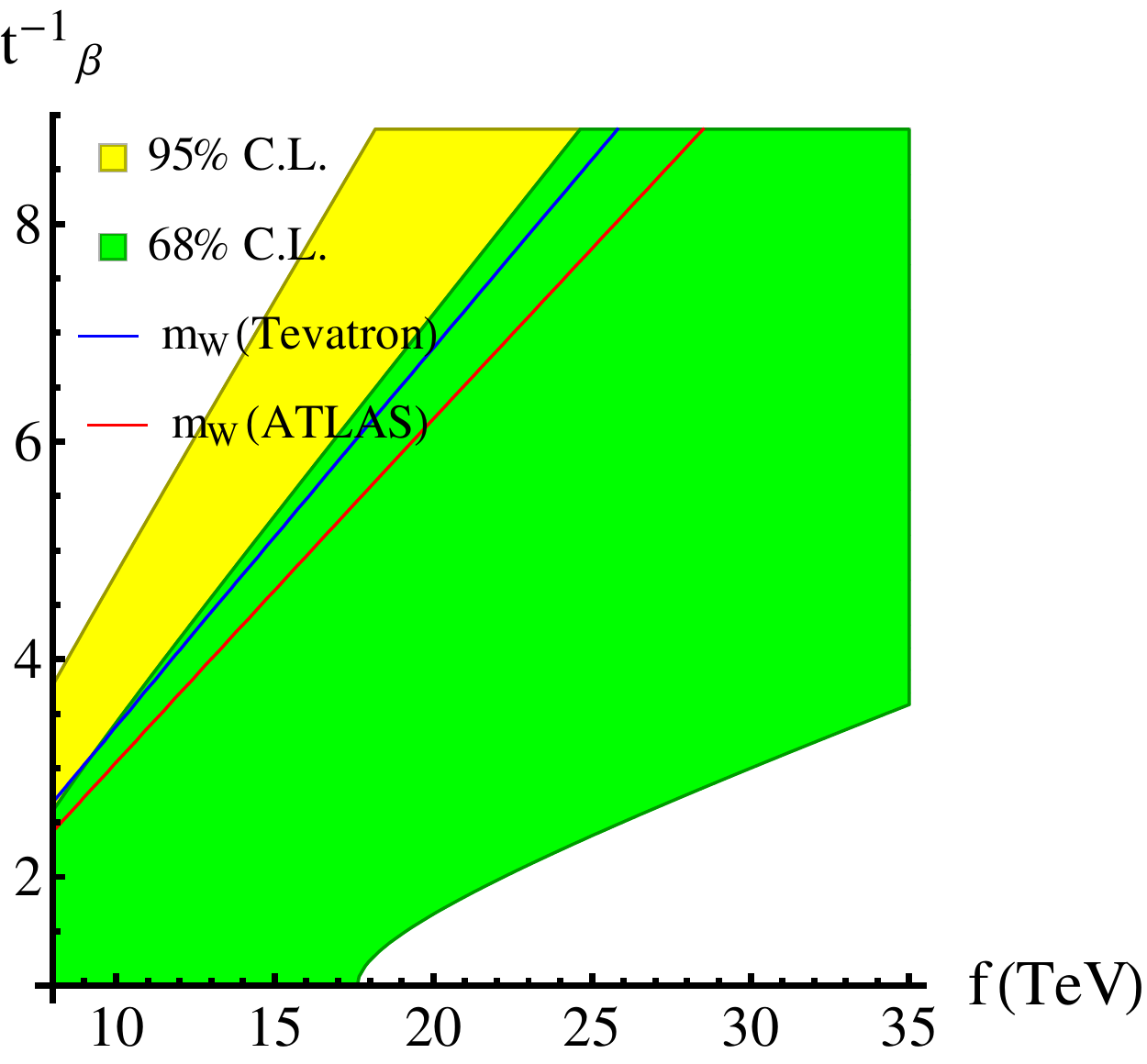}
\includegraphics[width=2.2in]{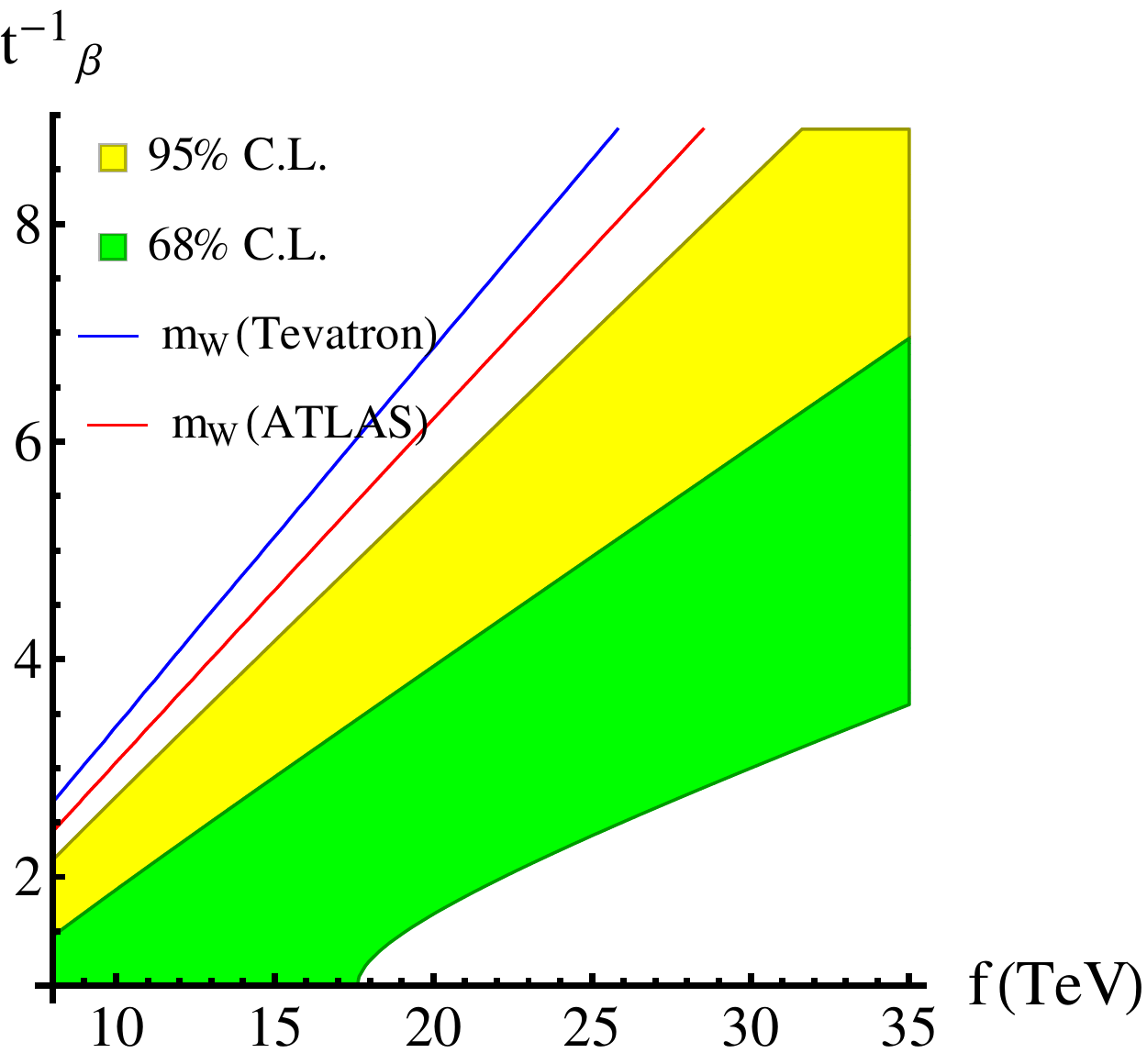}
\caption{\label{fig:ewpt}Constraints from $m_W$ and $R$ observables on the $f-t_\beta$ plane.
Upper left: $t_\beta\geq 1,\delta_{Dd}^+,\delta_{Ss}^+$,
upper middle: $t_\beta\geq 1,\delta_{Dd}^+,\delta_{Ss}^-$ or $t_\beta\geq 1,\delta_{Dd}^-,\delta_{Ss}^+$,
upper right: $t_\beta\geq 1,\delta_{Dd}^-,\delta_{Ss}^-$,
lower left: $t_\beta\leq 1,\delta_{Dd}^+,\delta_{Ss}^+$,
lower middle: $t_\beta\leq 1,\delta_{Dd}^+,\delta_{Ss}^-$ or $t_\beta\geq 1,\delta_{Dd}^-,\delta_{Ss}^+$,
lower right: $t_\beta\leq 1,\delta_{Dd}^-,\delta_{Ss}^-$. See the text for detailed description.}
\vspace{0.5cm}
\end{figure*}
In the above equation, $R_f$ denote the experimental values and
$\delta_{R_f}$ denotes the associated experimental uncertainty. Also, $R_{f,SM}$ is
the SM theory prediction and $\delta_{R_{f,SM}}$ denotes the associated theory uncertainty.
Their values are listed in Table ~\ref{table:rq}~\cite{Erler:2018rpp}.
\begin{table}[t!]
\begin{tabular}{|c|c|c|}
\hline
Quantity & Value & Standard Model \\
\hline
$R_e$ & $20.804\pm0.050$ & $20.737\pm0.010$ \\
\hline
$R_\mu$ & $20.785\pm0.033$ & $20.737\pm0.010$ \\
\hline
$R_\tau$ & $20.764\pm0.045$ & $20.782\pm0.010$ \\
\hline
$R_b$ & $0.21629\pm0.00066$ & $0.21582\pm0.00002$ \\
\hline
$R_c$ & $0.1721\pm0.0030$ & $0.17221\pm0.00003$ \\
\hline
\end{tabular}
\caption{Experimental values and the SM predictions of the $R$ observables.}
\label{table:rq}
\end{table}
As to the constraint from $W$ boson mass, we treat it separately and consider two
most precise measurements~\cite{Erler:2018rpp}
\begin{align}
m_W &=80.387\pm0.016\GeV\quad \text{(Tevatron)} \\
m_W &=80.370\pm0.019\GeV\quad \text{(ATLAS)}
\end{align}
while we note the SM prediction for $m_W$ is~\cite{Erler:2018rpp}
\begin{align}
m_{W,SM}=80.358\pm0.004\GeV
\end{align}

In Figure~\ref{fig:ewpt} the results of the electroweak precision analysis of $m_W$
and $R$ observables are shown. To clarify the situation we present the
results according to whether $t_\beta\geq 1$ and the sign combination
of the rotation parameters $\delta_{Dd},\delta_{Ss}$ (see Eq.~\eqref{eq:sc12}).
At first sight there are eight possibilities in total, however it is immediately
recognized that $\delta_{Dd}^+,\delta_{Ss}^-$ and $\delta_{Dd}^-,\delta_{Ss}^+$
make no difference in terms of constraints in the $f-t_\beta$ plane, reducing
the number of possibilities to six. Therefore we obtain the six panels in
Figure~\ref{fig:ewpt}, each panel showing one possibility as described in the
caption.

For all the panels, the green and yellow regions correspond to parameter points
that are allowed by $\chi^2$-fit of $R$ observables at $68\%$ and $95\%$ CL, respectively.
These allowed regions do not exhibit a $t_\beta\rightarrow t_\beta^{-1}$ symmetry (for example,
the allowed region in the upper right panel and the lower left panel still differ under the transformation
$t_\beta\rightarrow t_\beta^{-1}$), since in the computation of $R$ observables, the correction
of $s_W^2$ relative to its SM value has to be taken into account, as was pointed out previously.
When $f$ is larger than about $17\TeV$ there will be a lower theoretical bound (from the mass relation)
on $t_\beta$ or $t_\beta^{-1}$ which is larger than $1$, corresponding to the white region at large $f$ and small
$t_\beta$ or $t_\beta^{-1}$ in each panel. The $2\sigma$ constraints from $m_W$ measurements are simply
implemented by requiring
\begin{align}
|m_{W,SLH}-m_W|<2\sqrt{\delta_{m_W}^2+\delta_{m_{W,SM}}^2}
\end{align}
In the above equation $m_W$ denotes the experimentally measured $W$ boson mass and $\delta_{m_W}$ and
$\delta_{m_{W,SM}}$ denote the associated experimental and theoretical uncertainties, respectively.
We superimpose the constraint boundary on the six plots as blue or red lines, representing constraints
from Tevatron or ATLAS measurements, respectively. For all these $m_W$ constraint boundary lines,
the regions on the right side of the lines are allowed at $2\sigma$ level.
\begin{figure*}[ht]
\includegraphics[width=2.2in]{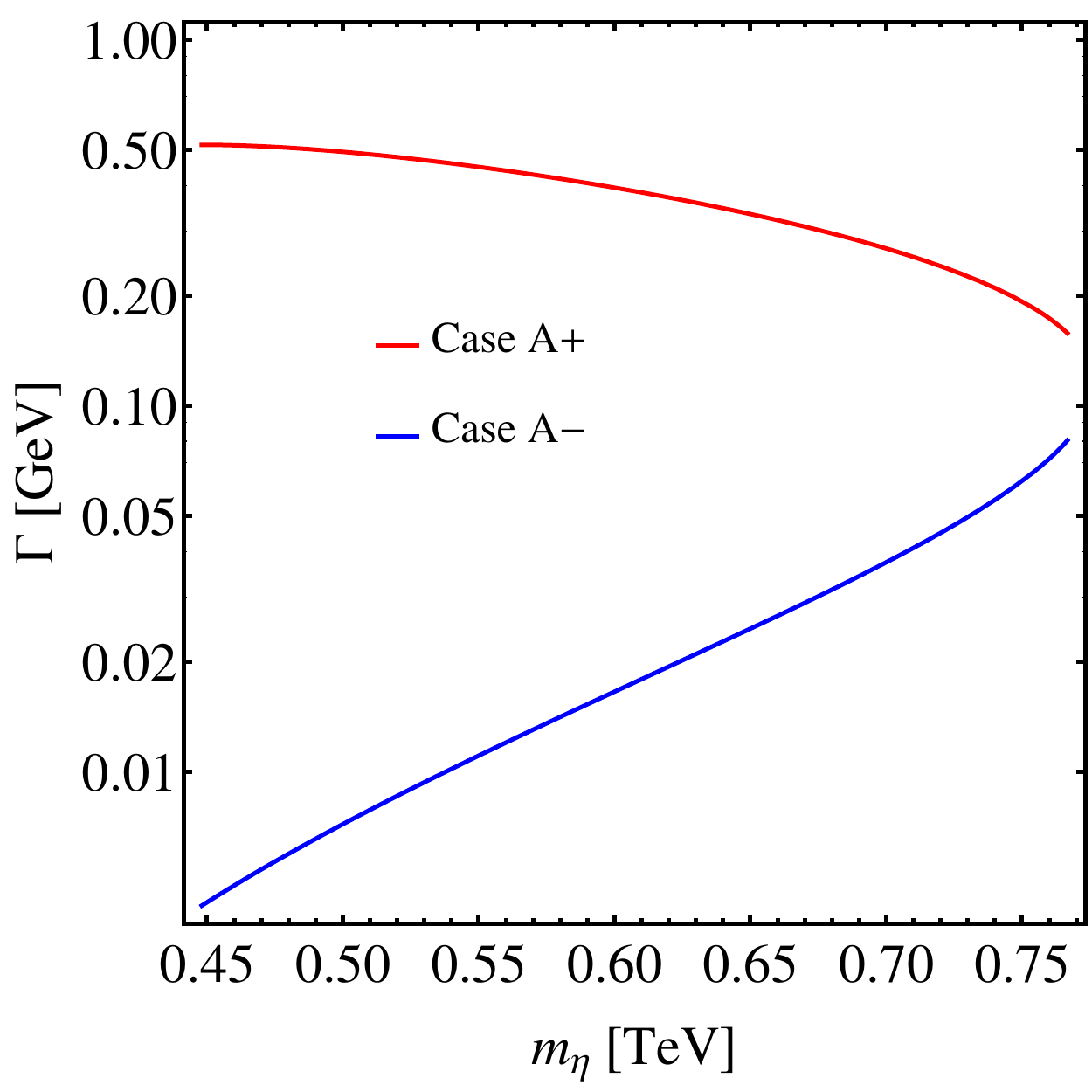}
\includegraphics[width=2.2in]{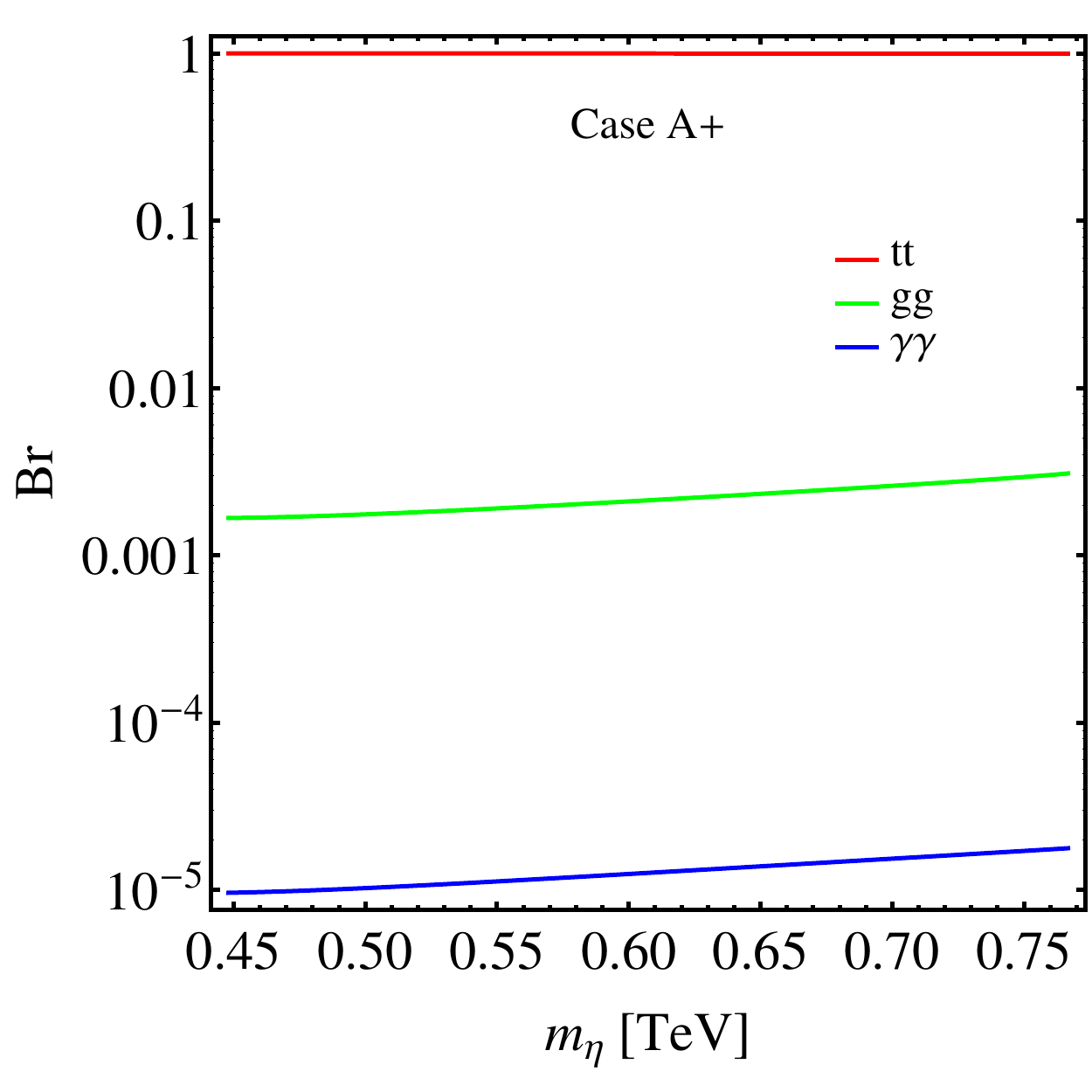}
\includegraphics[width=2.2in]{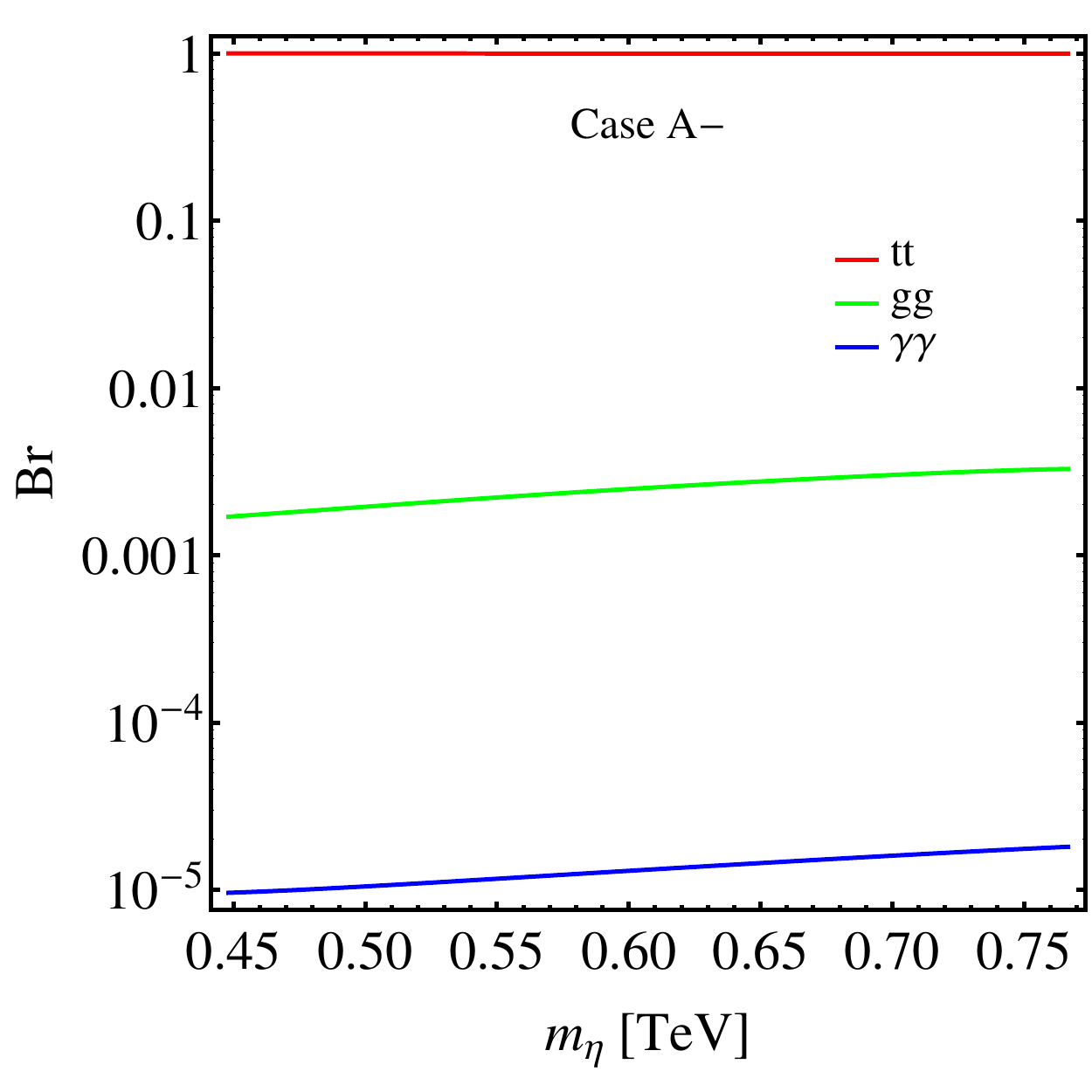}
\caption{\label{fig:etadA}Total width $\Gamma$ and decay branching ratios of $\eta$ in Case A.
}
\end{figure*}
\begin{figure*}[ht]
\includegraphics[width=2.2in]{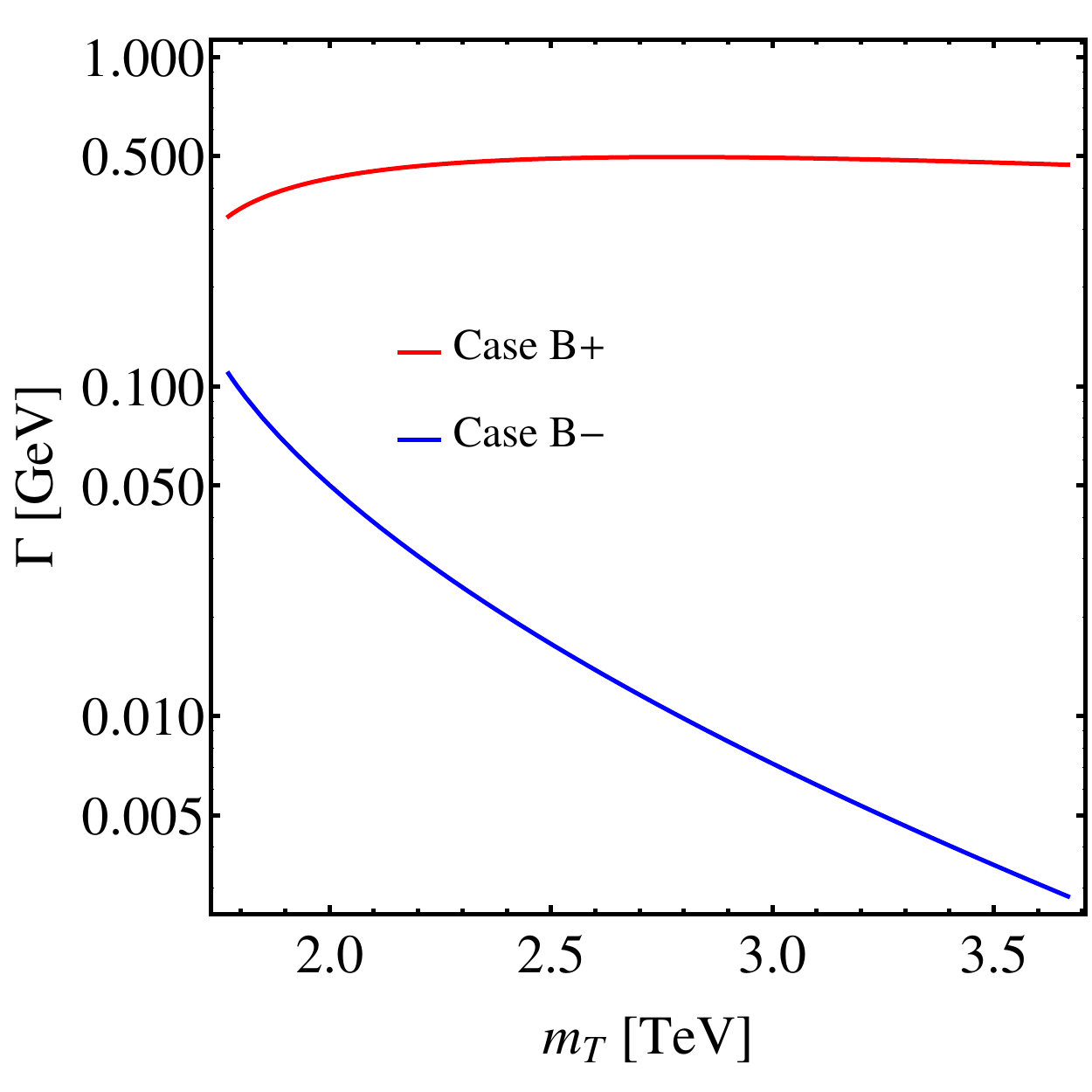}
\includegraphics[width=2.2in]{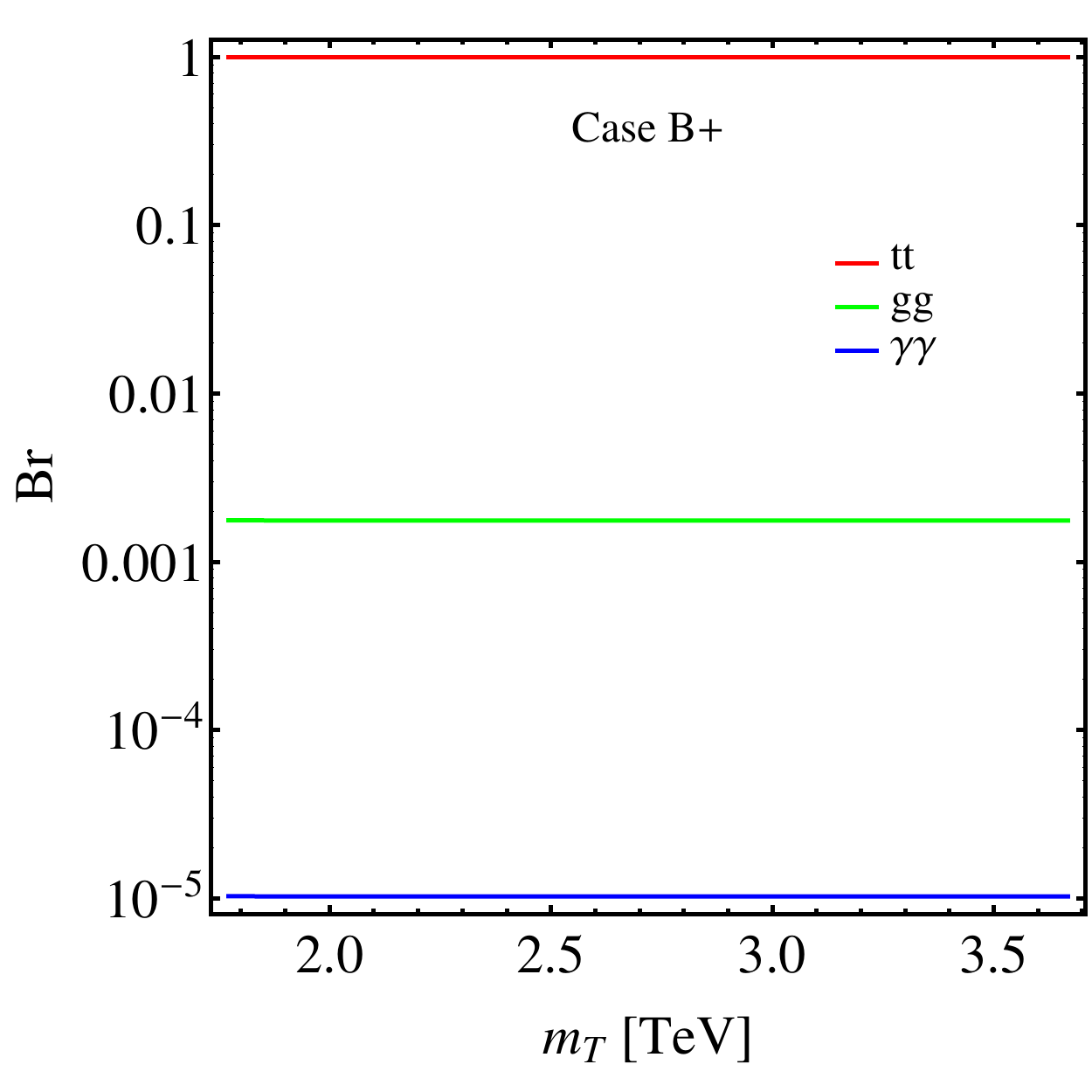}
\includegraphics[width=2.2in]{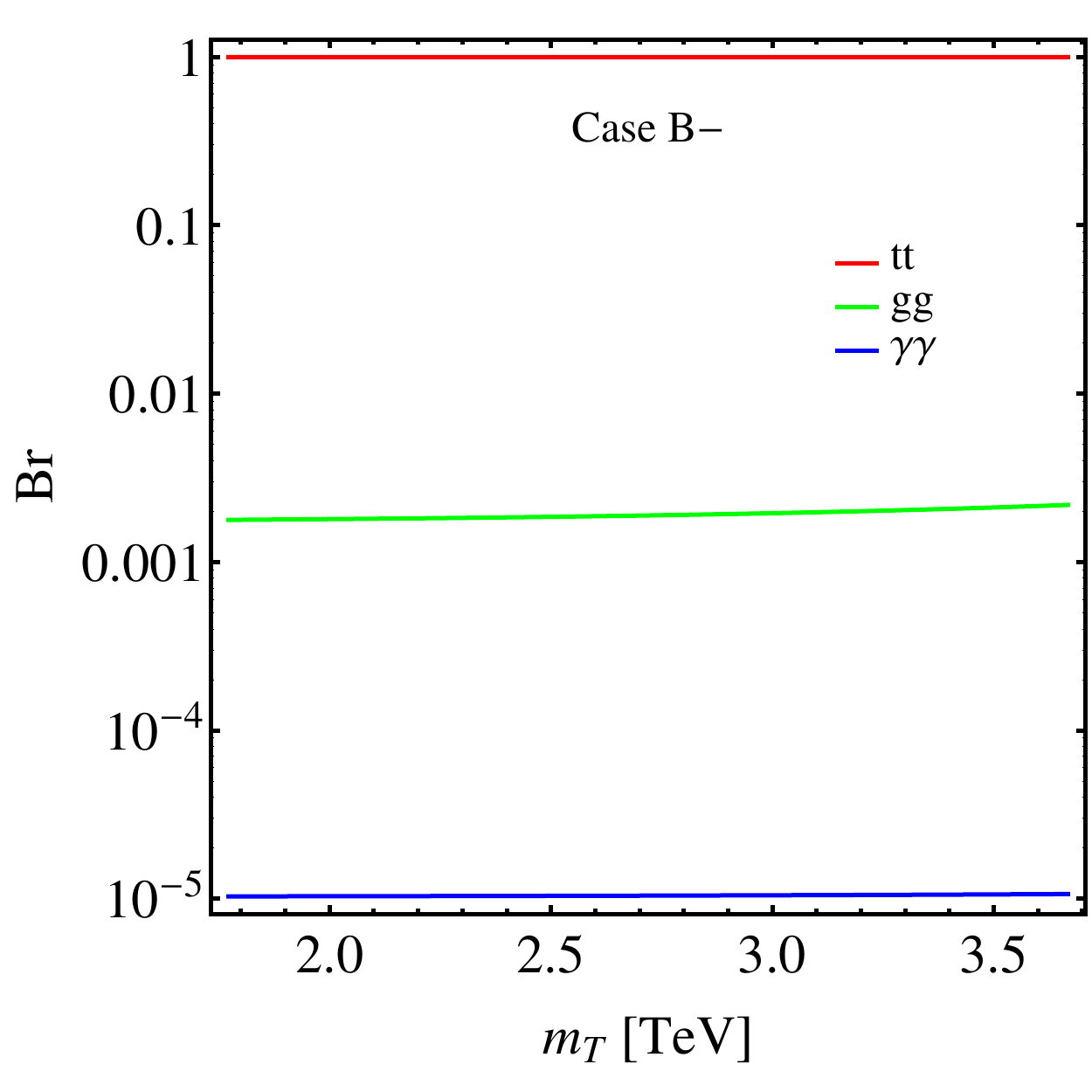}
\caption{\label{fig:etadB}Total width $\Gamma$ and decay branching ratios of $\eta$ in Case B.
}
\end{figure*}

As can be seen from Figure~\ref{fig:ewpt}, if $t_\beta<1$, then the region favored by naturalness
consideration is disfavored by constraints from both $R$ observables and $W$ boson mass measurements,
regardless of the sign combination of the rotation parameters $\delta_{Dd},\delta_{Ss}$. If $t_\beta\geq 1$,
then $W$ boson mass measurement does not constrain the parameter region favored by naturalness
consideration. However, in this
case constraints from $R$ observables are significant when any of the rotation parameters
$\delta_{Dd},\delta_{Ss}$ adopt the plus sign in Eq.~\eqref{eq:sc12}. This is because the choice of
plus sign leads to a large $t_\beta$ enhancement of the rotation parameter and therefore a larger deviation
of $Z$ couplings to the corresponding fermion. Although the lower bound on $f$ has been pushed to around
$7.5\TeV$ by LHC dilepton resonance searches, the $R$ observable constraints still force us
to avoid this $t_\beta$ enhancement, and consequently the only possibility left is
$\delta_{Dd}^-,\delta_{Ss}^-$ with $t_\beta\geq 1$. This result has important
consequences for the pseudo-axion phenomenology since the sign combinations of
$\delta_{Dd},\delta_{Ss}$ will determine how $\eta$ interacts with the $D,S$ quarks which
in turn influences the decay and production of the $\eta$ particle, as will be discussed
in more detail in the next section.

In previous literature on the SLH model the $t_\beta\geq 1$ and $t_\beta<1$ cases are usually
not distinguished, since a $t_\beta\rightarrow t_\beta^{-1}$ symmetry is tacitly assumed.
Then only the $t_\beta\geq 1$ case is considered. However strictly speaking this symmetry is
only valid when the leptonic sector is not considered. Here we established clearly that
if we consider the region favored by naturalness consideration, the $t_\beta<1$ case is
disfavored by measurements $m_W$ and $R$ observables. This is closely related to the
breakdown of the $t_\beta\rightarrow t_\beta^{-1}$ symmetry in the lepton sector. Moreover,
in previous literature~\cite{Han:2005ru,Reuter:2012sd}, the sign combination of the rotation parameter $\delta_{Dd},\delta_{Ss}$
was simply \emph{assumed} to be (effectively) $\delta_{Dd}^-,\delta_{Ss}^-$, in order to suppress
contribution to the electroweak precision observables. Here we also establish firmly this
choice based on constraints from $R$ observables, combined with $m_W$ and naturaless consideration,
keeping in mind that the constraint on $f$ has been pushed to around $7.5\TeV$ due to updated
LHC constraints.

\section{Production and Decay of the Pseudo-Axion}
\label{sec:prod}

With the preparation made in the previous three sections we are now ready to calculate
the production and decay of the pseudo-axion. We will restrict ourselves to
the region $2m_t\lesssim m_\eta\lesssim 1\TeV$, which is favored by naturalness consideration.
All the related partial widths formulae are given in Appendix~\ref{sec:pwf}.

\subsection{Decay of the Pseudo-Axion}
\begin{figure*}[ht]
\includegraphics[width=2.2in]{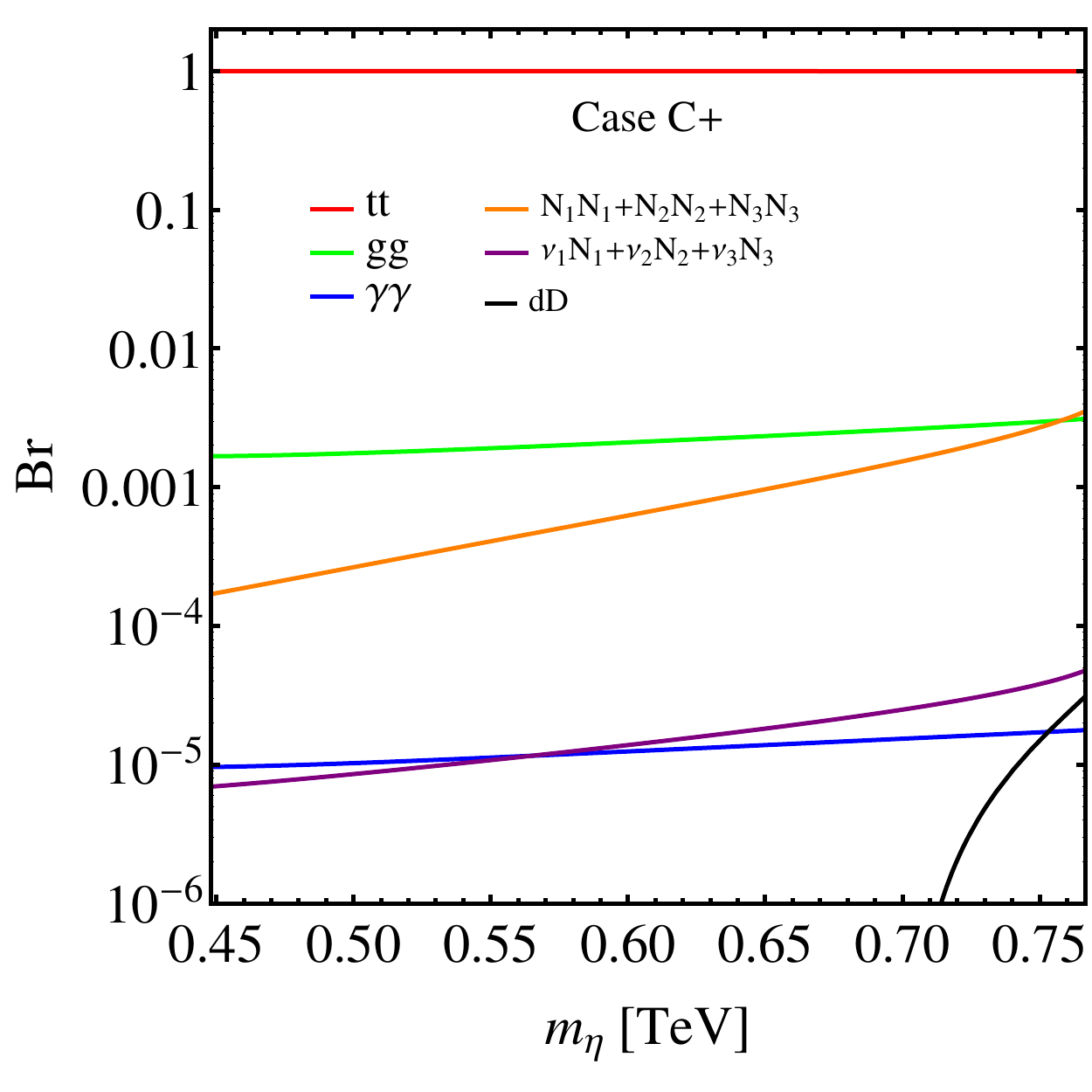}
\includegraphics[width=2.2in]{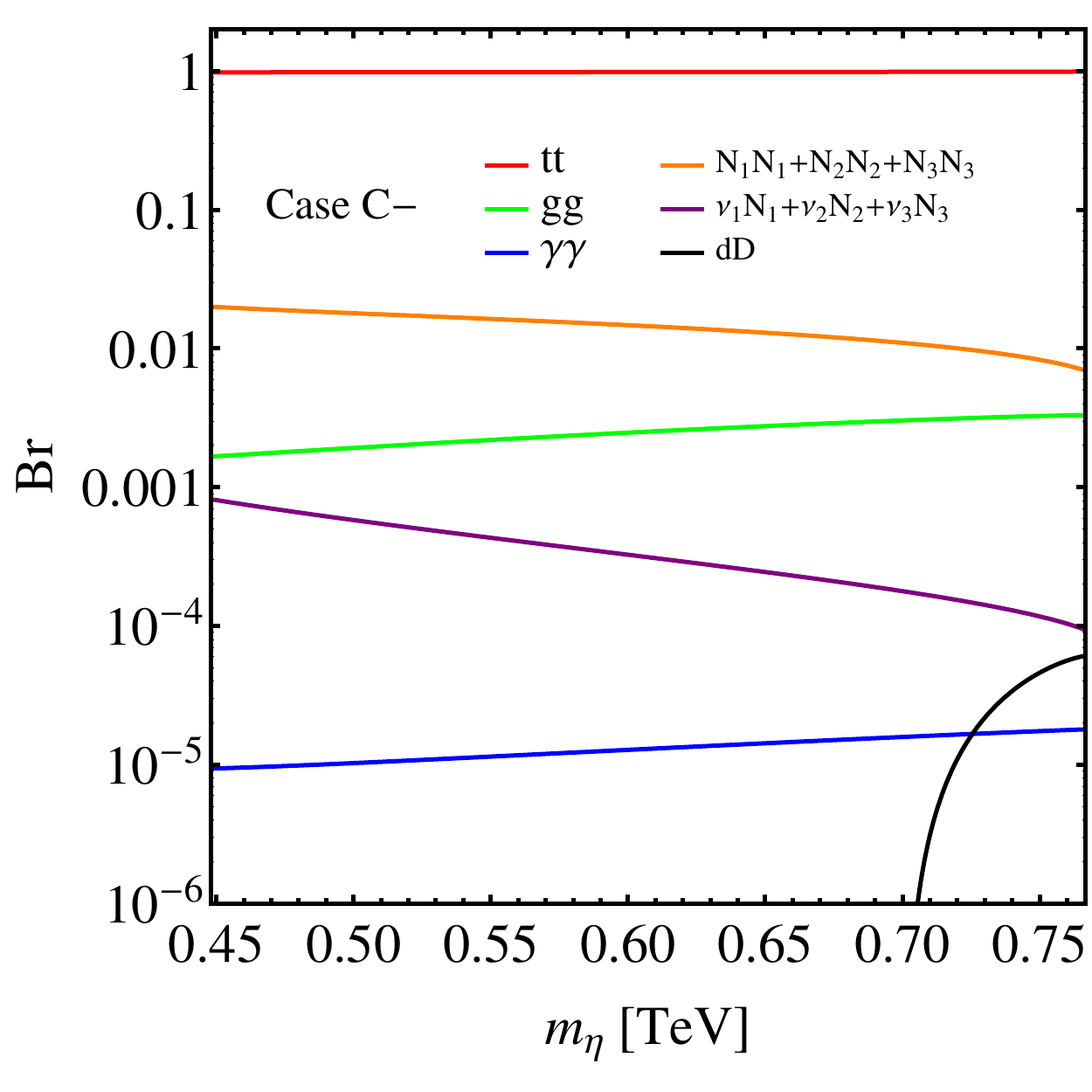} \\
\includegraphics[width=2.2in]{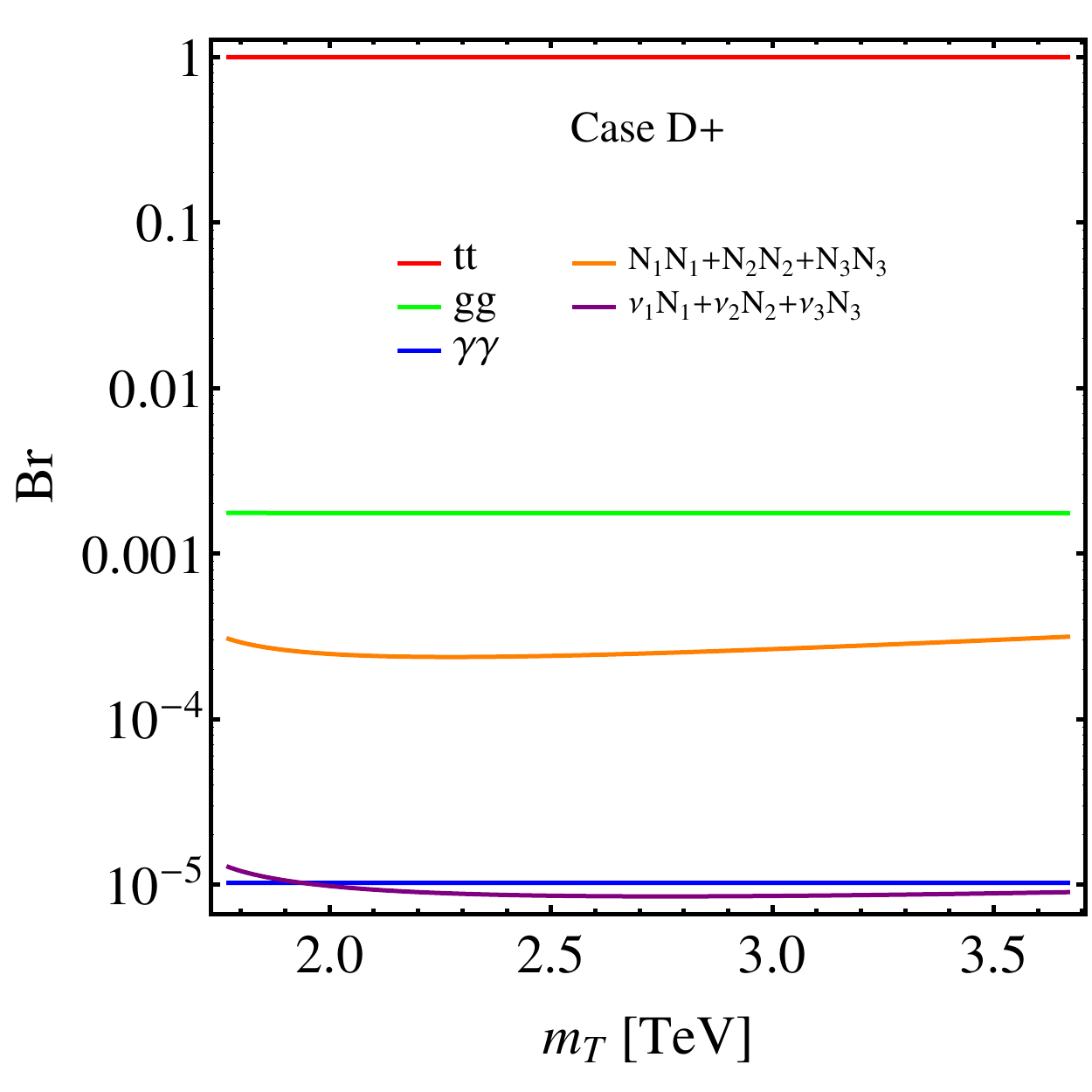}
\includegraphics[width=2.2in]{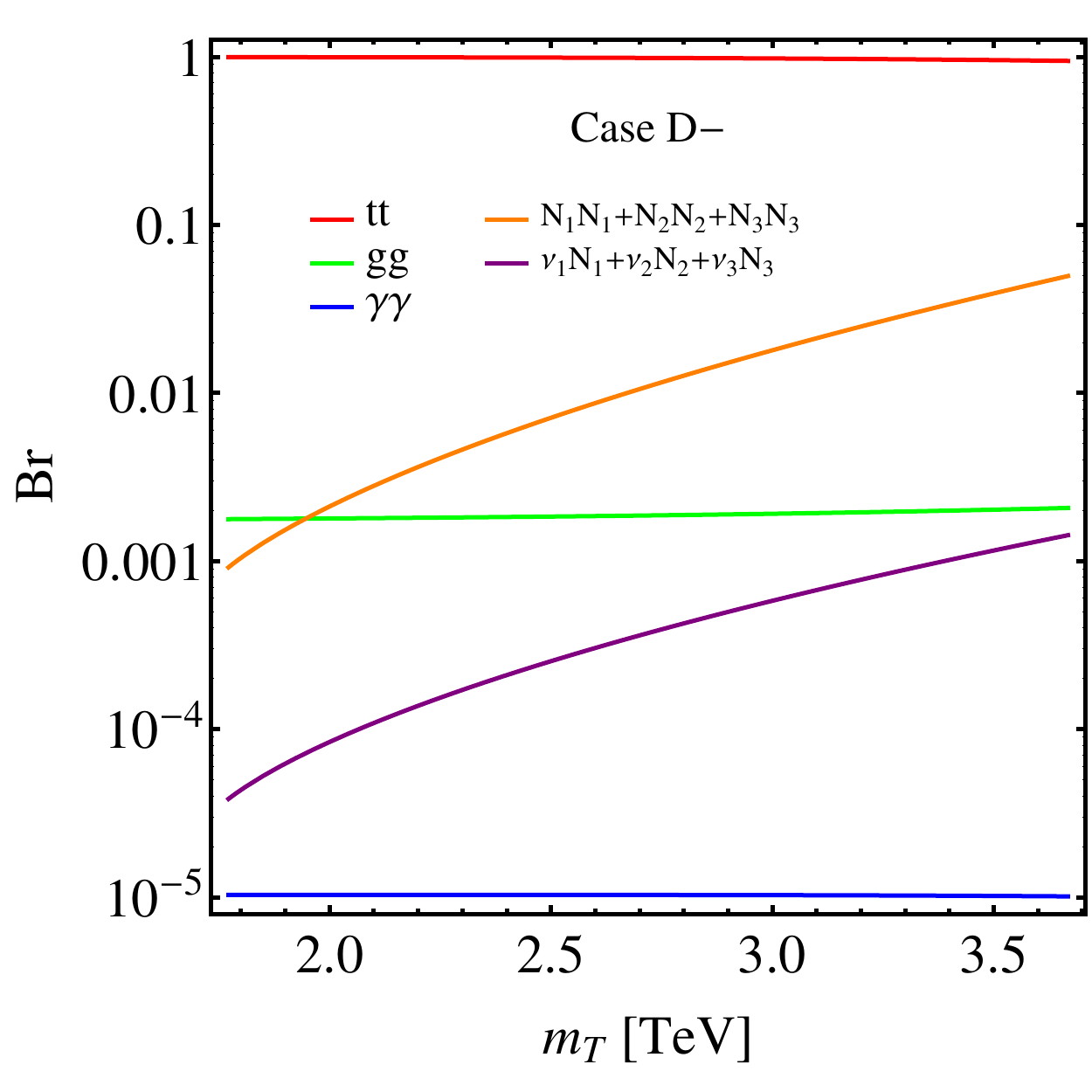}
\caption{\label{fig:etadCD}Decay branching ratios of $\eta$ in Case C and Case D.
}
\end{figure*}
For $\eta$ in the mass range $2m_t\lesssim m_\eta\lesssim 1\TeV$, it can always decay into
$t\bar{t},gg,\gamma\gamma$ channels. (The $WW,ZZ,Z\gamma$ channels are also possible
and may have comparable branching ratio compared to $\gamma\gamma$. However
from a detection viewpoint, it is preferrable to consider further decays into leptons in these
channels, leading to an additional suppression by the leptonic branching. For simplicity we
will not consider these channels further in this work.). $\eta\rightarrow ZH$ is highly suppressed,
since the antisymmetric $ZH\eta$ vertex is suppressed to $\ord\left(\frac{v^3}{f^3}\right)$ while
the symmetric $ZH\eta$ vertex does not contribute, as pointed out in Section~\ref{sec:vss}. If the
new fermions $D,S,N$ are heavy enough such that they cannot appear as decay products of $\eta$,
then we are left with only the $t\bar{t},gg,\gamma\gamma$ channels. Nevertheless we should keep in
mind that when $f$ and $m_\eta$ are given, the partial withds of these channels still depend on
the masses of the additional heavy quarks $T,D,S$ which do not appear as decay products of $\eta$.
First, the $\eta\rightarrow t\bar{t}$ decay is controlled by the rotation parameter $\delta_t$,
which in turn depends on the top partner mass. The loop-induced decays $\eta\rightarrow gg,\gamma\gamma$
have contributions from both the top quark and the heavy quark partners $T,D,S$. The top quark
contribution again depends on $\delta_t$ while the $T,D,S$ contributions depend on the $\eta T\bar{T},
\eta D\bar{D},\eta S\bar{S}$ couplings which are propotional to the corresponding rotation parameters
times the quark partner mass. Experimentally the current lower bound for light-flavor
quark partner $D$ and $S$ is around $700\GeV$~\cite{Sirunyan:2017lzl}. Thus for a heavy enough $\eta$
the $\eta\rightarrow Dd,Ss$ channels are still possible if the mass of $D$ or $S$ is close
to the lower bound. To be definite, we will consider four benchmark scenarios:
\begin{enumerate}
\item Case A: $f=8\TeV, m_T=m_D=m_S=3\TeV, \text{all }m_N>m_\eta$.
\item Case B: $f=8\TeV, m_\eta=500\GeV, m_D=m_S=m_T, \text{all }m_N>m_\eta$.
\item Case C: $f=8\TeV, m_T=3\TeV,m_D=700\GeV,m_S=1\TeV,\text{all }m_N=150\GeV$.
\item Case D: $f=8\TeV, m_\eta=500\GeV, m_D=m_S=m_T, \text{all }m_N=150\GeV$.
\end{enumerate}
For each case, there are two allowed sign combinations for the rotation parameters $(\delta_t,\delta_{Dd},\delta_{Ss})$:
$(+,-,-)$ and $(-,-,-)$. Other choices are excluded by electroweak precision measurements, if we are only
interested in parameter region favored by nartualness consideration. Therefore in the following we
will use Case A$+$, Case A$-$, etc. to indicate the sign choice of $\delta_t$ in each case (see Eq.~\eqref{eq:signchoice}).

\begin{figure*}[ht]
\includegraphics[width=2.2in]{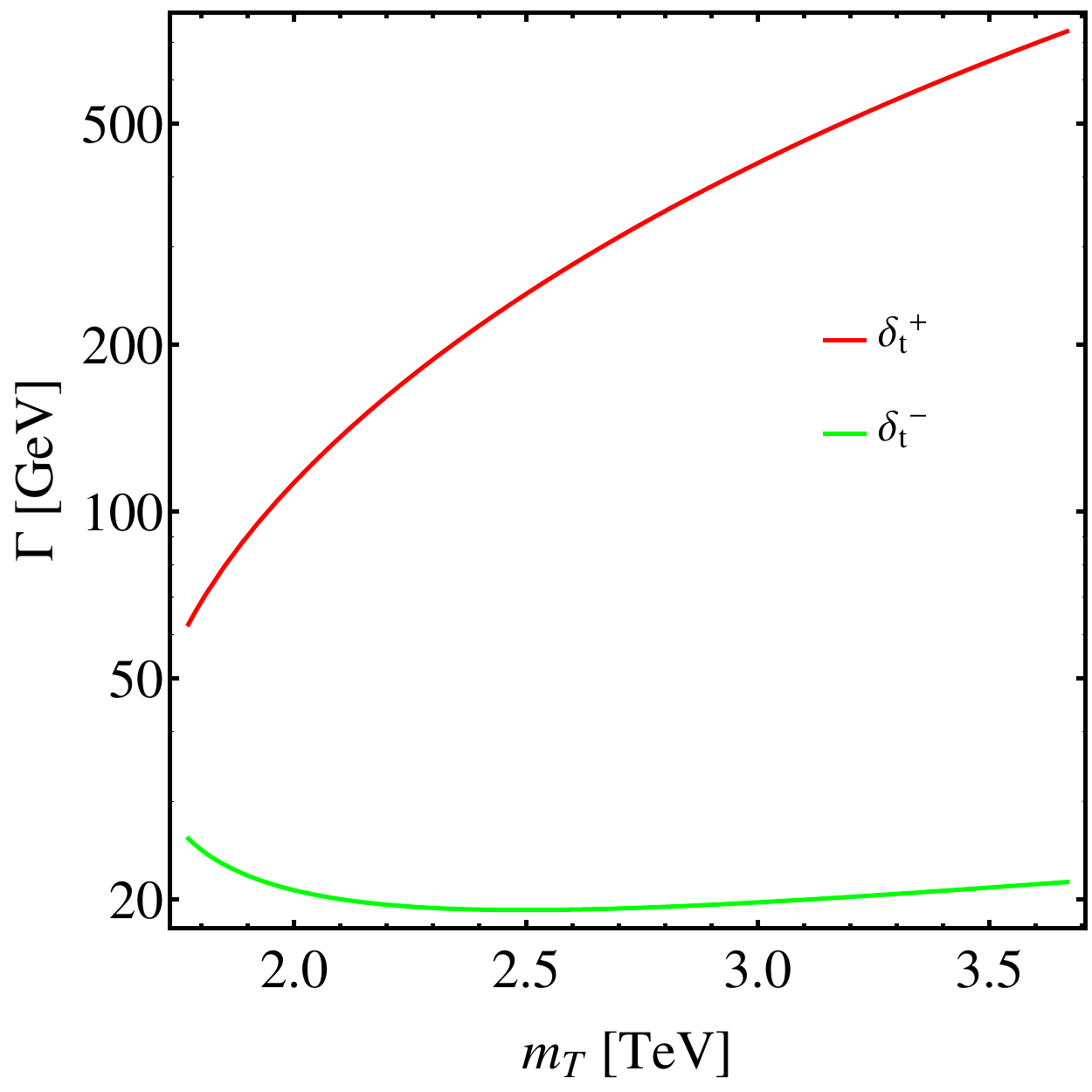}
\includegraphics[width=2.2in]{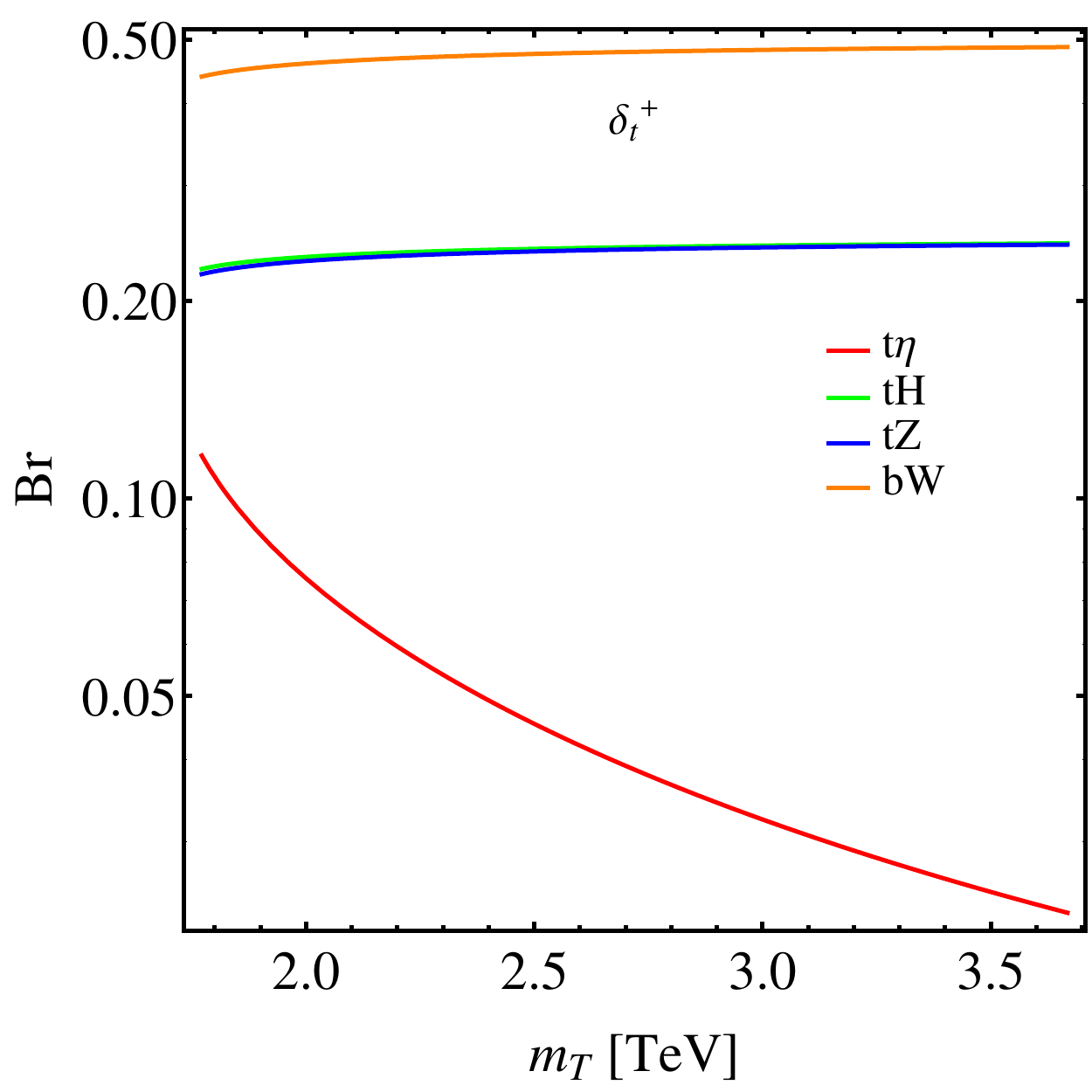}
\includegraphics[width=2.2in]{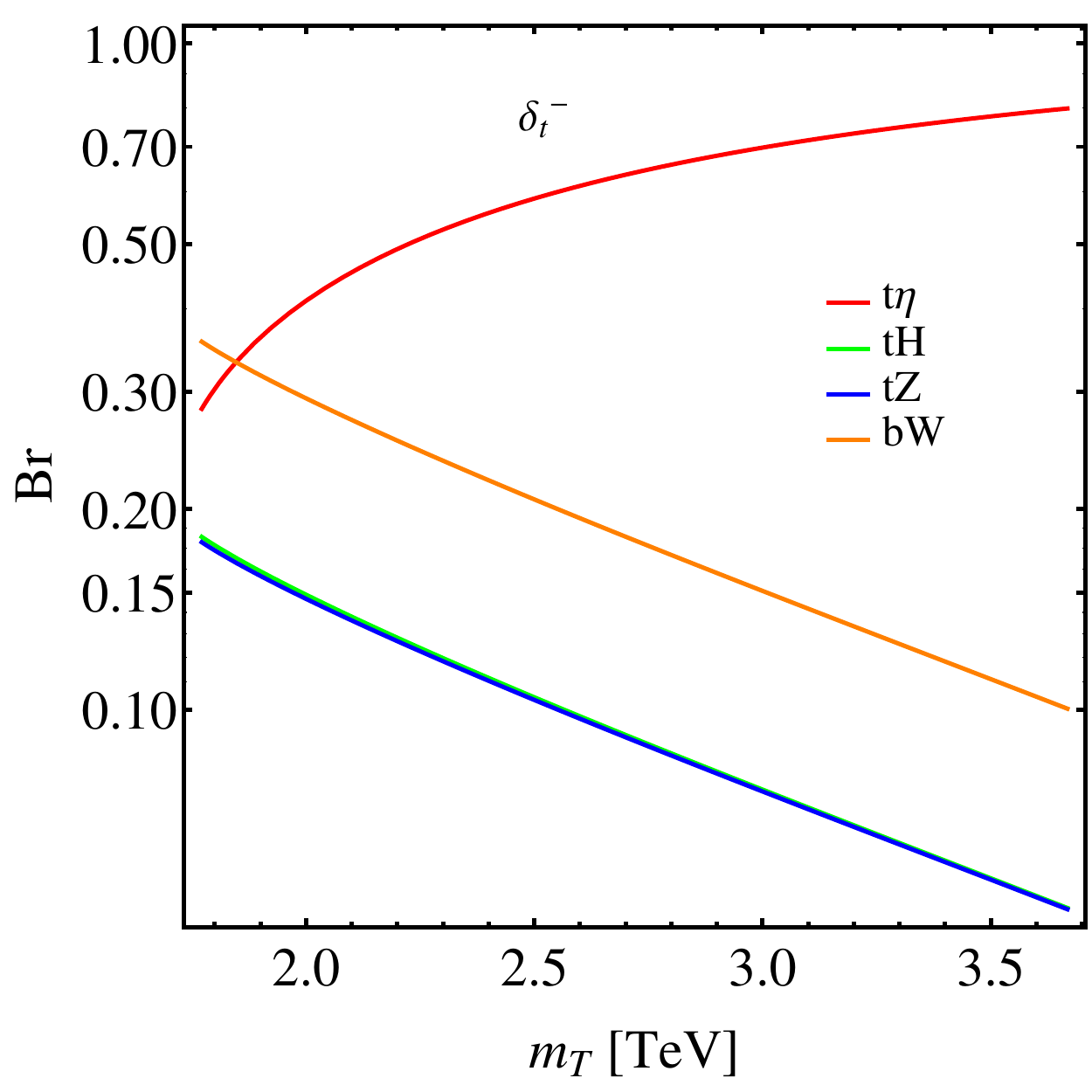}
\caption{\label{fig:Tdecay}Total width $\Gamma$ and decay branching ratios of $T$ in the SLH.
We assume $f=8\TeV$ and $m_\eta=500\GeV$. Note that in the considered mass range $T\rightarrow bX, tY$ channels do not open.}
\end{figure*}
\begin{figure*}[htbp]
\includegraphics[width=2.2in]{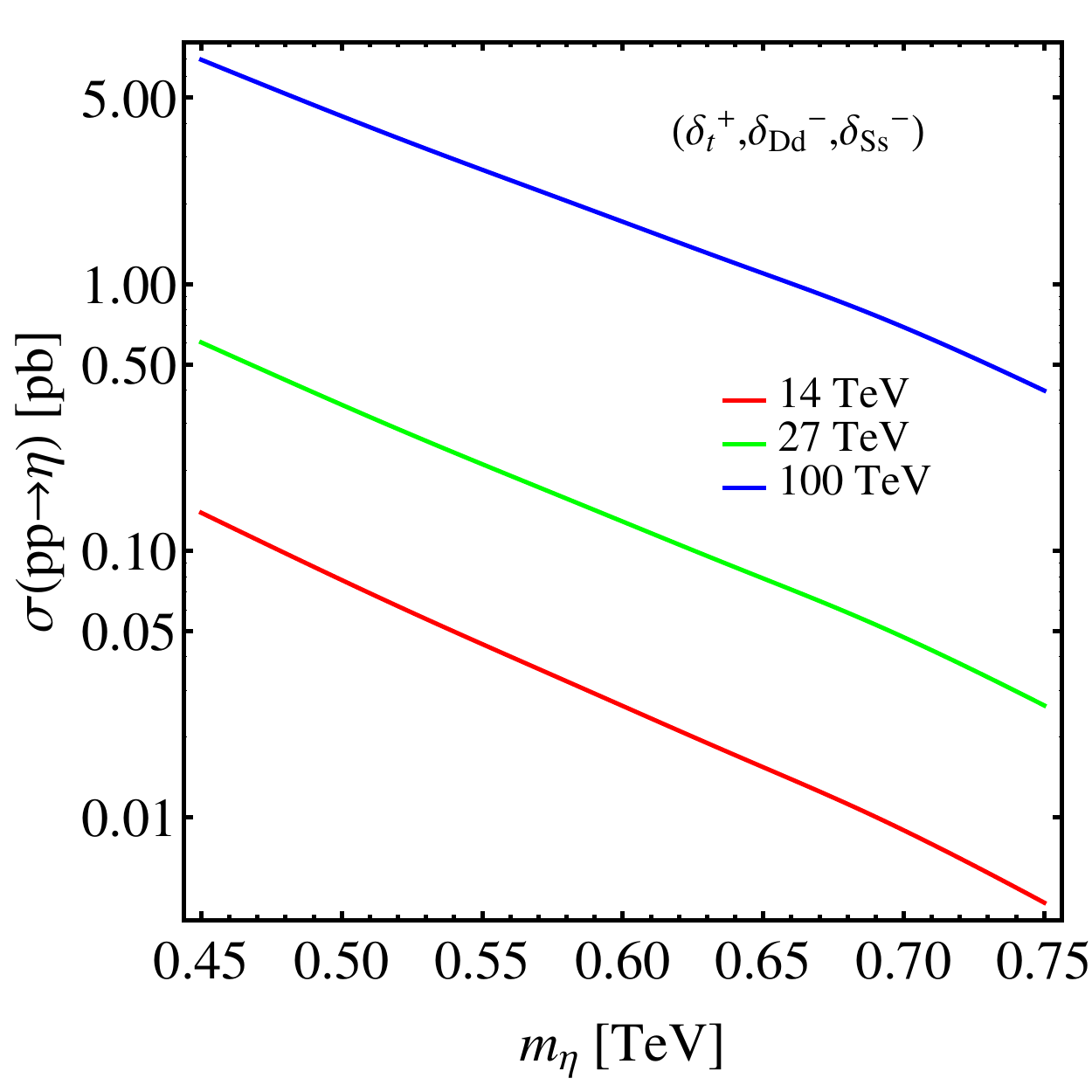}
\includegraphics[width=2.2in]{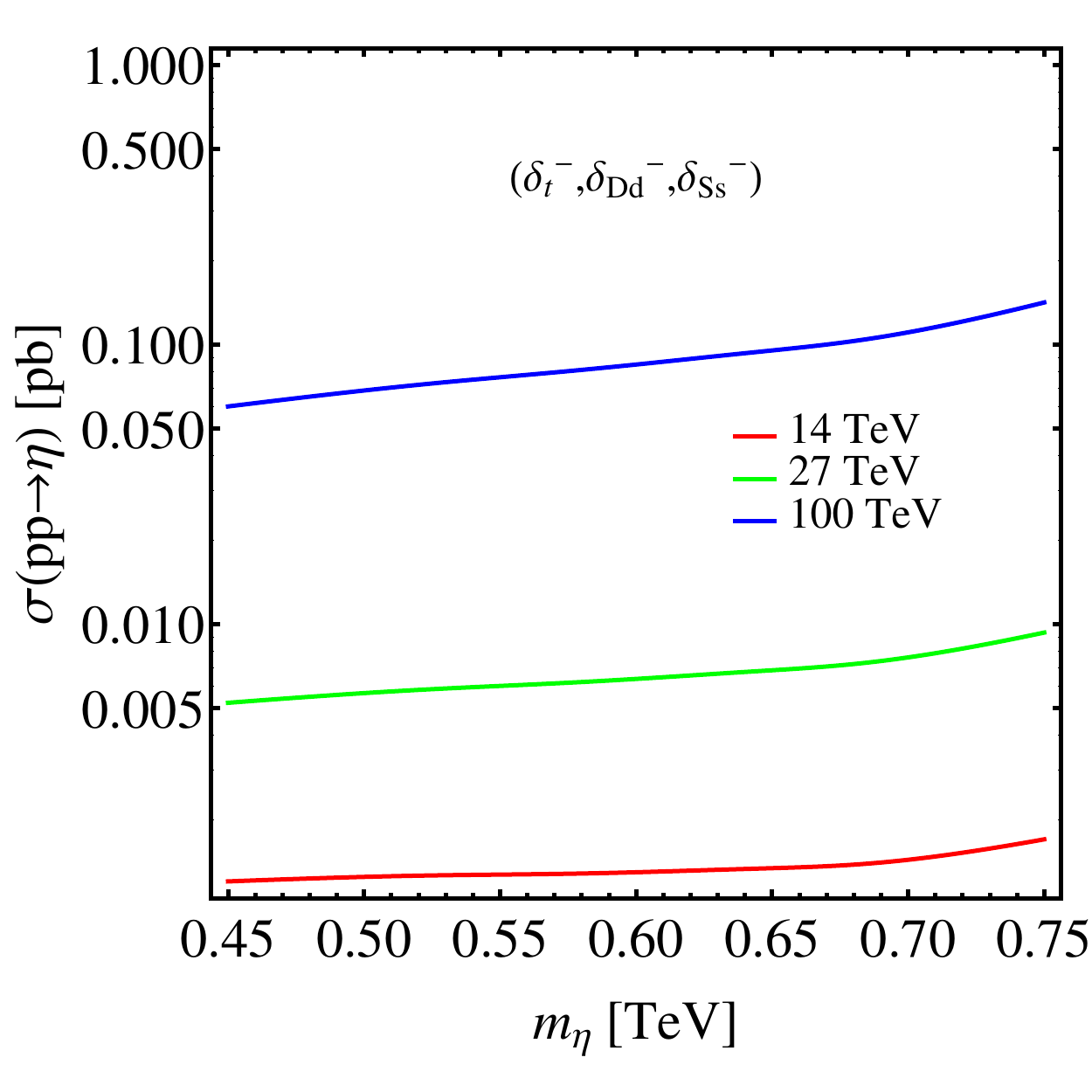} \\
\includegraphics[width=2.2in]{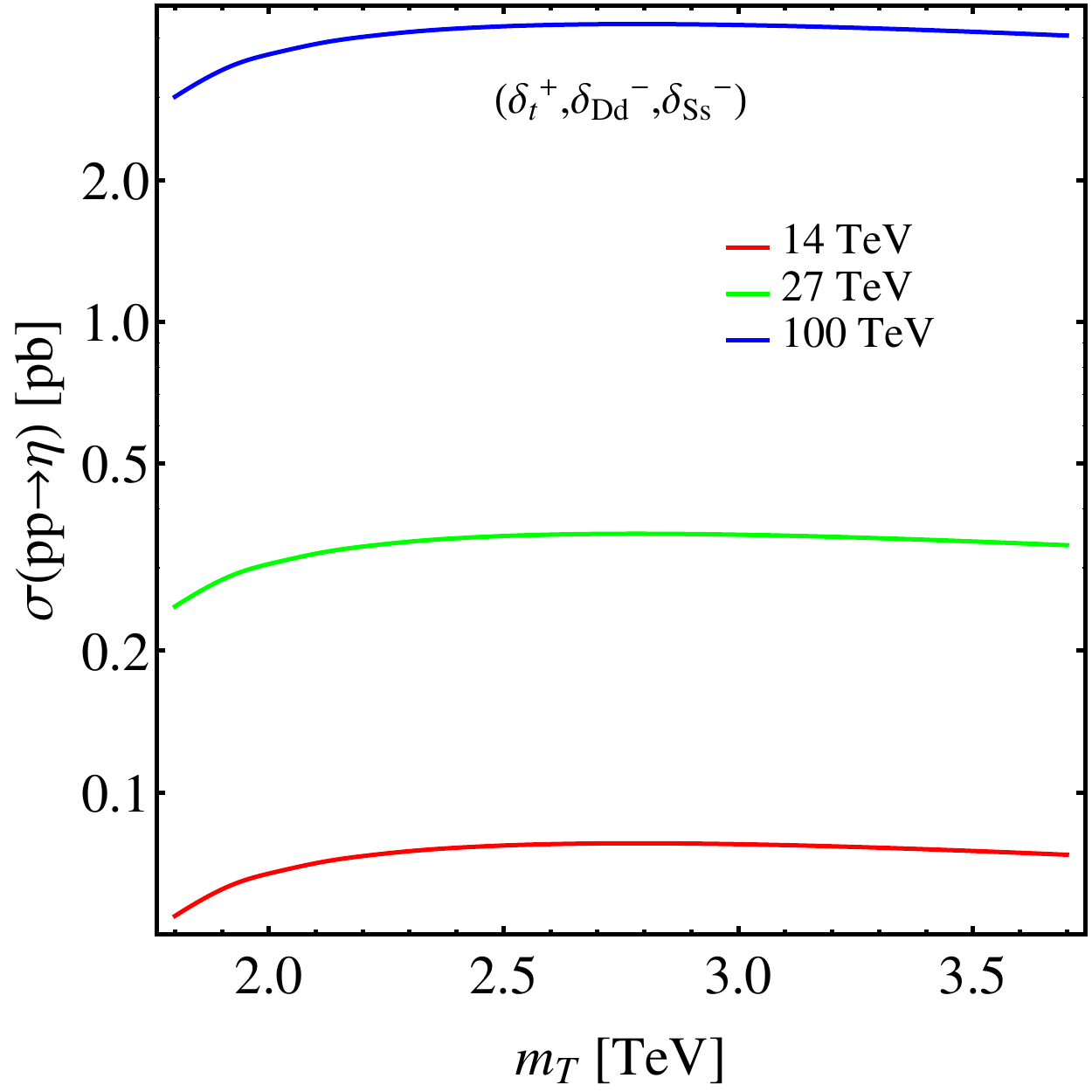}
\includegraphics[width=2.2in]{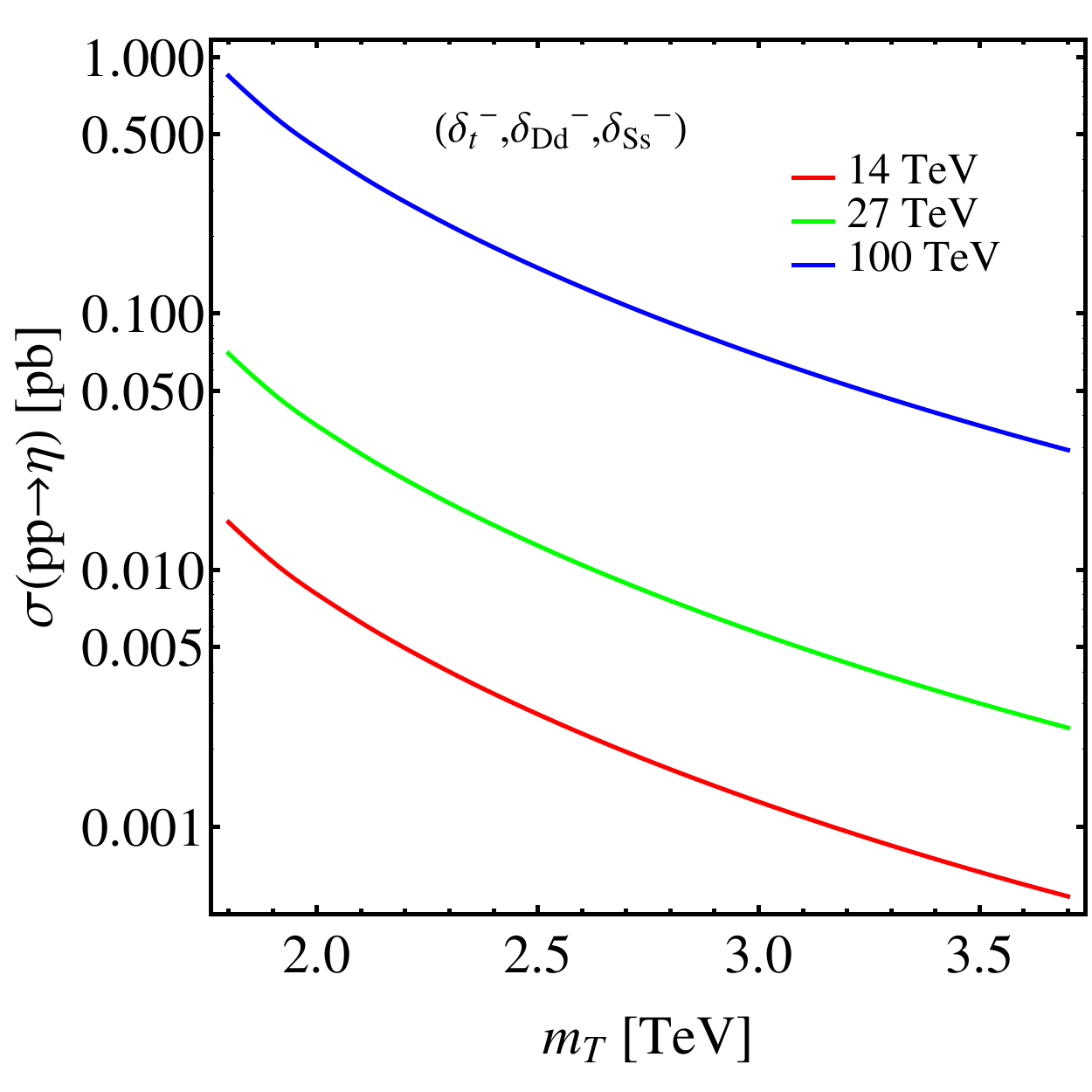}
\caption{\label{fig:ggFeta}Gluon fusion production cross section of $\eta$ as a function of
$m_\eta$ (upper panel, assuming $m_T=m_D=m_S=3\TeV$) or $m_T$ (lower panel,
assuming $m_D=m_S=m_T$ and $m_\eta=500\GeV$). The sign combination of $(\delta_t,\delta_{Dd},
\delta_{Ss})$ is indicated in each plot.
}
\end{figure*}

The total width and branching ratios of $\eta$ are shown in Figure~\ref{fig:etadA} and
Figure~\ref{fig:etadB} for Case A and Case B respectively. In these two cases, the additional
fermion partners $D,S,N$ are not light enough to appear as decay products of $\eta$ and therefore
we are left with the standard $\eta\rightarrow t\bar{t},gg,\gamma\gamma$ channels. From the figures
it is clear that $\eta$ can be viewed as a narrow width particle, however the width is not
small enough to give rise to displaced vertices. In both Case A and Case B and for both sign choices,
$\eta$ decays almost $100\%$ to $t\bar{t}$, with only very small branching ratios to $gg$ ($\ord(0.1\%)$)
and $\gamma\gamma$ ($\ord(0.001\%)$). Here (and in the following) all the partial widths are calculated
at LO, but it is obvious that the inclusion of higher order radiative corrections
has little effect on the whole picture. From a detection point of view this situation is somewhat
unfortunate since the dominant channel $t\bar{t}$ suffer from huge background at hadron colliders,
while the clean channel $\gamma\gamma$ has an extremely small branching ratio. It is natural to ask
how the situation will change if any of $D,S,N$ is light enough, such that exotic channels
like $\eta\rightarrow NN, N\nu, Dd, Ss$ could be open. This is embodied in Cases C and D and we show
the corresponding branching ratio plots in Figure~\ref{fig:etadCD}. Nevertheless the exotic channels
contribute at most a few percent in terms of branching ratio, therefore are of little use for $\eta$ detection even if
any of $D,S,N$ is light enough. This can be understood from the interaction Lagrangian containing
the $\eta Dd, \eta Ss$ and $\eta N\nu,\eta NN$ vertices. The $\eta Dd$ vertex is shown in
Eq.~\eqref{eq:dYu}. When $\eta\rightarrow Dd$ is open, $\frac{M_D}{v}$ is an $\ord(1)$ quantity,
and therefore from Eq.~\eqref{eq:dYu} we may recognize that  the $\eta Dd$ coupling can be
considered as being relatively suppressed by $\ord(\frac{v}{f})$ compared to $\eta t\bar{t}$
vertex. This leads to the suppression of $\eta\rightarrow Dd$ channel. The $\eta N\nu$ coupling
is relatively suppressed by $\ord(\frac{v}{f})$ compared to $\eta NN$ coupling, as can be seen
from Eq.~\eqref{eq:lYu}. However, when $\eta\rightarrow NN$ is open, $M_{Nn}$ can be at most
$\ord(v)$. Moreover, the $\eta NN$ coupling suffers from a $t_\beta$ suppression. Therefore numerically
$\eta\rightarrow NN$ channel is much suppressed compared to $\eta\rightarrow t\bar{t}$ channel.

\subsection{Decay of the Top Partner}
\begin{figure*}[ht]
\includegraphics[width=2.2in]{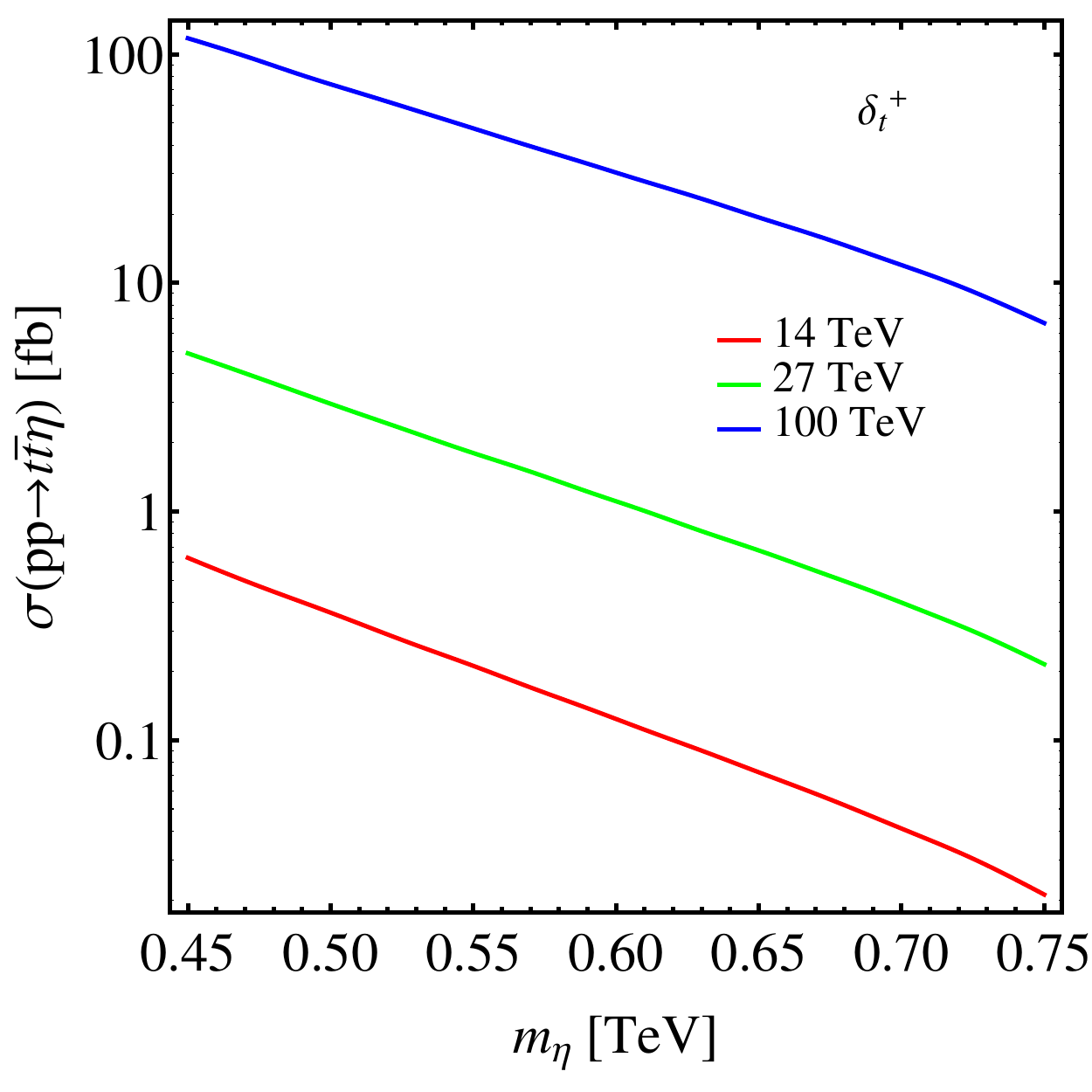}
\includegraphics[width=2.2in]{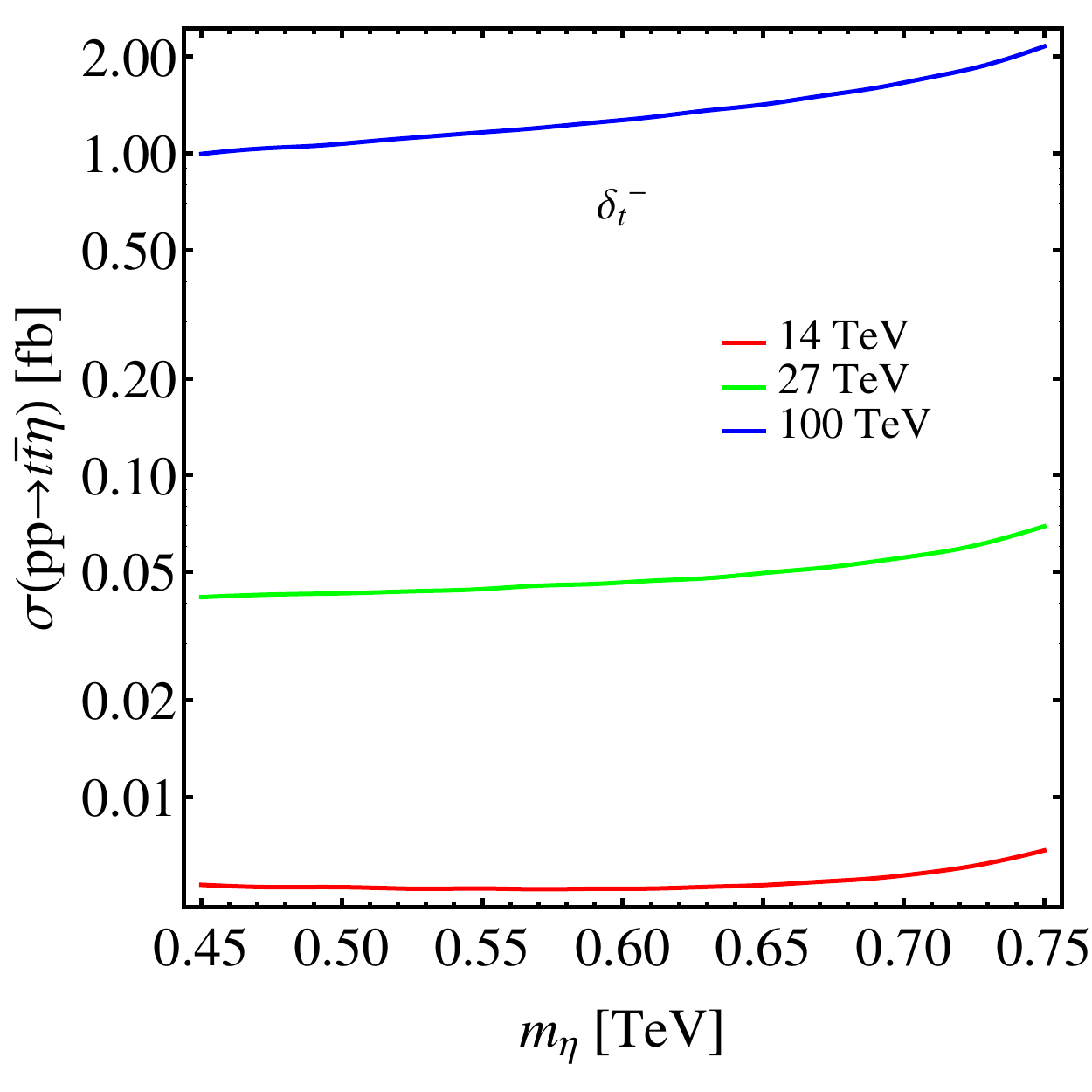}
\caption{\label{fig:tteta}Production cross section of $pp\rightarrow t\bar{t}\eta$ as a function of $m_\eta$.
We assume $f=8\TeV$ and $m_T=3\TeV$.}
\end{figure*}
The pseudo-axion may appear as a decay product of some additional heavy particles in the model.
Among the additional particles in the SLH only $Z'$ and $T$ are closely related to EWSB and
naturalness favors small $Z'$ and $T$ masses within theoretical constraints. In this subsection
we consider the decay of the top partner. The possibility of $T\rightarrow t+a$ where $T$ and $a$
denote the top partner and a pNGB in the context of composite Higgs models have been investigated
in the literature~\cite{Bizot:2018tds,Serra:2015xfa,Kearney:2013cca}. Here we focus on the situation
in the SLH. To be specific we fix $f=8\TeV$ and $m_\eta=500\GeV$
and then plot the total width and branching ratios of $T$ as a function of the top partner mass
$m_T$ in Figure~\ref{fig:Tdecay}. Both $\delta_t^+$ and $\delta_t^-$ possibilities are considered.
Note that when $m_T$ is also given, then according to the mass relation, $t_\beta$ can be calculated,
which in turn determines the total width and branching ratios. The relation $\Br(T\rightarrow bW)=
2\Br(T\rightarrow tH)=2\Br(T\rightarrow tZ)$ holds to a good approximation. In the $\delta_t^+$ case,
$\Br(T\rightarrow t\eta)$ is small (not larger than $10\%$ for $m_T>2\TeV$) and decreases with
the increase of $m_T$. In the $\delta_t^-$ case, $\Br(T\rightarrow t\eta)$ is sizable and becomes
dominant (larger than $50\%$) for $m_T\gtrsim 2.2\TeV$. Another interesting and important feature
is about the total width of $T$. In the $\delta_t^-$ case, the total width is around $20\GeV$ which
makes the narrow width approximation valid to high precision. In the $\delta_t^+$ case, the total width
increases with $m_T$. For $m_T\approx 3.5\TeV$ the total width increases to around $500\GeV$. In this case
$\Gamma/M\lesssim 20\%$ and the narrow width approximation still roughly holds, if the phase space is
large enough. The width will however leave appreciable impact on the invariant mass distribution
of the $T$ decay products.
\subsection{Direct Production of the Pseudo-Axion}

The pseudo-axion can be directly produced via the gluon fusion mechanism at hadron colliders. The particles
running in the loop now contain $t,T,D,S$. In the calculation of the production cross section\footnote{For simplicity, in this work, all
the cross sections are calculated at LO using MadGraph5\_aMC@NLO~\cite{Alwall:2014hca} and FeynRules~\cite{Alloul:2013bka}. We use the MSTW2008lo68cl PDF~\cite{Martin:2009iq}. For $2\rightarrow 1$ production, the renormalization and factorization scale is taken to be the rest mass of the s-channel resonance. Otherwise, the renormalization and factorization scale is taken to be the sum of transverse mass of final state particles (before resonance decay) divided by two.}, we consider
the $14\TeV$ (HL-)LHC, the $27\TeV$ HE-LHC and also the $100\TeV$ FCC-hh. The production cross sections are plotted in Figure~\ref{fig:ggFeta} as a function of $m_\eta$ or $m_T$, with other parameters described in the figure caption.
Although the production cross section may reach $\ord(\text{pb})$ in certain region of parameter space, unfortunately
when combined with $\eta$ decay it turns out very difficult to detect in the gluon fusion channel. The dominant
$t\bar{t}$ decay mode suffers from huge background, while the $\gamma\gamma$ decay mode has only $\ord(10^{-5})$
branching ratio.
\begin{figure*}[ht]
\includegraphics[width=2.2in]{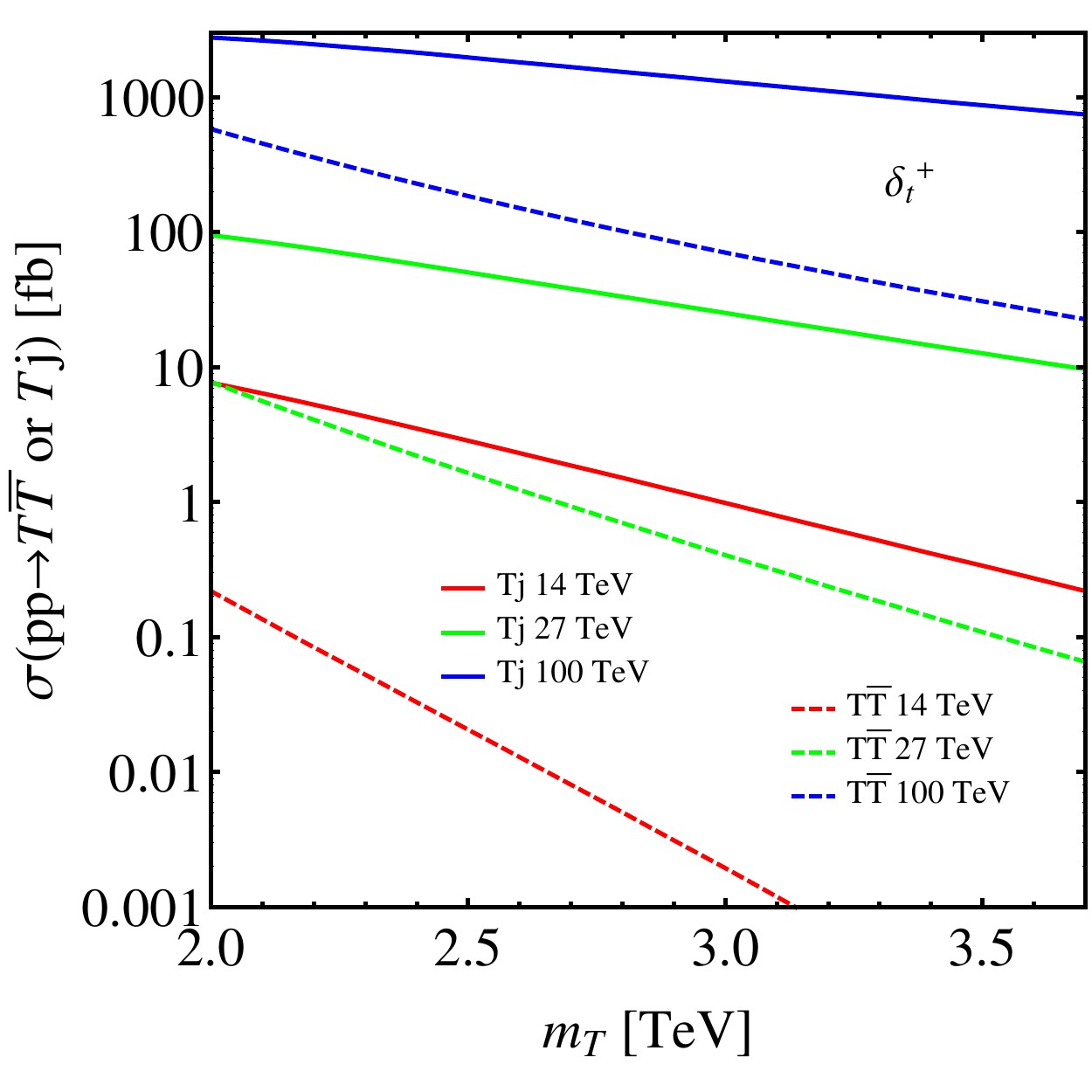}
\includegraphics[width=2.25in]{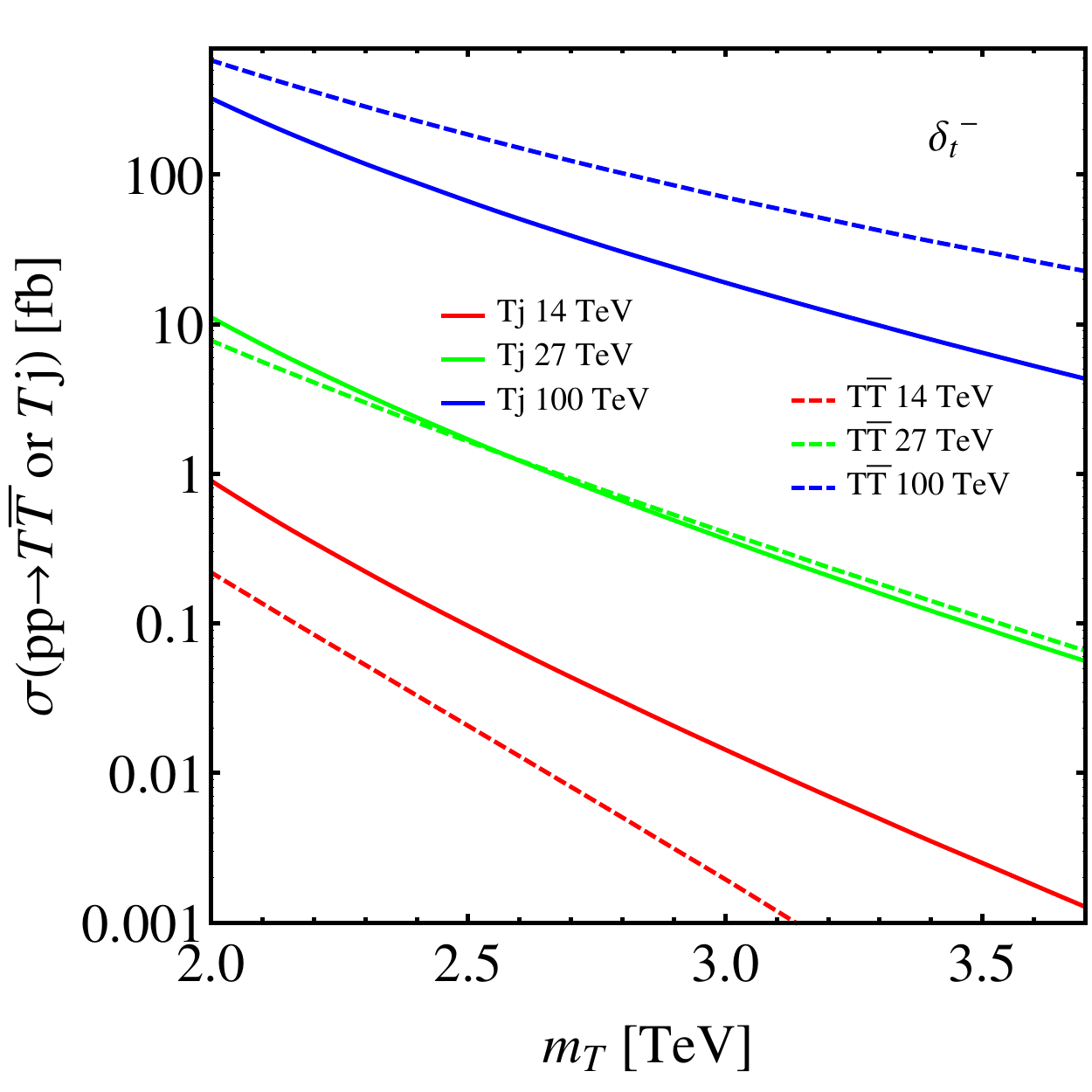}
\caption{\label{fig:TjTT}Production cross section of $pp\rightarrow T\bar{T}$ and $pp\rightarrow Tj$ as a function of $m_T$.
We assume $f=8\TeV$ and $m_\eta=500\GeV$. For $pp\rightarrow Tj$, the contribution from $pp\rightarrow\bar{T}j$ is also included.}
\end{figure*}

Another way to directly produce $\eta$ is through the $pp\rightarrow t\bar{t}\eta$ channel. We plot the
production cross section as a function of $m_\eta$ in Figure~\ref{fig:tteta}, for three center of mass energies
and both $\delta_t^+$ and $\delta_t^-$. Here we fix $f=8\TeV$ and $m_T=3\TeV$, and therefore for given $m_\eta$,
$t_\beta$ (and $\delta_t^\pm$)is also determined. The cross section in the $\delta_t^-$ case is much smaller than
that of the $\delta_t^+$ case. Even in the $\delta_t^+$ case the detection of $pp\rightarrow t\bar{t}\eta$ process
is still very difficult. For instance, if we take $m_\eta=450\GeV$, then in the $\delta_t^+$ case the cross section
reaches only about $0.6\fb$ at $14\TeV$ and $100\fb$ at $100\TeV$. When we consider $\eta\rightarrow t\bar{t}$ decay,
there exists the SM four-top production as an irreducible background, with cross section of about $10\fb$ at $14\TeV$ and
$5000\fb$ at $100\TeV$. Unfortunately, since $m_\eta$ is not far above the $2m_t$ threshold, we don't expect
large differences of kinematical features between the $pp\rightarrow t\bar{t}\eta$ signal and the SM four-top background,
making the discrimination very difficult. With larger $m_\eta$ (say $1\TeV$), the top pair from $\eta$ decay
can be boosted, with invariant mass distribution peaked around a high value, which can facilitate the discrimination
from SM backgrounds. However, the cross section for such a heavy $\eta$ becomes very small. Therefore we don't expect $pp\rightarrow t\bar{t}\eta$ to be a promising channel for future $\eta$ detection in the SLH.
\begin{figure*}[htbp]
\includegraphics[width=2.2in]{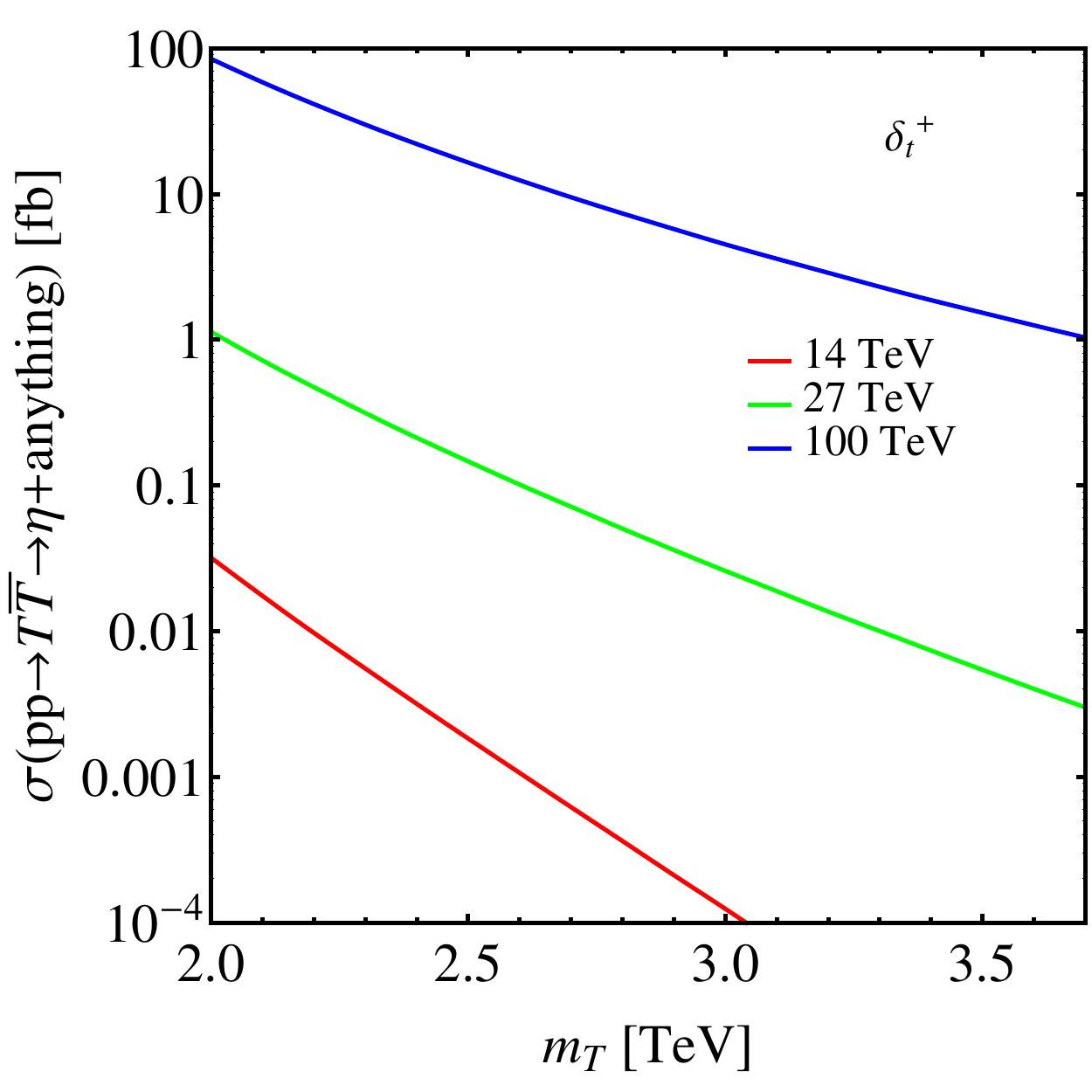}
\includegraphics[width=2.2in]{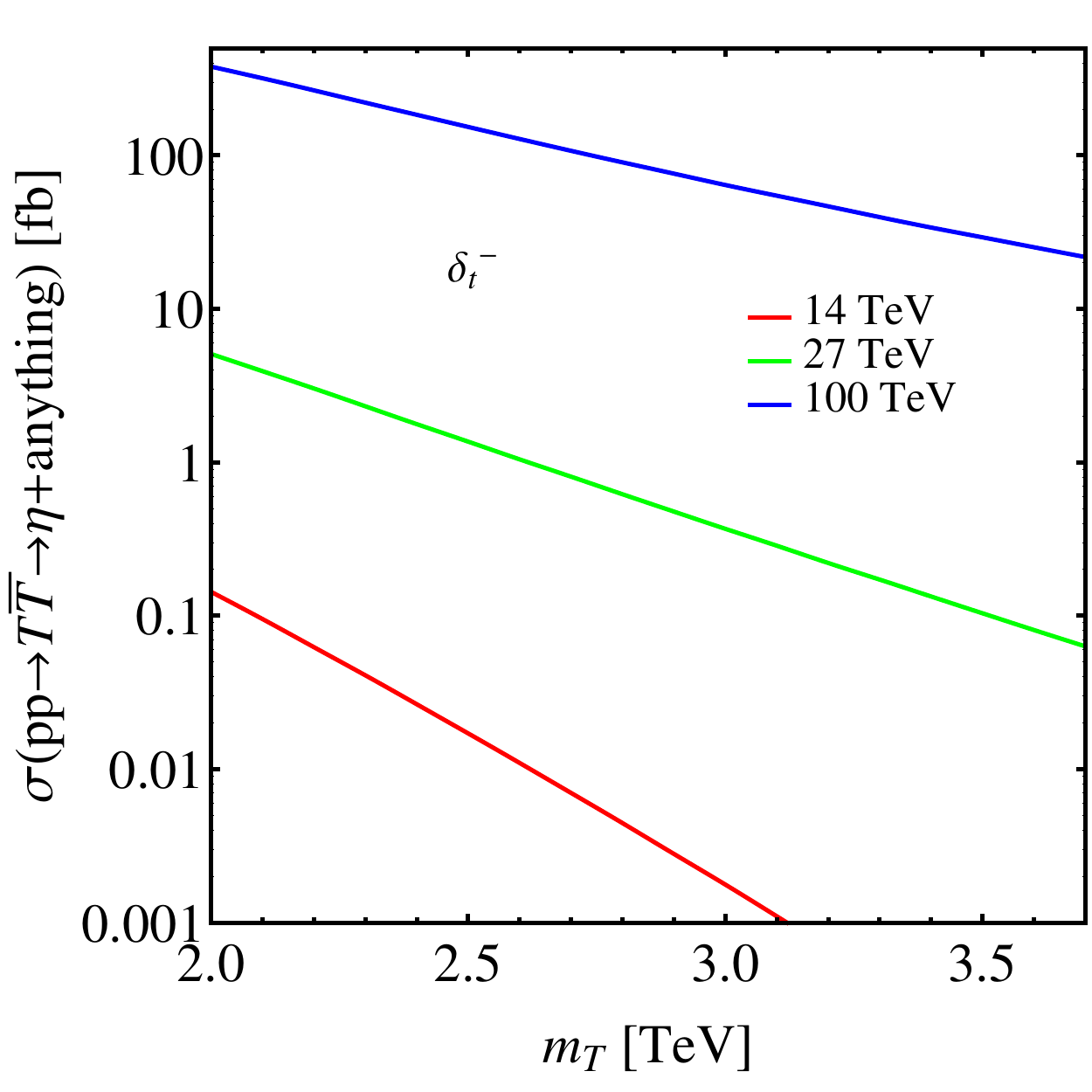} \\
\includegraphics[width=2.2in]{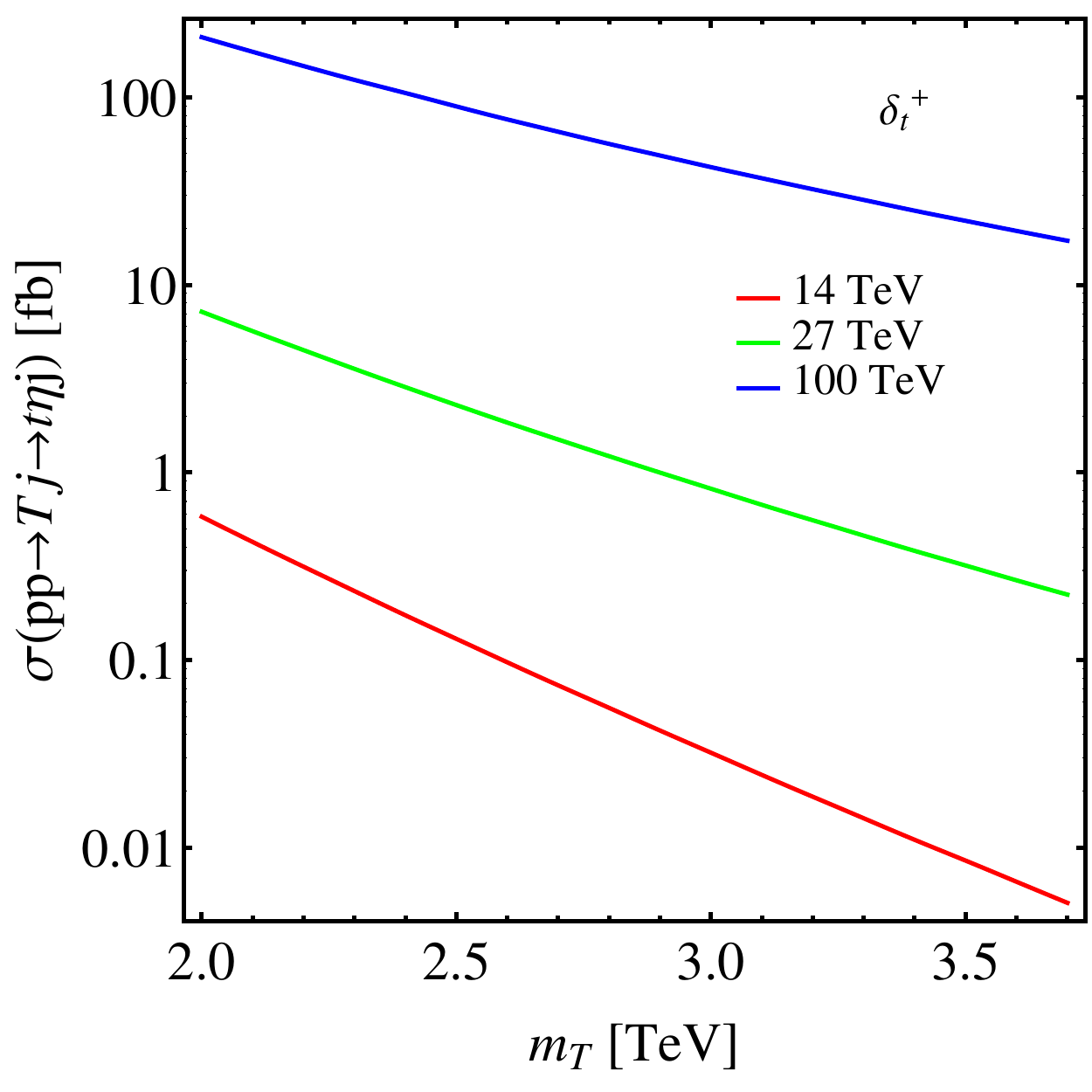}
\includegraphics[width=2.2in]{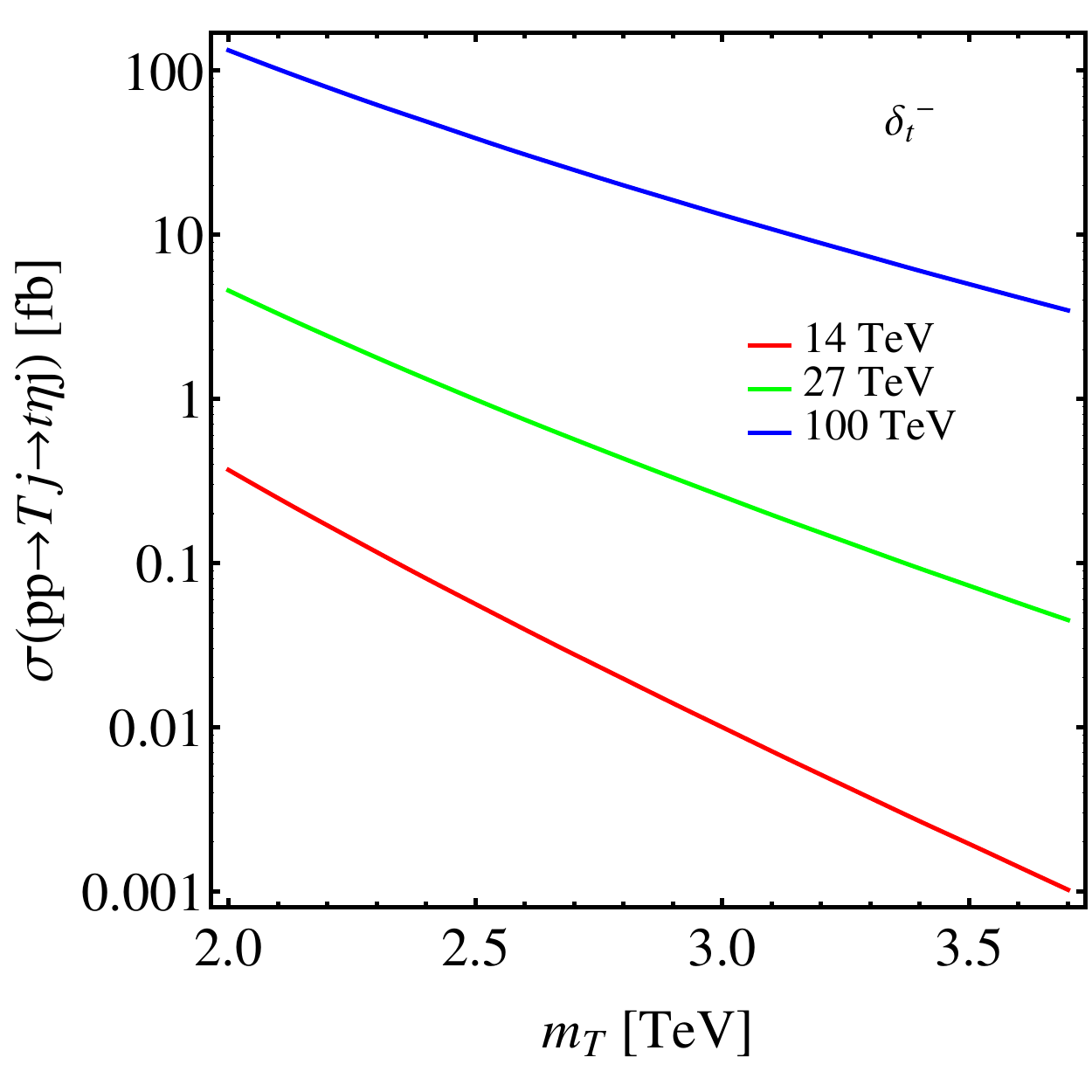}
\caption{\label{fig:TTTjtoeta} Cross section of $pp\rightarrow T\bar{T}\rightarrow\eta+\text{anything}$
and $pp\rightarrow Tj\rightarrow\eta+\text{anything}$ as a function of $m_T$. We assume $f=8\TeV$ and $m_\eta=500\GeV$.
For $pp\rightarrow Tj$, the contribution from $pp\rightarrow\bar{T}j$ is also included.
}
\end{figure*}

\subsection{Pseudo-Axion Production from Top Partner Decay}

The above discussion shows that it is very difficult to detect $\eta$ via the gluon fusion and
$t\bar{t}\eta$ associated production channels. It is therefore natural to consider alternative
$\eta$ production mechanisms, such as decay from heavier particles. In the SLH, particles that
can be heavier than $\eta$ are $T,D,S,N,Z',X$ and $Y$. Here we will concentrate on $T$, which
is most tightly connected to EWSB. We will briefly comment on the possibility of detecting $\eta$
from other heavy particle decays in the next subsection.

Under current constraints, the lower bound on $m_T$ is already larger than the largest possible
value of $m_\eta$ plus $m_t$, therefore the exotic decay channel $T\rightarrow t\eta$ will always
open. The branching fraction of $T\rightarrow t\eta$ has been discussed (see Figure~\ref{fig:Tdecay}).
Here we focus on top partner production. Two major production mechanisms are pair production through
QCD interaction, and single production through the $TbW$ vertex. Pair production has the virtue
of being model-independent, while single production depends on the value of $\delta_t$. In Figure~\ref{fig:TjTT}
we present the cross section of $pp\rightarrow T\bar{T}$ and $pp\rightarrow Tj+\bar{T}j$ for both $\delta_t^+$
and $\delta_t^-$, as a function of $m_T$ while we fix $f=8\TeV,m_\eta=500\GeV$. Three center of mass energies ($14,27,100\TeV$)
are considered. Whether pair or single production delivers a larger cross section depends on the sign choice
for $\delta_t$ and the center of mass energy. In the $\delta_t^+$ case, for all three center of mass energies
the single production cross section is larger. In the $\delta_t^-$ case, at $14\TeV$ single production is
larger since pair production is highly suppressed by phase space. At $27\TeV$ pair production and single production
become comparable while for $100\TeV$ collider energy pair production dominates.
\begin{figure*}[ht]
\includegraphics[width=2.2in]{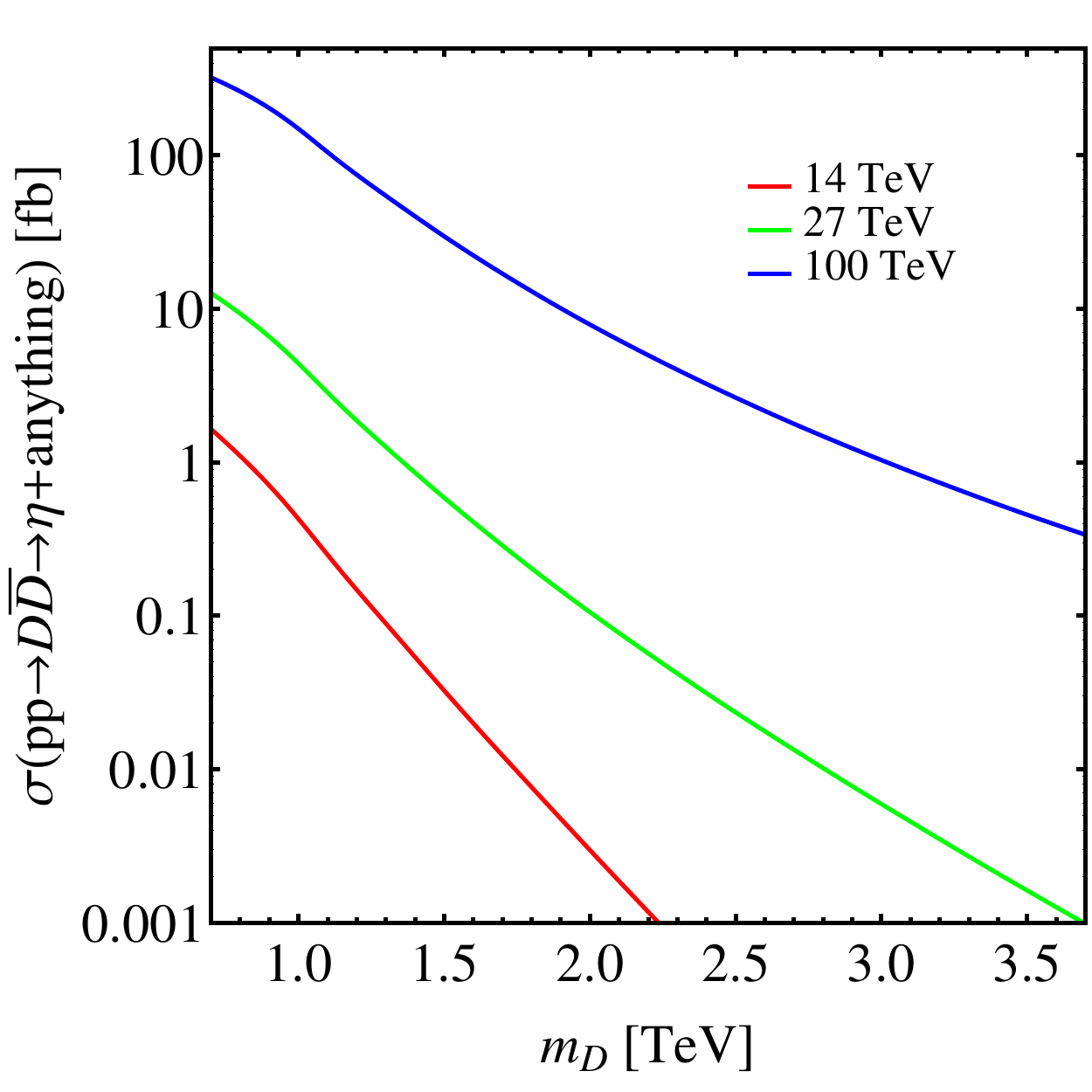}
\includegraphics[width=2.25in]{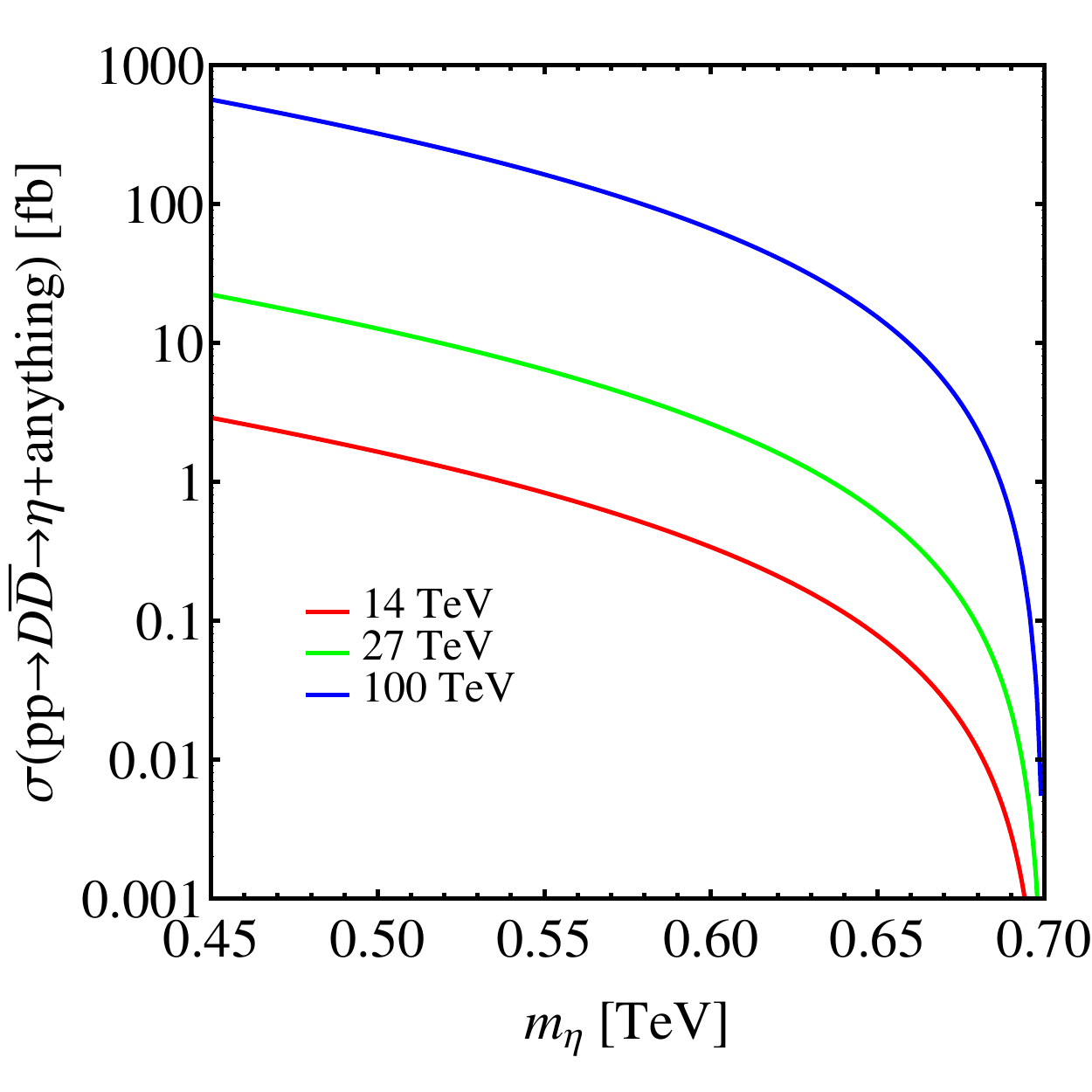}
\caption{\label{fig:Deta}Production cross section of $pp\rightarrow D\bar{D}\rightarrow\eta+\text{anything}$
as a function of $m_D$ (left) and $m_\eta$ (right). For the left plot, we assume $f=8\TeV,m_T=3\TeV,m_\eta=500\GeV$.
For the right plot, we assume $f=8\TeV,m_T=3\TeV,m_D=700\GeV$.}
\end{figure*}

To detect $\eta$ we would also like to consider the top partner decay $T\rightarrow t\eta$ that follows the
pair or single production of $T$. The associated cross sections are plotted as a function of $m_T$ in
Figure~\ref{fig:TTTjtoeta}, using narrow width approximation, for both $\delta_t^+$ and $\delta_t^-$.
For definiteness we take $f=8\TeV,m_\eta=500\GeV$. To be precise, the plotted cross sections are defined by (for $pp\rightarrow Tj$,
the contribution from $pp\rightarrow\bar{T}j$ is also included)
\bwt
\begin{align}
\sigma(pp\rightarrow Tj\rightarrow\eta+\text{anything})
& =\sigma(pp\rightarrow Tj)\times\Br(T\rightarrow t\eta) \\
\sigma(pp\rightarrow T\bar{T}\rightarrow\eta+\text{anything})
& =2\sigma(pp\rightarrow T\bar{T})\times\Br(T\rightarrow t\eta)(1-\Br(T\rightarrow t\eta)) \nonumber \\
& +\sigma(pp\rightarrow T\bar{T})\times\Br^2(T\rightarrow t\eta)
\end{align}
\ewt
For the purpose of $\eta$ detection, let us consider using the $\eta\rightarrow t\bar{t}$ channel, which
has almost $100\%$ branching fraction. Then the $\eta$ production from top partner decays generically
leads to a multi-top ($\geq 3$) signature. Moreover, the top quarks will be boosted since
$m_T\gg m_t+m_\eta$. For example, suppose a $2\TeV$ top partner is produced with little boost
in the lab frame and then decays into $t+\eta$. At this step $t$ and $\eta$ roughly shares the rest energy
of the top partner and therefore will each have about $1\TeV$ energy.
The $\eta$ boson then further decays into $t$ and $\bar{t}$,
each of which roughly has an energy about $0.5\TeV$. All three
top quarks are boosted: the first one will have the decay
($t \to b W$) cone size approximated by
$\sim 2 m_t / E_t \simeq 0.4$ while the second and third top have
$\sim 2 m_t / E_t \simeq 0.8$.  Furthermore, the second and the
third top quark decaying from $\eta$ is close to each other,
of separation approximated by $\sim 2 m_\eta / E_\eta \simeq 0.8$.

In the single production case, the signature will be $3t+j$, in
which the first top is highly boosted while the second and third
are still somewhat boosted and close to each other. One can make use of such kinematics to discriminate from
QCD backgrounds. The most serious background is perhaps multi-top
production. One may be able to reduce the background using
the {\it boosted} techniques~\cite{Kaplan:2008ie}. In the pair production case,
if we consider one top partner decaying into $t\eta$ with the
other decaying into $bW$, then we obtain a signature of $3t+b+W$ in which the top quarks and also the $W$ boson will
be boosted. In both single and pair production channels, the invariant mass peaks at $m_T$ and $m_\eta$ will also be helpful in discriminating between the signal and background. Nevertheless, a full
signal-background analysis using boosted-top techniques is beyond the scope of the present work.

From Figure~\ref{fig:TTTjtoeta}, we see that the cross sections at $14\TeV$ (HL-)LHC
for all these channels are very small ($<1\fb$), making the detection very difficult. Nevertheless, with
the increase of collider energy, the signal cross sections increase significantly. For example,
at the $100\TeV$ FCC-hh, for both $\delta_t^+$ and $\delta_t^-$ and pair and single production channels,
at relatively small $m_T$ the cross sections could reach $\ord(100\fb)$. In the $\delta_t^+$ case,
the single production (with the top partner decaying to $t\eta$) turns out to deliver a cross section of
about $200\fb$, which is larger than the pair production channel. In the $\delta_t^-$ case,
the pair production (with one top partner decaying to $t\eta$) turns out to deliver a cross section of
about $400\fb$, which is however larger than the single production channel.

In principle, top partner production and decay provide a way to measure $t_\beta$ (which is important for testing the
SLH mass relation) and also discriminate
between the $\delta_t^+$ and $\delta_t^-$ cases. In practice, we may consider the partial width ratio
$R_\eta\equiv\frac{\Gamma(T\rightarrow t\eta)}{\Gamma(T\rightarrow bW)}$ as both an indicator of the sign
choice for $\delta_t$ and a way to measure $\delta_t$, which in turn determines $t_\beta$. $\delta_t$ can
also be determined from $pp\rightarrow Tj$ production since the cross section is proportional to $\delta_t^2$.
Furthermore, in the $\delta_t^+$ case the total width of $T$ could reach $\ord(100\GeV)$, which may have impact on
the invariant mass distribution of $T$ decay products (e.g. $bW$). Measurement of the $T$ total width
in principle could also help determine the value of $\delta_t$. If $\delta_t$ is determined (including the
sign choice), we should note however the determination of $t_\beta$ and the test of mass relation still
requires the measurement of $f$ and $m_\eta$, which can be obtained if we are able to measure the masses
of $Z'$ and $\eta$ particles.

\subsection{Comments on Other Channels}
Currently the SLH is stringently constrained by the LHC $Z'\rightarrow ll$ search, nevertheless it also means that
if the SLH was realized in nature, the $Z'\rightarrow ll$ signature would be the first place that we might expect
the appearance of new physics. Then it would also be important to consider whether we may detect $\eta$ as a decay product
of $Z'$. Two channels might be conceived: $Z'\rightarrow\eta H$ and $Z'\rightarrow\eta Y$. However, it turns out
they give too small branching fractions: $\Br(Z'\rightarrow\eta H)<0.01$ and $\Br(Z'\rightarrow\eta Y)<10^{-4}$.
This is regardless of whether the $Z'\rightarrow DD, SS, NN$ channels are kinematically allowed. Therefore it
is not preferable to consider detecting $\eta$ from $Z'$ decay.

If kinematically allowed, we might also consider $D\rightarrow d\eta, S\rightarrow s\eta, N\rightarrow\nu\eta$ decays.
However, these decay channels also suffer from small branching fractions, since the $\eta Dd,\eta Ss,\eta N\nu$ couplings
are $\ord(\frac{v}{f})$ suppressed compared to $HDd,HSs,HN\nu$ couplings (see Eq.~\eqref{eq:lYu} and Eq.~\eqref{eq:dYu}).
For example, $D$ will dominantly decay to $uW,dZ,dH$, with only $\Br(D\rightarrow d\eta)<1\%$, for the benchmark point
$f=8\TeV,m_T=2\TeV,m_\eta=0.5\TeV$ and any value of $m_D$. Here $\delta_{Dd}^-$ is assumed, to be consistent with
electroweak precision constraints. As to $D$ production, for the $\delta_{Dd}^-$ case, there is a $t_\beta^{-1}$ suppresion
for single $D$ production, therefore $D$ pair production is more promising. Moreover, current collider constraint
on $D$ mass is not stringent, such that $m_D=700\GeV$ is still allowed~\cite{Sirunyan:2017lzl}. Therefore, if $m_D$
is as light as $700\GeV$, the large $pp\rightarrow D\bar{D}$ production cross section could compensate for the small
$D\rightarrow d\eta$ branching fraction, leading to sizable $\eta$ production rate. At the $100\TeV$ FCC-hh, the $\eta$
production cross section from $D$ decay, $\sigma(pp\rightarrow D\bar{D}\rightarrow\eta+\text{anything})$ could also
reach more than $100\fb$ for $m_D$ not much larger than $700\GeV$(see Figure~\ref{fig:Deta}). This is comparable with $\eta$ cross section
from top partner production, and in principle could also be used to measure $t_\beta$. The expected signature would be $t\bar{t}+2j+W/Z/H$, in which the $W/Z/H$ should be boosted.
The existence of various intermediate resonances would be helpful in discriminating signal and background. Nevertheless, we should
be aware that naturalness does not offer any guidance on the preferred value of $m_D$. This is different from the case of $m_T$,
in which naturalness clearly favors a lighter top partner. The case of $pp\rightarrow S\bar{S}$ production with $S\rightarrow s\eta$
decay is completely similar to the above discussion of $D$ production and decay. For $N$, $Br(N\rightarrow\nu\eta)$ is also very small
(less than $1\%$ for the benchmark point $f=8\TeV,m_T=2\TeV,m_\eta=0.5\TeV$ and any value of $m_N$. Moreover, $N$ does not have
QCD pair production channels like $D,S$, therefore it is difficult to detect $\eta$ from $N$ decay at hadron colliders.

The $X,Y$ gauge bosons in the SLH may have decays like $X\rightarrow\eta W$ and $Y\rightarrow H\eta$. However, single production cross section of $X,Y$ at hadron colliders are highly suppressed, and we need to rely on production with other heavy particles (heavy gauge bosons
or quark partners)~\cite{Han:2005ru}. Since $X,Y$ bosons are quite heavy (with masses of about $0.8m_{Z'}$), their production with other
heavy particles would be limited by phase space while their decays are expected to be dominated by fermionic final states.
Therefore we don't consider $\eta$ production from $X,Y$ decays as promising channels for $\eta$ detection.

\section{Discussion and Conclusions}
\label{sec:dnc}
\begin{table*}[ht]
\begin{tabular}{|c|c|c|}
\hline
Channel & Cross section at the benchmark point ($\sqrt{s}=100\TeV$)(fb) & Signature \\
\hline
$pp\rightarrow T\bar{T}\rightarrow\eta+\text{anything}$ & $84(\delta_t^+),\quad\quad379(\delta_t^-)$ & $3t+W+b$ or $4t+Z/H$ \\
\hline
$pp\rightarrow Tj\rightarrow\eta+\text{anything}$ & $209(\delta_t^+),\quad\quad133(\delta_t^-)$ & $3t+j$ \\
\hline
$pp\rightarrow D\bar{D}\rightarrow\eta+\text{anything}$ & 322 & $2t+W/Z/H+2j$ \\
\hline
\end{tabular}
\caption{Summary of $\eta$ production from $T,D(S)$ decays at the $100\TeV$ FCC-hh. For $pp\rightarrow Tj$, the contribution from
$pp\rightarrow\bar{T}j$ is also taken into account. For $T\bar{T},Tj$ channels, the benchmark point is $f=8\TeV,m_T=2\TeV,
m_\eta=500\GeV$ while for $D\bar{D}$ channels, the benchmark point is $f=8\TeV,m_T=3\TeV,m_\eta=500\GeV,
m_D=700\GeV$. When listing the signatures for $T\bar{T},D\bar{D}$ channels we don't consider the situation
in which both quark partners decay into $\eta+t$ or $\eta+j$, but this possibility is taken into account
in the cross section values and plots.}
\label{table:sumeta}
\end{table*}
The Simplest Little Higgs model provides a most simple manner to concretely realize the
collective symmetry breaking mechanism, in order to alleviate the Higgs mass naturalness
problem. In the scalar sector, its particle content is very economical, since besides
the CP-even Higgs which should serve as the $125\GeV$ Higgs-like particle, the only additional
scalar particle is the pseudo-Nambu-Goldstone particle $\eta$ associated with a remnant
global $U(1)$ symmetry. The detection of $\eta$ is important since its mass enters into the crucial
SLH mass relation and it will also play an important role in discriminating SLH from other new physics
scenarios. In this work we are concerned with the production and decay of $\eta$
particle at future hadron colliders. We found that for natural region of parameter space,
$m_\eta$ is larger than $2m_t$ and decays almost exclusively to $t\bar{t}$, and $\Br(\eta\rightarrow\gamma\gamma)$
is too small to be considered promising for detection. Also it is very difficult to detect $\eta$ in direct
production channels $pp\rightarrow\eta$ (gluon fusion) and $pp\rightarrow t\bar{t}\eta$. Channels that
are worth further consideration include $\eta$ production from heavy quark partner ($T,D,S$) decays, in which
the heavy quark partner might be singly (for $T$) or pair produced. The corresponding $\eta$ production cross section at $100\TeV$ FCC-hh could reach $\ord(100\fb)$ for certain range of parameter space that is allowed by current constraints, while at $14\TeV$ (HL-)LHC
the rate might be too small for detection. However, the detection prospects in these channels (at $100\TeV$) might still be challenging since the final states are quite complicated, including multi-top associated production with other objects, in which one
or more of them could be boosted, requiring sophisticated tagging techniques. At the same time the SM
background also enjoys a large increase with the collider energy, with more complicated hadronic environment.
The aim of this paper is to examine the $\eta$ production channels with a LO estimate of the $\eta$ cross sections
in the relatively promising ones as a function of model parameters, keeping in mind the most up-to-date theoretical
and experimental constraints (see Table~\ref{table:sumeta} for a summary). We do not attempt here to give a quantitative assessment of the collider sensitivities in these channels.

Phenomenology of the $\eta$ particle in the SLH was studied long time ago by several papers (e.g.~\cite{Kilian:2004pp,Kilian:2006eh,
Cheung:2006nk,Cheung:2008zu}). Compared to all the previous studies, the present paper is different in a few crucial aspects:
\begin{enumerate}
\item Instead of working with the ad hoc assumption of no direct contribution to the scalar potential
from the physics at the cutoff, we take into account in all calculations the crucial SLH mass relation Eq.~\eqref{eq:mr2} which is
a reliable prediction of the SLH. Therefore our prediction preserves all the correlation required by theoretical consistency
but does not depend on the choice of any fixed cutoff value such as $4\pi f$.
\item We have focused our attention on the parameter region favored by naturalness consideration. This region is characterized
by small $m_T$ and large $t_\beta$ or $t_\beta^{-1}$. The favored $\eta$ mass is larger than $2m_t$.
\item We have taken into account the recent collider constraint on $f$ ($f\gtrsim 7.5\TeV$) which is much more stringent than
the constraints obtained long time ago. We also take into account the constraint from perturbative unitarity which sets
an upper bound on the allowed value of $t_\beta$ or $t_\beta^{-1}$. These two factors determine the current lower bound on
$m_T$ and crucially affect the largest cross section that can be achieved in all channels.
\item Our study is based on an appropriate treatment of the diagonalization of the vector-scalar system in the SLH, and especially
the field redefinition related to $\eta$. This affect the derivation of $ZH\eta$ vertices
and also $\eta$ coupling to fermions, which are not treated properly in previous works until ref.~\cite{He:2017jjx}.
\item We also clarify the role played by the symmetric VSS vertices that appear in the Lagrangian and how they are compatible with
the general principle like field redefinition invariance and gauge independence.
\end{enumerate}
From our study it turns out that the detection of $\eta$ at the $14\TeV$ (HL-)LHC will be very difficult, and therefore a $pp$ collider
with higher energy and larger luminosity, such as the $27\TeV$ HE-LHC or even the $100\TeV$ FCC-hh or SppC, is motivated
to capture the trace of such an elusive particle. Moreover, generally we would expect some other SLH signatures (e.g. $Z'\rightarrow ll,
T\rightarrow bW$ or $D\rightarrow uW$) to show up earlier than $\eta$ signatures since $\eta$ signatures are usually
very complicated (with multiple top quarks) and suffer from small rates). It is nonetheless important to study $\eta$ properties
since they are crucial in testing the SLH mass relation and also provide a basis for model discrimination.

\subsection*{Acknowledgements}

We thank Yue-Lin Sming Tsai for helpful discussion.
P.Y.T. was supported by World Premier International Research Center Initiative (WPI), MEXT, Japan.
This work was supported in part by the Natural Science
Foundation of China (Grants No. 11635001 and No. 11875072), the China Postdoctoral Science
Foundation (Grant No. 2017M610992) and the MoST of Taiwan under
the grant no.:105-2112-M-007-028-MY3 and 107-2112-M-007-029-MY3.

\appendix

\section{Convention Conversion}
\label{sec:cc}
In previous literature, Ref.~\cite{delAguila:2011wk} and Ref.~\cite{Han:2005ru}
contain detailed treatment of the anomaly-free SLH model. However, they use
different conventions and it is useful to establish a conversion rule
to relate formulae in the two conventions. Ref.~\cite{delAguila:2011wk} uses
the following covariant derivative expression:
\begin{equation}
D_\mu=\partial_\mu-igA_\mu^a T^a+ig_xQ_xB_\mu^x,\quad
g_x=\frac{gt_W}{\sqrt{1-t_W^2/3}}
\end{equation}
while Ref.~\cite{Han:2005ru} uses
\begin{equation}
D_\mu=\partial_\mu+igA_\mu^a T^a+ig_xQ_xB_\mu^x,\quad
g_x=\frac{gt_W}{\sqrt{1-t_W^2/3}}
\end{equation}
Therefore to convert between the two conventions, we need
\begin{align}
g\leftrightarrow -g,\quad t_W\leftrightarrow -t_W
\end{align}
if we assume $g_x\leftrightarrow g_x$ and $A_\mu^a\leftrightarrow A_\mu^a,B_\mu^x\leftrightarrow B_\mu^x,
T^a\leftrightarrow T^a, Q_x\leftrightarrow Q_x$. The transformation
of $s_W$ and $c_W$ are still not determined. For convenience we would like to
identify the first-order gauge boson mass eigenstates $Z,Z',A$ in both conventions, namely
\begin{align}
Z\leftrightarrow Z,\quad Z'\leftrightarrow Z',\quad A\leftrightarrow A
\end{align}
Then by comparing the first-order gauge boson mixing formulae in the two papers
we are led to the following conversion rule for $s_W$ and $c_W$:
\begin{align}
c_W\leftrightarrow c_W,\quad s_W\leftrightarrow -s_W
\end{align}
Using these rules it is straightforward to convert between the two conventions.
(Our present work adopts the same convention as Ref.~\cite{delAguila:2011wk}.)
Then for example, the Lagrangian coefficient of $Z'f\bar{f}$ couplings will
acquire a minus sign during conversion since $g\leftrightarrow -g$. However,
the expression for $\delta_Z$ (see Eq.~\eqref{eq:deltaZ}) remains the same
since $c_W\leftrightarrow c_W$.

\section{Partial Width Formulae}
\label{sec:pwf}
Let us define
\begin{equation}
F(x,y)\equiv\sqrt{(1+x+y)(1-x-y)(1+x-y)(1-x+y)}
\end{equation}
In particular we have
\begin{equation}
F(0,x)=1-x^2,\quad \text{for } |x|\leq 1
\end{equation}
The partial width formulae related to $\eta,T,D,S,N,Z'$ decays are listed as follows:
\begin{enumerate}
\item $\eta$ decay:
Tree-level decay channels (to fermion final states):
\bwt
\begin{align}
\Gamma_{\eta\rightarrow t\bar{t}}&=\frac{3m_{\eta}}{8\pi}\left(\frac{m_t\delta_t}{v}\right)^2\sqrt{1-\frac{4m^2_t}{m^2_{\eta}}}.\\
\Gamma_{\eta\rightarrow dD}&=\frac{3m_{\eta}}{8\pi}\left(\frac{m_D}{v}\right)^2\left(\delta^2_{Dd}+\frac{v^2}{2f^2}\right)^2
\left(1-\frac{m^2_D}{m^2_{\eta}}\right)^2.\\
\Gamma_{\eta\rightarrow N\bar{N}}&=\frac{m_{\eta}m^2_N}{16\pi f^2t^2_{\beta}}\sqrt{1-\frac{4m^2_N}{m^2_{\eta}}}.\\
\Gamma_{\eta\rightarrow \nu N}&=\frac{m_{\eta}}{8\pi}\left(\frac{vm_N}{2f^2s^2_{\beta}}\right)^2\left(1-\frac{m^2_N}{m^2_{\eta}}\right)^2.
\end{align}
Here we adopt the notation $\Gamma_{\eta\rightarrow dD}\equiv\Gamma_{\eta\rightarrow d\bar{D}}+\Gamma_{\eta\rightarrow D\bar{d}}$
and $\Gamma_{\eta\rightarrow \nu N}\equiv\Gamma_{\eta\rightarrow\nu\bar{N}}+\Gamma_{\eta\rightarrow\bar{\nu}N}$.
Loop-induced decay channels:
\begin{eqnarray}
\Gamma_{\eta\rightarrow gg}&=&\frac{m^3_{\eta}\alpha_s^2}{128\pi^3v^2}\left|-\delta_tA_{\frac{1}{2}}(\tau_t)+\delta_tA_{\frac{1}{2}}(\tau_T)+
\delta_{Dd}A_{\frac{1}{2}}(\tau_D)+\delta_{Ss}A_{\frac{1}{2}}(\tau_S)\right|^2.\\
\Gamma_{\eta\rightarrow\gamma\gamma}&=&\frac{m^3_{\eta}\alpha_{em}^2}{2304\pi^3v^2}\left|-4\delta_tA_{\frac{1}{2}}(\tau_t)+4\delta_tA_{\frac{1}{2}}(\tau_T)+
\delta_{Dd}A_{\frac{1}{2}}(\tau_D)+\delta_{Ss}A_{\frac{1}{2}}(\tau_S)\right|^2.
\end{eqnarray}
Here $\tau_f\equiv m^2_{\eta}/4m^2_f$ and for $f=T,D,S$ we have $\tau_f\ll1$. The function $A_{\frac{1}{2}}(\tau)\equiv2f(\tau)/\tau$ where
\begin{equation}
f(\tau)=\left\{\begin{array}{cc}\arcsin^2\sqrt{\tau},&(\tau\leq1)\\
-\frac{1}{4}\left(\ln\frac{1+\sqrt{1-1/\tau}}{1-\sqrt{1-1/\tau}}-\textrm{i}\pi\right)^2&(\tau>1).\end{array}\right.
\end{equation}
\item $T$ decay:
\begin{eqnarray}
\Gamma_{T\rightarrow Wb}&=&\frac{g^2\delta_t^2m^3_T}{64\pi m^2_W}\left(1+\frac{m^2_W}{m^2_T}-\frac{2m^4_W}{m^4_T}\right)F\left(0,\frac{m_W}{m_T}\right)\simeq\frac{\delta_t^2m^3_T}{16\pi v^2}.\\
\Gamma_{T\rightarrow Zt}&=&\frac{g^2\delta_t^2m^3_T}{128\pi c^2_Wm^2_Z}\left(1+\frac{m^2_Z-2m^2_t}{m^2_T}+\frac{m^4_t+m^2_tm^2_Z-2m^4_Z}{m^4_T}\right)
F\left(\frac{m_t}{m_T},\frac{m_Z}{m_T}\right)\simeq\frac{\delta_t^2m^3_T}{32\pi v^2}.\\
\Gamma_{T\rightarrow Ht}&=&\frac{m_T^3\delta_t^2}{32\pi v^2}\left(1+\frac{m^2_t-m^2_H}{m^2_T}\right)F\left(\frac{m_t}{m_T},\frac{m_H}{m_T}\right)\simeq\frac{\delta_t^2m^3_T}{32\pi v^2}.\\
\Gamma_{T\rightarrow\eta t}&=&\frac{m_Tm^2_t}{32\pi v^2}\left(1+\frac{m^2_t-m^2_{\eta}}{m^2_T}\right)F\left(\frac{m_t}{m_T},\frac{m_\eta}{m_T}\right)\simeq\frac{m_Tm^2_t}{32\pi v^2}\left(1-\frac{m^2_{\eta}}{m^2_T}\right)^2.
\end{eqnarray}
\item $D,S,N$ decays
\begin{eqnarray}
\Gamma_{D\rightarrow Wu}&=&\frac{g^2\delta_{Dd}^2m^3_D}{64\pi
m^2_W}\left(1+\frac{m^2_W}{m^2_D}-\frac{2m^4_W}{m^4_D}\right)F\left(0,\frac{m_W}{m_D}\right)\simeq\frac{\delta_{Dd}^2m^3_D}{16\pi v^2}.\\
\Gamma_{D\rightarrow Zd}&=&\frac{g^2\delta_{Dd}^2m^3_T}{128\pi c^2_Wm^2_Z}\left(1+\frac{m^2_Z}{m^2_D}-\frac{2m^4_Z}{m^4_D}\right)
F\left(0,\frac{m_Z}{m_D}\right)\simeq\frac{\delta_{Dd}^2m^3_D}{32\pi v^2}.\\
\Gamma_{D\rightarrow Hd}&=&\frac{m_D^3\delta_{Dd}^2}{32\pi v^2}\left(1-\frac{m^2_H}{m^2_D}\right)F\left(0,\frac{m_H}{m_D}\right)
\simeq\frac{\delta_{Dd}^2m^3_D}{32\pi v^2}.\\
\Gamma_{D\rightarrow\eta d}&=&\frac{m^3_D}{32\pi v^2}\left(\frac{v^2}{2f^2}+\delta_{Dd}^2\right)^2\left(1-\frac{m^2_{\eta}}{m^2_D}\right)^2
\ll\Gamma_{D\rightarrow Wu,Zd,Hd}.
\end{eqnarray}
The same formulae hold for $S$ decay channels with the replacements $\delta_{Dd}\rightarrow\delta_{Ss},m_D\rightarrow m_S,
D\rightarrow S,d\rightarrow s,u\rightarrow c$. They also hold for $N$ decay channels with the replacements $m_D\rightarrow m_N,D\rightarrow N,
d\rightarrow\nu,u\rightarrow\ell$ and $\delta_{Dd}\rightarrow\delta_{Dd}^-=\frac{v}{\sqrt{2}ft_{\beta}}$.
\item $Z'$ decay

For $Z^{\prime}\rightarrow f\bar{f}$ decay modes, assuming the interaction Lagrangian $\mathcal{L}\supset\sum_fg(a_L^f\bar{f}_L\gamma^{\mu}f_L+a_R^f\bar{f}_R\gamma^{\mu}f_R)Z^{\prime}_{\mu}$,
then the decay width is given by
\begin{equation}
\Gamma_{Z'\rightarrow f\bar{f}}=\frac{N_cg^2m_{Z'}}{24\pi}\left(\left(\left(a_L^f\right)^2+\left(a_R^f\right)^2\right)\left(1-\frac{m^2_f}{m^2_{Z'}}\right)
+6a_L^fa_R^f\frac{m^2_f}{m^2_{Z'}}\right)\sqrt{1-\frac{4m^2_f}{m^2_{Z'}}}.
\end{equation}
$N_c=1$ for leptons and $N_c=3$ for quarks. For SM quarks, we can take $m_f=0$ since $m_{Z'}\sim\mathcal{O}(f)$. $a_L^f$ and $a_R^f$
can be extracted from Eq.~\eqref{eq:zpl}, Eq.~\eqref{eq:zp3q} and Eq.~\eqref{eq:zp12q}. Thus the $Z^{\prime}\rightarrow f\bar{f}$ decay widths are
\begin{eqnarray}
\Gamma_{Z'\rightarrow\ell^+\ell^-}&=&\frac{g^2m_{Z'}\left((1-t^2_W)^2+4t^4_W\right)}{96\pi(3-t^2_W)}.\\
\Gamma_{Z'\rightarrow\nu\bar{\nu}}&=&\frac{g^2m_{Z'}(1-t^2_W)^2}{96\pi(3-t^2_W)}.\\
\Gamma_{Z'\rightarrow u\bar{u}}=\Gamma_{Z'\rightarrow c\bar{c}}&=&\frac{g^2m_{Z'}}{72\pi}\left(\frac{3-t^2_W}{4}+\frac{4t^4_W}{3-t^2_W}\right).\\
\Gamma_{Z'\rightarrow d\bar{d}}=\Gamma_{Z'\rightarrow s\bar{s}}&=&\frac{g^2m_{Z'}}{72\pi}\left(\frac{3-t^2_W}{4}+\frac{t^4_W}{3-t^2_W}\right).\\
\Gamma_{Z'\rightarrow b\bar{b}}&=&\frac{g^2m_{Z'}}{72\pi(3-t^2_W)}\left(\frac{(3+t^2_W)^2}{4}+t^4_W\right).\\
\Gamma_{Z'\rightarrow t\bar{t}}&=&\frac{g^2m_{Z'}}{72\pi(3-t^2_W)}\left(\frac{(3+t^2_W)^2}{4}+4t^4_W\right).
\end{eqnarray}
\begin{eqnarray}
\Gamma_{Z'\rightarrow N\bar{N}}&=&\frac{g^2m_{Z'}}{24\pi(3-t^2_W)}\left(1-\frac{m^2_N}{m^2_{Z'}}\right)\sqrt{1-\frac{4m^2_N}{m^2_{Z'}}}.\\
\Gamma_{Z'\rightarrow D\bar{D},S\bar{S}}&=&\frac{g^2m_{Z'}}{72\pi(3-t^2_W)}\left[\left((3-t^2_W)^2+t^4_W\right)\left(1-\frac{m^2_{D,S}}{m^2_{Z'}}\right)
-6t^2_W(3-t^2_W)\frac{m^2_{D,S}}{m^2_{Z'}}\right]\sqrt{1-\frac{4m^2_{D,S}}{m^2_{Z'}}}.\\
\Gamma_{Z'\rightarrow T\bar{T}}&=&\frac{g^2m_{Z'}}{72\pi(3-t^2_W)}\left[\left((3-2t^2_W)^2+4t^4_W\right)\left(1-\frac{m^2_T}{m^2_{Z'}}\right)
-12t^2_W(3-2t^2_W)\frac{m^2_T}{m^2_{Z'}}\right]\sqrt{1-\frac{4m^2_T}{m^2_{Z'}}}.
\end{eqnarray}
Decay widths in bosonic channels:
\begin{eqnarray}
\Gamma_{Z'\rightarrow W^+W^-}&=&\frac{g^2m_{Z'}(1-t^2_W)^2}{192\pi(3-t^2_W)}\left(1-\frac{4m^2_W}{m^2_{Z'}}\right)^{\frac{3}{2}}
\left(1+\frac{20m^2_W}{m^2_{Z'}}+\frac{12m^4_W}{m^4_{Z'}}\right)\simeq\frac{g^2m_{Z'}(1-t^2_W)^2}{192\pi(3-t^2_W)}.\\
\Gamma_{Z'\rightarrow ZH}&=&\frac{g^2m_{Z'}(1-t^2_W)^2}{192\pi(3-t^2_W)}F\left(\frac{m_Z}{m_{Z'}},\frac{m_H}{m_{Z'}}\right)
\left[F^2\left(\frac{m_Z}{m_{Z'}},\frac{m_H}{m_{Z'}}\right)+\frac{12m^2_Z}{m^2_{Z'}}\right]\simeq\frac{g^2m_{Z'}(1-t^2_W)^2}{192\pi(3-t^2_W)}.\\
\Gamma_{Z'\rightarrow H\eta}&=&\frac{g^2m_{Z'}(v/f)^2}{24\pi(3-t^2_W)t^2_{2\beta}}F^3\left(\frac{m_H}{m_{Z'}},\frac{m_{\eta}}{m_{Z'}}\right).\\
\Gamma_{Z'\rightarrow Y\eta}&=&\frac{g^2m_{Z'}(v/f)^2}{384\pi c^4_W}F\left(\frac{m_Y}{m_{Z'}},\frac{m_{\eta}}{m_{Z'}}\right)
\left(2+\frac{(m^2_{Z'}+m^2_Y-m^2_{\eta})^2}{4m^2_{Z'}m^2_Y}\right).
\end{eqnarray}
\ewt
\end{enumerate}


\bibliography{paph_v6}
\bibliographystyle{h-physrev}

\end{document}